\documentclass[a4paper,fleqn,usenatbib,twocolumn]{mnras}

\usepackage{graphicx}	
\usepackage{amsmath}	
\usepackage{amssymb}	
\usepackage{savesym}
\usepackage{natbib}
\usepackage{times}

\usepackage{mathrsfs,dsfont}
\usepackage{multirow}
\usepackage{cleveref}
\usepackage{captcont}
\usepackage{float}
\usepackage{booktabs}
\usepackage{epstopdf}
\usepackage{romannum}
\usepackage{color}
\usepackage{hyperref}
\usepackage{verbatim}

\usepackage[none]{hyphenat}
\usepackage{soul}

\graphicspath{{figures/}}

\def\be{\begin{equation}}
\def\ee{\end{equation}}
\def\kms{{\rm \,km\,s^{-1}}}

\def\Gyr{{\rm \,Gyr}}

\def\Mpc{{\rm \,Mpc}}
\def\kpc{{\rm \,kpc}}

\def\ckpc{{\rm \,ckpc}}

\def\msun{{\,M_\odot}}

\newcommand{\dd}{{\rm d}}

\usepackage{xcolor}

\title[Mergers of Self-Similar Clusters]{Evolution of Splashback Boundaries and Gaseous Outskirts: Insights from Mergers of Self-similar Galaxy Clusters}

\author[Congyao Zhang et al.]{
Congyao Zhang,$^{1}$\thanks{E-mail: cyzhang@astro.uchicago.edu}
Irina Zhuravleva,$^1$
Andrey Kravtsov,$^{1,2,3}$ and
Eugene Churazov$^{4,5}$
\vspace{1mm}
\\
$^1$~Department of Astronomy and Astrophysics, The University of Chicago, Chicago, IL 60637, USA \\
$^2$~Kavli Institute for Cosmological Physics, The University of Chicago, Chicago, IL 60637, USA \\
$^3$~Enrico Fermi Institute, The University of Chicago, Chicago, IL 60637, USA \\
$^4$~Max Planck Institute for Astrophysics, Karl-Schwarzschild-Str. 1, D-85741 Garching, Germany  \\
$^5$~Space Research Institute (IKI), Profsoyuznaya 84/32, Moscow 117997, Russia \\
\vspace{-7mm}
}

\date{Accepted XXX. Received YYY; in original form ZZZ}

\pubyear{2021}

\begin{document}
\label{firstpage}
\pagerange{\pageref{firstpage}--\pageref{lastpage}}
\maketitle

\begin{abstract}

A self-similar spherical collapse model predicts a dark matter (DM) splashback and accretion shock in the outskirts of galaxy clusters while misses a key ingredient of structure formation -- processes associated with mergers. To fill this gap, we perform simulations of merging self-similar clusters and investigate their DM and gas evolution in an idealized cosmological context. Our simulations show that the cluster rapidly contracts during the major merger and the splashback radius $r_{\rm sp}$ decreases, approaching the virial radius $r_{\rm vir}$. While $r_{\rm sp}$ correlates with a smooth mass accretion rate (MAR) parameter $\Gamma_{\rm s}$ in the self-similar model, our simulations show a similar trend with the total MAR $\Gamma_{\rm vir}$ (includes both mergers and $\Gamma_{\rm s}$). The scatter of the $\Gamma_{\rm vir}-r_{\rm sp}/r_{\rm vir}$ relation indicates a generally low $\Gamma_{\rm s}\sim1$ in clusters in cosmological simulations. In contrast to the DM, the hot gaseous atmospheres significantly expand by the merger-accelerated (MA-) shocks formed when the runaway merger shocks overtake the outer accretion shock. After a major merger, the MA-shock radius is larger than $r_{\rm sp}$ by a factor of up to $\sim1.7$ for $\Gamma_{\rm s}\lesssim1$ and is $\sim r_{\rm sp}$ for $\Gamma_{\rm s}\gtrsim3$. This implies that (1) mergers could easily generate the MA-shock-splashback offset measured in cosmological simulations, and (2) the smooth MAR is small in regions away from filaments where MA-shocks reside. We further discuss various shocks and contact discontinuities formed at different epochs of the merger, the ram pressure stripping in cluster outskirts, and the dependence of member galaxies' splashback feature on their orbital parameters.

\end{abstract}



\section{Introduction} \label{sec:introduction}

Galaxy clusters are important laboratories for studying galaxy and large-scale structure formation theories (see, e.g., \citealt{Kravtsov2012,Vikhlinin2014} for reviews). Observations reveal that the properties of stellar and hot gaseous components of many clusters are quite regular within $r_{500}$\footnote{The $r_{500}$ represents the radius within which the mass density is 500 times the critical density at the cluster redshift.}, especially outside of the innermost region (e.g., $\gtrsim 0.15r_{500}$). This regularity is reflected in the low scatter of scaling relations between the cluster's observables, e.g., X-ray luminosity, temperature, and Sunyaev-Zel'dovich (SZ) decrement (see \citealt{Kravtsov2012} and \citealt{Giodini2013} for reviews), providing a basis for cluster cosmology (see, e.g., \citealt{Vikhlinin2009,Mantz2014} and also \citealt{Allen2011,Pratt2019}).

From a theoretical point of view, the baseline expectation of cluster's scaling relations and other properties (e.g., entropy profiles; see \citealt{Tozzi2001,Voit2005}) is provided by the self-similar spherical collapse model \citep{Gunn1972,Fillmore1984,Bertschinger1985}. Several predictions of the model have been qualitatively confirmed by cosmological simulations of clusters, including the existence of caustics in the collisionless dark matter (DM) density distribution and accretion shock at the boundary of the intracluster medium (ICM). However, a key non-linear process accompanying the cluster formation, namely mergers of clusters, is missing. This caveat significantly limits the utility of the self-similar model in interpreting observations and/or cosmological simulations.

In the hierarchical formation scenario, both mergers and smooth accretion contribute significantly to the cluster mass assembly \citep[e.g.,][]{Genel2010}. By definition, the mergers involve accretion of objects having comparable or moderately smaller masses and, therefore, are accompanied by (1) non-linear perturbations of mass and potential distributions on a cluster dynamical time-scale and (2) supersonic gas flows inside the ICM. The former results in significant matter redistributions, while the latter creates shocks and contact discontinuities (CDs) often observed in merging clusters (see \citealt{Markevitch2007} for a review).

Since mergers involve highly non-linear processes, their theoretical study requires numerical modeling. Both cosmological and idealized (non-cosmological) merger simulations have been used extensively in such studies in the past. The former gives an opportunity to understand the merger process in ``realistic'' cosmological contexts \citep[e.g.,][]{Nagai2003,Paul2011,Iapichino2017,Zinger2018b,Wittor2019,Shi2020} and study the mergers' statistical properties (e.g., merger rate, see \citealt{Fakhouri2008}; pairwise-velocity distribution, see \citealt{Dolag2013}).
The idealized simulations, in contrast, offer a unique way to study merger processes in a controlled setting \citep[e.g.,][]{Roettiger1997,Ricker1998,Poole2006}. The clusters therein are usually assumed to be spherical and in equilibrium in the initial conditions and the parameters of the merger, such as its mass ratio and orbital parameters, can be set by hand (i.e., the merger process is isolated from the cluster formation history). Therefore, this approach is suitable for (1) modeling individual observed clusters (e.g., the ``Bullet'' and ``El Gordo'' clusters; see \citealt{Springel2007,Zhang2015}, respectively) to figure out their merger configurations; and (2) exploring some aspects of the merger physics, including, e.g., merger-driven gas mixture \citep[e.g.,][]{Mitchell2009,Zuhone2011,Vijayaraghavan2015}, baryonic effects on merger time-scale \citep{Zhang2016}.

Although cluster mergers have been extensively studied using the aforementioned approaches, they mostly focused on the inner regions of the clusters, within the virial radius $r_{\rm vir}\ (\simeq2r_{500}$). The role of the mergers in shaping the cluster outskirts, however, is still poorly understood. Both simulation approaches have limitations in this respect. In cosmological simulations, the spatial resolution is usually poor in low-density regions far from the cluster center (e.g., $\gtrsim r_{\rm vir}$). It is also non-trivial to trace an individual merger event to study how it affects the cluster outskirts, where the environment is highly chaotic. The idealized merger simulations, however, face a more serious issue: the initial density profiles of the clusters are always oversimplified at $r\gtrsim r_{\rm vir}$ \citep[see, e.g.,][]{Kazantzidis2004b}. The smooth mass accretion that might be dynamically important in the outer cluster regions is also absent in such models.

A plethora of physical processes accompanying mergers can affect the properties of DM and gas in the cluster outskirts. For example, \citet{Zhang2019b} found that bow shocks that arise during mergers detach from the merging clusters that initially drive them and propagate all the way to larger cluster radii. A ``habitable zone'' populated by these runaway merger shocks exists beyond $\sim r_{500}$ (see \citealt{XZhang2020} for an observational example). These runaway shocks eventually collide with the external accretion shock and dramatically change the morphology of the ICM \citep{Zhang2020a}. Specifically, a salient forward shock, known as the {\it Merger-accelerated Accretion shock} (or MA-shock), is formed in such process. It overtakes and replaces the accretion shock as the new outer boundary of the ICM. A giant CD is also expected to appear near the cluster virial radius. In observations, such a CD-candidate has been recently identified in the outskirts of the Perseus cluster \citep{Walker2020,Zhang2020b}. However, we note that the theoretical scenario described above is mostly obtained from the one-dimensional (1D) models and simulations. For robust predictions, it is important to perform three-dimensional (3D) simulations and to quantitatively explore how a merger shapes the cluster's atmosphere near and beyond the virial radius.

Cluster mergers also affect the DM distribution in the cluster periphery. The radius of the outermost DM caustic, or the splashback radius $r_{\rm sp}$, has been proposed as a new definition of the boundary of the DM halos \citep[e.g.,][]{Diemer2014,Adhikari2014,More2015}. These and subsequent studies of cosmological simulations showed that the splashback radius strongly depends on the total mass accretion rate (MAR; see also \citealt{Diemer2017,Mansfield2017}) that includes mass accreted from both mergers and periods of smooth accretion. Calculations based on the self-similar model predict that a stronger smooth accretion tends to diminish the splashback radius \citep{Fillmore1984,Adhikari2014,Shi2016}. It shows the same trend to that of the cosmological simulation results, but the quantitative dependence predicted by the self-similar model is considerably stronger than found in the simulations \citep{Diemer2017}. It is unclear what role the mergers, in particular, major mergers play in this process.

In addition, cosmological simulations show that the boundaries of the gaseous and DM halos are at different radii in galaxy clusters. The accretion shock radii are on average larger than the splashback radii by a factor of $\simeq1.5-2.5$ \citep{Molnar2009,Lau2015,Zinger2018a,Walker2019,Aung2020}. This is very different from the self-similar model predictions \citep{Shi2016}, where the splashback and external shock radii are very close to each other. Recently, \citet{Zhang2020a} suggested that the mergers and the formation of MA-shocks may cause an offset between external shock and splashback boundary. However, their arguments have not been directly tested yet in full 3D simulations.

In this study, we propose a novel approach to numerically explore the effects of mergers on the outskirts of clusters, whereby self-similar profiles are used to initialize clusters before the merger. Such \textit{idealized cosmological simulations} have several advantages when studying how and to what extent the mergers reshape the outskirts of galaxy clusters. First, the self-similar solution provides a self-consistent initial setup from the core to the outer region. The smooth accretion, as well as accretion shocks, are automatically included in the model. It thus provides a unique chance to investigate an individual merger event in a setting closest to the real cosmological conditions. Secondly, the splashback radius is well defined in our undisturbed cluster, providing an important baseline for quantifying how it changes during the merger. In this sense, our simulations bridge the spherical self-similar models and cosmological simulations of cluster formation and shed light on how nonlinear processes during mergers redistribute mass and break the self-similarity of the initial setup.

In our study, we will distinguish the smooth MAR of galaxy clusters that corresponds to the accretion-rate parameter involved in the self-similar model ($\Gamma_{\rm s}$; see Eq.~\ref{eq:mar}, and also \citealt{Shi2016}) and a conventional total MAR ($\Gamma_{\rm vir}$; see Eq.~\ref{eq:mar_vir}) commonly used in the literature \citep[e.g.,][]{Diemer2014,Mansfield2017}. The latter includes contributions of the smooth accretion, mergers, and pseudo-evolution of mass \citep[e.g.,][]{Diemer2013}. We find that the mergers (particularly major mergers) and the smooth accretion play very different roles in shaping properties of galaxy clusters and argue that distinguishing their effects is crucial for studying the evolution of the splashback and accretion/MA- shock radii and their dependence on the MAR.

This paper is organized as follows. In Section~\ref{sec:method}, we describe the model and simulation method adopted in this work. The main results from our simulations are presented in Sections~\labelcref{sec:results:dm,sec:results:gas}, where we investigate how the DM halos and gaseous atmospheres evolve in the merging clusters, respectively. In Section~\ref{sec:discussions}, we discuss the implications of our merger simulations in galaxy clusters (i.e., constraints on the smooth MAR) and their member galaxies (i.e., $r_{\rm sp}$ measurements in galaxy observations and ram-pressure stripping at large cluster radii). In Section~\ref{sec:conclusions}, we summarize our conclusions.

Throughout the paper, we assume an Einstein-de Sitter universe with $\Omega_{\rm m}=1$, $\Omega_{\rm b}=0.1$, and $H_0=70\kms\Mpc^{-1}$, unless stated otherwise. Correspondingly, the virial radius $r_{\rm vir}$ is defined as the radius enclosing the density contrast of $\Delta_{\rm vir}=18\pi^2$ relative to the mean density of the universe, $\rho_{\rm m}$, which in this cosmology is equal to the critical density of the universe $\rho_{\rm c}=\rho_{\rm m}$. Although the adopted Einstein-de Sitter cosmology is far from the parameters describing the real universe, the focus in this study is on the universal dynamical processes  accompanying mergers of clusters with the structure predicted by the self-similar model. The results apply to mergers in any cosmological model, and the use of the Einstein-de Sitter cosmological model is appropriate.

\section{Simulating mergers of self-similar clusters} \label{sec:method}

We simulate mergers between two self-similar galaxy clusters in the cosmological background. A detailed description of our simulation method is presented in this section. Specifically, we show how we model self-similar clusters and set the initial conditions for the merger simulations in Sections~\ref{sec:method:cluster} and \ref{sec:method:ic}, respectively.

All our simulations are performed using the moving-mesh code \textsc{Arepo} \citep{Springel2010,Weinberger2020}, which provides advantages of high computational efficiency and sufficient treatments of the fluid instabilities and discontinuities. In these simulations, we ignore any astrophysical non-adiabatic processes (e.g., radiative cooling) and focus solely on the gravitational and hydrodynamic effects. We stored $40$ snapshots in each run to guarantee a sufficient time resolution for our post-analysis. They are uniformly distributed in the cosmic time with the time interval of $\simeq0.2\Gyr$.

\subsection{Modeling self-similar clusters} \label{sec:method:cluster}

\begin{figure}
\centering
\includegraphics[width=0.9\linewidth]{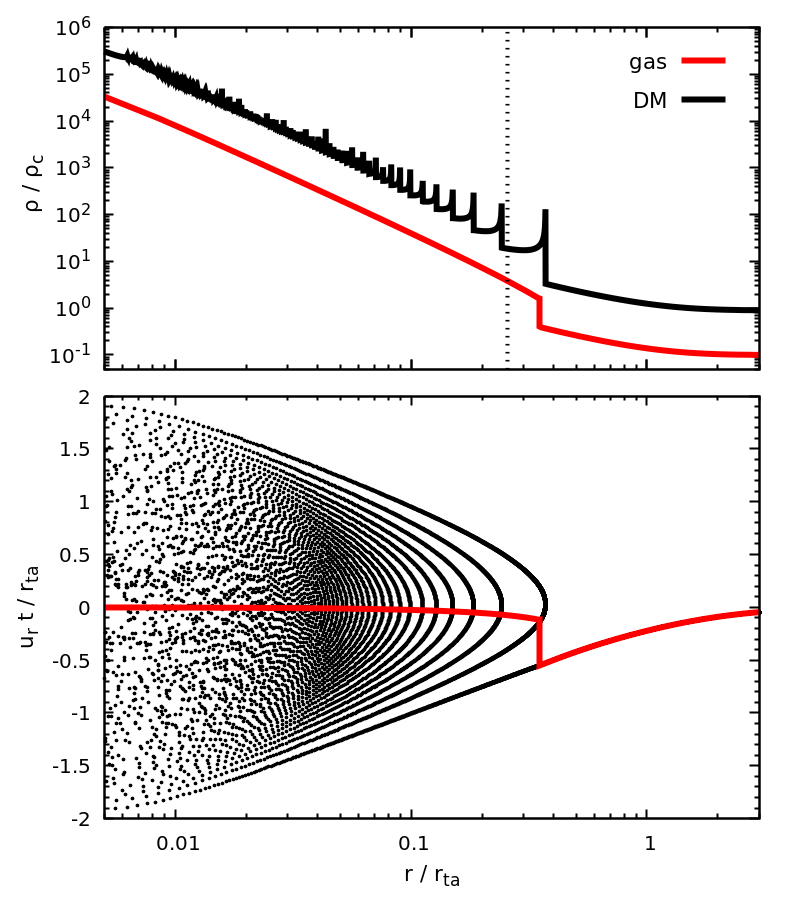}
\caption{\textit{Top panel:} DM and gas density profiles (in units of the critical density of the universe $\rho_{\rm c}$) in the self-similar spherical collapse model with $\Gamma_{\rm s}=1$. The vertical dotted line indicates the cluster virial radius. \textit{Bottom panel:} phase-space distribution (radius vs. radial velocity) of the DM particles overlaid with the gas radial velocity profile (red line). This figure shows an example of the initial profiles of our self-similar cluster implemented in the simulations (see Section~\ref{sec:method:cluster}).}
\label{fig:init_profs}
\end{figure}

Each cluster in our simulations is initially modeled by a self-similar solution of the spherical collapse model \citep{Fillmore1984,Bertschinger1985,Adhikari2014,Shi2016}. In such solutions, a cluster is determined by two parameters: a constant smooth MAR parameter $\Gamma_{\rm s}$ and an initial turnaround radius $r_{\rm ta}(z_{\rm ini})$ at the starting redshift $z_{\rm ini}$ of the simulation. The evolution of the cluster mass $M(z)$ with redshift $z$ therefore follows
\be
M(z) = M(z_{\rm ini})\left[\frac{a(z)}{a(z_{\rm ini})}\right]^{\Gamma_{\rm s}},
\label{eq:mar}
\ee
where $a(z)$ is the scale factor. We note that in Eq.~(\ref{eq:mar}), the cluster grows through only smooth mass accretion (no contribution from mergers at all). Fig.~\ref{fig:init_profs} shows an example of the scaled gas and DM density (and radial velocity) profiles in a cluster with $\Gamma_{\rm s}=1$. The cluster's self-similarity guarantees the phase-space distribution of its DM particles is fully determined in the model (see the bottom panel in Fig.~\ref{fig:init_profs} and more technical details in Appendix~\ref{sec:appendix:simulation:dm}). In all our simulations, we fix the isothermal sound speed of the intergalactic medium to $30\kms$, much lower than that of the ICM.

\begin{figure*}
\centering
\includegraphics[width=0.8\linewidth]{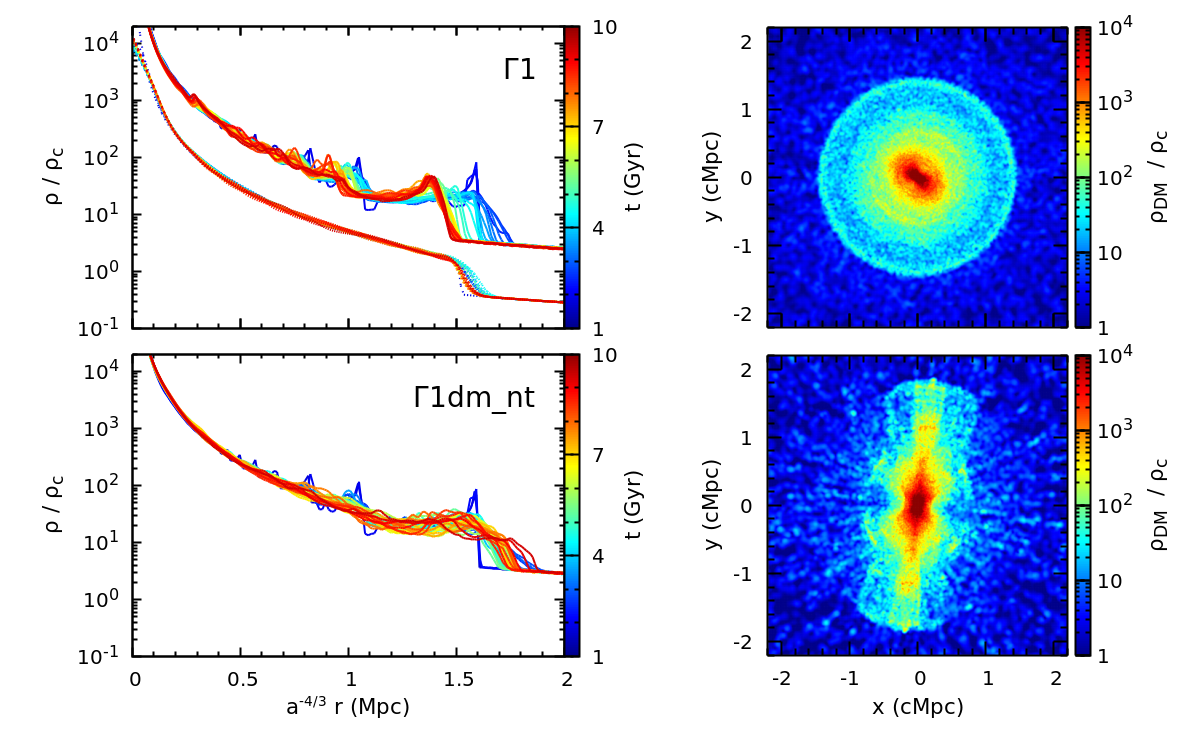}
\caption{Comparison of the isolated clusters evolved in the simulations $\rm\Gamma1$ (with initial tangential velocity components; top panels) and $\rm\Gamma1dm\_nt$ (without initial tangential velocity components; bottom panels). Left panels show their DM (solid lines) and gas (dotted lines) density profiles. Line color encodes the cosmic time. The radius is scaled by $a(t)^{1+\Gamma_{\rm s}/3}$ to highlight the deviations of the clusters from the self-similarity. Right panels show the DM-density slices of the clusters in the $x-y$ plane ($h=0$) at redshift $z=0$. This figure shows that the initial tangential velocity of the DM particles efficiently suppresses the radial-orbit instabilities and maintains the cluster's self-similarity (see Section~\ref{sec:method:cluster}). }
\label{fig:prof_evo_one_halo}
\end{figure*}

The 1D self-similar model accounts for only radial components of both the DM and gas velocities. For the initial conditions in 3D simulations, we artificially include an additional tangential velocity for the DM particles as
\be
u_{\rm t}(r)= \sqrt{\eta \frac{GM(r_{\rm ta})}{r_{\rm ta}}\Big[\zeta(r/r_{\rm ta})^2+1-\zeta\Big]},
\label{eq:init_vt}
\ee
where $G$ is the gravitational constant, $M(r_{\rm ta})$ represents the cluster mass within $r_{\rm ta}$ at redshift $z_{\rm ini}$, to avoid radial-orbit instabilities that may develop otherwise \citep{Antonov1973,Vogelsberger2009}. The velocity direction of each particle is set randomly in the tangential plane so that the total angular momentum of the DM halo is zero. We fix $\eta=0.2$ and $\zeta=0.9$ when $r<r_{\rm ta}$ otherwise $\zeta=0$ in our simulations. We have tested different parameter settings (e.g., $\eta=0.1$ and $0.3$ and/or $\zeta=0.8$ when $r<r_{\rm ta}$) and confirmed that the simulation results do not change significantly.

Fig.~\ref{fig:prof_evo_one_halo} compares the evolution of the isolated clusters with (top panels) and without (bottom panels) the initial DM tangential velocity. One can see that the DM halos evolve into very different shapes in these two cases (also cf., fig.~4 in \citealt{Vogelsberger2009}). The initial tangential velocity is important for maintaining the cluster's stability, particularly in the outer regions. In the test run $\rm\Gamma1$ (top panels; for simulation runs, see Table~\ref{tab:sim_params}), the scaled gas profile evolves closely to the self-similar prediction throughout the simulation. The DM profile shows some deviations from self-similarity during the early stages, but the evolution becomes close to self-similar prediction after $t\simeq 5$ Gyrs. The early deviations from self-similarity are mainly due to the added tangential velocities of the DM particles and the corresponding dynamical adjustment at the start of the simulation.

\begin{figure*}
\centering
\includegraphics[width=0.9\linewidth]{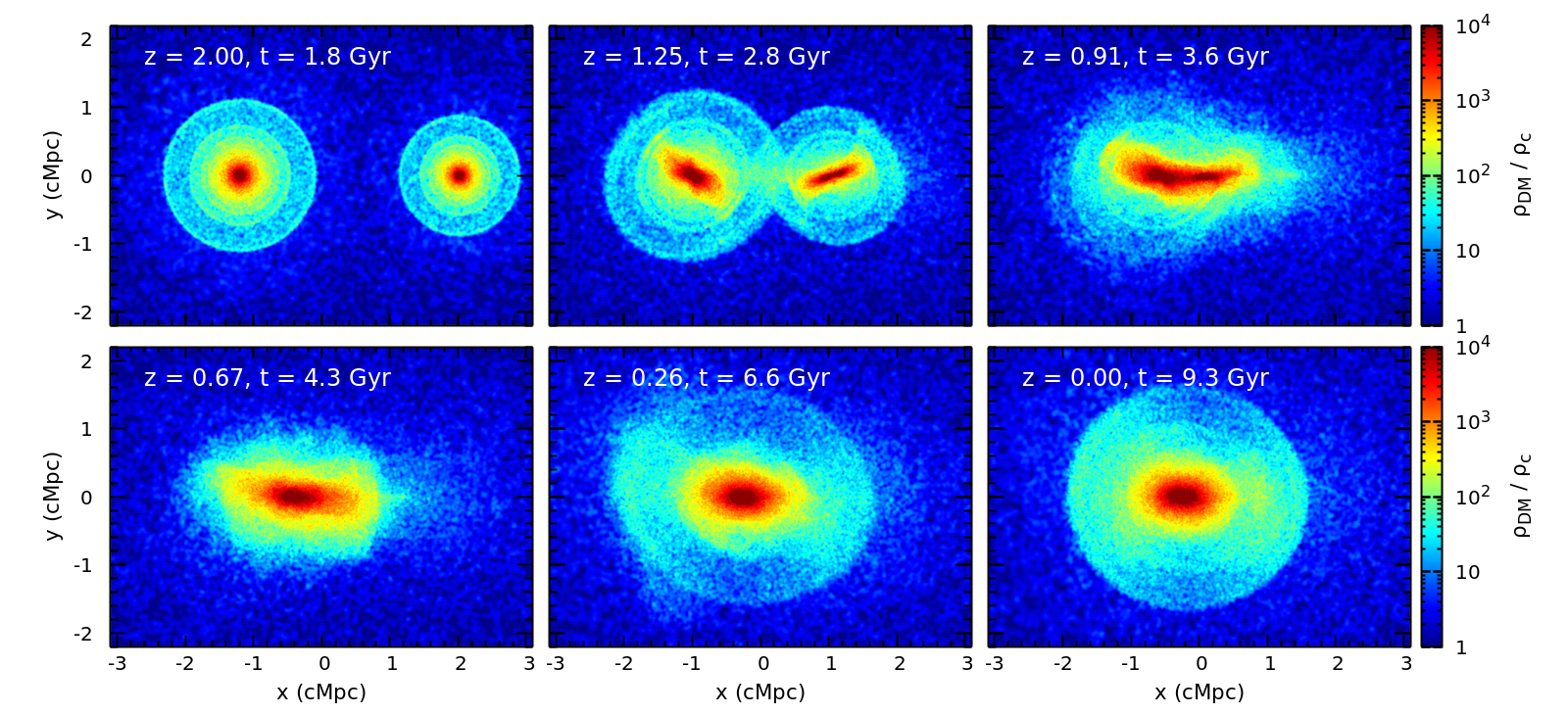}
\caption{Evolution of the DM density slices (in units of $\rho_{\rm c}$) in the $x-y$ plane ($h=0$) in the DM-only simulation $\rm\xi2V10\Gamma1dm$. This figure illustrates a major head-on merger between two DM halos. While approaching each other, these two (ellipsoidal) halos tend to be aligned along the merger axis by the gravitational torque. They experience a rapid contraction after the core passage due to the dramatic change of the gravitational potential. A genuine DM outermost caustic surface  emerges after $t\simeq5\Gyr$ (see Section~\ref{sec:results:dm}). }
\label{fig:slice_dm_pden}
\end{figure*}

The cluster's outermost DM caustic and accretion shock are well-resolved in this run. The location of the splashback radius matches the analytical prediction during later stages of evolution to better than $\sim10$ per cent. Morphologically, the cluster core has a bar-like shape (see the top-right panel), which emerges as a result of gravitational instability at $\Delta t\lesssim0.5\Gyr$ after the start of the simulation. Its orientation, however, is uncontrollable and rather random in our simulations (see also Fig.~\ref{fig:slice_dm_pden}). The cluster's outer region is not affected by this instability and maintains its spherical shape throughout the simulation (compare to the bottom panels in Fig.~\ref{fig:prof_evo_one_halo}). These results demonstrate the robustness of our simulations for modeling the outskirts of the self-similar clusters.

\begin{table*}
\centering
\begin{minipage}{0.95\textwidth}
\centering
\caption{Parameters of simulations (see Section~\ref{sec:method}).}
\label{tab:sim_params}
\begin{tabular}{lcccccc@{}c@{}c@{}c}
  \hline
  IDs\footnote{The IDs of our simulation runs, which are written as $\xi a$V$bc$$\Gamma d$, where $a$ indicates the merger mass ratio, $b$ and $c$ represent the initial relative velocity in the $x-$ and $y-$ axes (`h' is short for 0.5), $d$ is the smooth MAR parameter $\Gamma_{\rm s}$. The suffixes `dm' and `hres' indicate that the simulations are DM-only and in high gas-mass resolution, respectively. } &
  $\Gamma_{\rm s}$\footnote{The cluster's smooth MAR parameter (see Eq.~\ref{eq:mar}).} &
  \begin{tabular}{@{}c@{}}$r_{\rm ta,1/2}$\footnote{The initial turnaround radii of the main and sub merging clusters.} \\ (Mpc) \end{tabular} &
  $\xi$\footnote{The merger mass ratio between the main and sub clusters.}  &
  $\textbf{v}_0$\footnote{The initial relative velocity between two merging clusters in units of $u_{\rm r}(\kappa_{\rm sep}r_{\rm ta,1}+\kappa_{\rm sep}r_{\rm ta,2})$, where $u_{\rm r}(r)$ is the gas radial velocity profile given by the self-similar model.} &
  $\kappa_{\rm max,1/2}$\footnote{Maximum initial radii of the main and sub clusters in units of their turnaround radii (see a sketch in Fig.~\ref{fig:ic_sketch}).}&
  \begin{tabular}{@{}c@{}}$N_{\rm DM}$\footnote{The total number of DM particles used in the simulations.} \\ ($10^6$) \end{tabular} &
  \begin{tabular}{@{}c@{}}$m_{\rm gas}$\footnote{The mass resolution of gas cells.} \\ ($10^6\msun$) \end{tabular} &
  \begin{tabular}{@{}c@{}}$L_{\rm box}$\footnote{The box size for the simulations in units of comoving Mpc (cMpc).} \\ (${\rm cMpc}$) \end{tabular} &
  \begin{tabular}{@{}c@{}}$M_{\rm vir}(z=0)$\footnote{The final virial mass of the merger remnant at $z=0$.} \\ ($10^{14}\msun$) \end{tabular} \\ \hline
  $\rm\Gamma1dm$        & $1.0$ & $1.0\,/\,$--   & --  & --             & $3.0\,/\,$--  & 7.4 & -- & 28 & 1.31 \\
  $\rm\Gamma1dm\_nt$\footnote{This run uses the same initial parameters as those in the `$\rm\Gamma1dm$', except that the DM particles have zero initial tangential velocity (see a comparison between the runs `$\rm\Gamma1dm$' and `$\rm\Gamma1dm\_nt$' in Fig.~\ref{fig:prof_evo_one_halo}).}
              & $1.0$ & $1.0\,/\,$--   & --  & --             & $3.0\,/\,$--  & 7.4 & -- & 28 & -- \\
  $\rm\xi1V10\Gamma1dm$   & $1.0$ & $1.0\,/\,1.0$  & $1$ & $(1,\,0,\,0)$  & $3.0\,/\,3.0$ & 15  & -- & 28 & 2.00 \\
  $\rm\xi2V10\Gamma1dm$   & $1.0$ & $1.0\,/\,0.79$ & $2$ & $(1,\,0,\,0)$  & $3.0\,/\,3.0$ & 12  & -- & 28 & 1.58 \\
  $\rm\xi4V10\Gamma1dm$   & $1.0$ & $1.0\,/\,0.58$ & $4$ & $(1,\,0,\,0)$  & $3.0\,/\,3.0$ & 9.4 & -- & 28 & 1.35 \\
\hline
  $\rm\Gamma1$         & 1.0 & $1.0\,/\,$--    & --   & --              & $3.0\,/\,$--  & 7.4  & 14 & 28 & 1.32 \\
  $\rm\xi2V10\Gamma1$    & 1.0 & $1.0\,/\,0.79$  & $2$  & $(1,\,0,\,0)$   & $3.0\,/\,3.0$ & 6.8  & 14 & 28 & 1.59 \\
  $\rm\xi2V1h\Gamma1$    & 1.0 & $1.0\,/\,0.79$  & $2$  & $(1,\,0.5,\,0)$ & $3.0\,/\,3.0$ & 6.8  & 14 & 28 & 1.59 \\
  $\rm\xi2V1h\Gamma1\_hres$
             & 1.0 & $1.0\,/\,0.79$  & $2$  & $(1,\,0.5,\,0)$ & $3.0\,/\,3.0$ & 14   & 1.4 & 28 & 1.59 \\
  $\rm\xi2V21\Gamma1$    & 1.0 & $1.0\,/\,0.79$  & $2$  & $(2,\,1,\,0)$   & $3.0\,/\,3.0$ & 6.8  & 14 & 28 & 1.65 \\
  $\rm\xi10V1h\Gamma1$   & 1.0 & $1.0\,/\,0.46$  & $10$ & $(1,\,0.5,\,0)$ & $3.0\,/\,0.6$ & 7.6  & 14 & 28 & 1.37 \\
  $\rm\xi10V1h\Gamma1\_hres$
             & 1.0 & $1.0\,/\,0.46$  & $10$ & $(1,\,0.5,\,0)$ & $3.0\,/\,0.6$ & 9.0  & 1.4 & 28 & 1.39 \\
  $\rm\xi10V21\Gamma1$   & 1.0 & $1.0\,/\,0.46$  & $10$ & $(2,\,1,\,0)$   & $3.0\,/\,0.6$ & 7.6  & 14 & 28 & 1.37 \\
  $\rm\Gamma0.7$         & 0.7 & $0.95\,/\,$--   & --   & --              & $3.0\,/\,$--  & 4.1  & 14 & 28 & 0.942 \\
  $\rm\xi2V1h\Gamma0.7$  & 0.7 & $0.95\,/\,0.75$ & $2$  & $(1,\,0.5,\,0)$ & $3.0\,/\,3.0$ & 6.4  & 14 & 28 & 1.17 \\
  $\rm\xi10V1h\Gamma0.7$ & 0.7 & $0.95\,/\,0.44$ & $10$ & $(1,\,0.5,\,0)$ & $3.0\,/\,0.6$ & 4.2  & 14 & 28 & 0.991 \\
  $\rm\Gamma3$           & 3.0 & $1.2\,/\,$--    & --   & --              & $4.0\,/\,$--  & 8.5  & 14 & 43 & 10.6 \\
  $\rm\xi2V1h\Gamma3$    & 3.0 & $1.2\,/\,0.95$  & $2$  & $(1,\,0.5,\,0)$ & $4.0\,/\,4.0$ & 12   & 14 & 43 & 10.8 \\
  $\rm\xi10V1h\Gamma3$   & 3.0 & $1.2\,/\,0.56$  & $10$ & $(1,\,0.5,\,0)$ & $4.0\,/\,0.6$ & 8.5  & 14 & 43 & 10.8 \\
\hline
\vspace{-7mm}
\end{tabular}
\end{minipage}
\end{table*}

\subsection{Setup of merger initial conditions} \label{sec:method:ic}

We include two self-similar clusters with the same MAR parameter $\Gamma_{\rm s}$ in each of our cluster merger simulations. Their masses and sizes are determined by their initial turnaround radii $r_{\rm ta,1}$ and $r_{\rm ta,2}$ ($r_{\rm ta,1}\geq r_{\rm ta,2}$). The merger mass ratio $\xi(>1)$ in the simulations is defined as the ratio of the initial virial masses of the primary (main) and secondary (sub) merging clusters.

We adopt a Cartesian coordinate system $(x,\ y,\ h)$ while setting up the initial conditions. The centers of the two clusters are initially put on the $x$-axis. Their initial distance is $\kappa_{\rm sep}(r_{\rm ta,1}+r_{\rm ta,2})$, where $\kappa_{\rm sep}$ is fixed to $0.6$. In the initial conditions, the main cluster is set at rest, while the subcluster moves in the $x-y$ plane with the relative velocity $\bf v_0$. The two merging clusters overlap inevitably. Therefore, we set up their gas and DM distributions following two different strategies for major and minor mergers (see an illustration in Fig.~\ref{fig:ic_sketch}). In the minor merger case (i.e., $\xi>4$), we discard the outer part of the subcluster ($r>\kappa_{\rm sep}r_{\rm ta,2}$) and embed the remaining part in the main cluster. In the major merger scenario ($\xi\leq4$), we define a ``background'' region (i.e., the area beyond $\kappa_{\rm sep}r_{\rm ta,1}$ and $\kappa_{\rm sep}r_{\rm ta,2}$ from the centers of the main and sub clusters, respectively), where the gas and DM distributions are a linear combination of those of the two individual clusters (see Appendix~\ref{sec:appendix:simulation:ic} for more details). Note that due to such modifications, the clusters generally need a few $0.1\Gyr$ to relax in the simulations. This does not affect the mergers we are discussing here since the primary pericentric passage usually occurs at $\sim1-2\Gyr$ after the start of the simulation.

To trace the gas mixing during the mergers, we include a scalar field $f_{\rm dye}$ in the simulation. It is initially set as $f_{\rm dye}=1$ in the gas cells belonging to the subcluster ICM and $f_{\rm dye}=0$ in all other areas, and is passively advected with the gas flow during the simulation.

All our merger simulations start from the redshift $z_{\rm ini}=2$ that corresponds to the cosmic time $t=1.8\Gyr$. Non-periodic boundary conditions are adopted. The gravitational softening length of the DM particles is fixed to $5$ comoving kpc (ckpc) throughout the simulations. In the runs including hydrodynamics, the gas-mass resolution reaches $1.4\times10^7\msun$. We have also performed two additional high-resolution simulations with a ten times higher gas mass resolution to (1) check the convergence of our numerical results and (2) explore the fine gaseous structures formed in the cluster outskirts during the merger. In all our simulations, we limit the gas spatial resolution $\gtrsim5{\rm\,ckpc}$.

In this study, we explore a broad range of parameter space of the initial conditions, including the cluster's smooth MAR parameter, $\Gamma_{\rm s}$, and the parameters determining the merger configurations (i.e., $\xi$, $\bf v_0$). Both major and minor mergers with different initial orbital angular momentum are covered in this study. For a robust comparison of these simulations, we fix the initial virial mass of the primary cluster as $\simeq5\times10^{13}\msun$ in all the runs. Three $\Gamma_{\rm s}$ values ($=0.7,\ 1,\ 3$) explored in this work cover typical smooth MARs of galaxy clusters (see Sections~\ref{sec:discussions:mar} and \ref{sec:discussions:mar_eps} for discussion). Table~\ref{tab:sim_params} summarizes the main parameters used in our simulations. Each simulation is labeled by an ID of the form $\xi aVbc\Gamma d$, where $a$ indicates the merger mass ratio, $b$ and $c$ represent the initial relative velocity in the $x-$ and $y-$ axes, $d$ is the smooth MAR parameter $\Gamma_{\rm s}$. The suffixes `dm' and `hres' indicate that the simulations are run with only DM and that in high gas-mass resolution, respectively.

\subsection{Measuring splashback and MA-shock radii} \label{sec:method:radii}

In this study, we determine the splashback and MA- (and accretion) shock radii of merging clusters, $r_{\rm sp}$ and $r_{\rm mas}$, based on their DM-density and gas-entropy profiles, respectively. The center of each cluster is fixed at the location of the deepest potential of the system. Since the merging clusters deviate from a spherical shape, we measure the radii along $10^3$ directions uniformly distributed on the surface of a unit sphere. Along each direction, we estimate the locations of the steepest slopes of the DM-density and gas-entropy profiles to identify $r_{\rm sp}$ and $r_{\rm mas}$, respectively. The bin sizes of the profiles are fixed to $20\kpc$ for all our estimations. To suppress the noise (e.g., caused by the small-scale structures), we smooth the radial profiles with the Gaussian and Savitzky–Golay filters before calculating their gradients (see also \citealt{Mansfield2017}). For each cluster, we present the averaged $r_{\rm sp}$ and $r_{\rm mas}$ over the $10^3$ directions and their scatters (10th--90th percentiles) in this work. The latter reflects the deviations of the caustic/shock surface from spherical geometry (see Figs.~\labelcref{fig:bnd_evo_s1,fig:bnd_evo_s3} for examples, where the shaded area indicates the scatter).

\section{Evolution of DM halos} \label{sec:results:dm}

We begin by exploring the evolution of the DM halos in merging clusters. Since the impact of a minor merger on the DM distribution (including the splashback radius) of the main cluster is minor (see, e.g., Figs.~\labelcref{fig:bnd_evo_dm_mar,fig:slice_m10_V21_all,fig:bnd_evo_s1}), we only focus on the major mergers in this section. However, we note that multiple minor mergers taking place in a short time period may deepen the gravitational potential similarly to a major merger, and, thus, significantly change the DM distribution of the main cluster. The processes we discuss are, therefore, at least in part relevant to such a situation.

\begin{figure*}
\centering
\begin{minipage}[c]{\textwidth}
\centering
\includegraphics[width=0.9\linewidth]{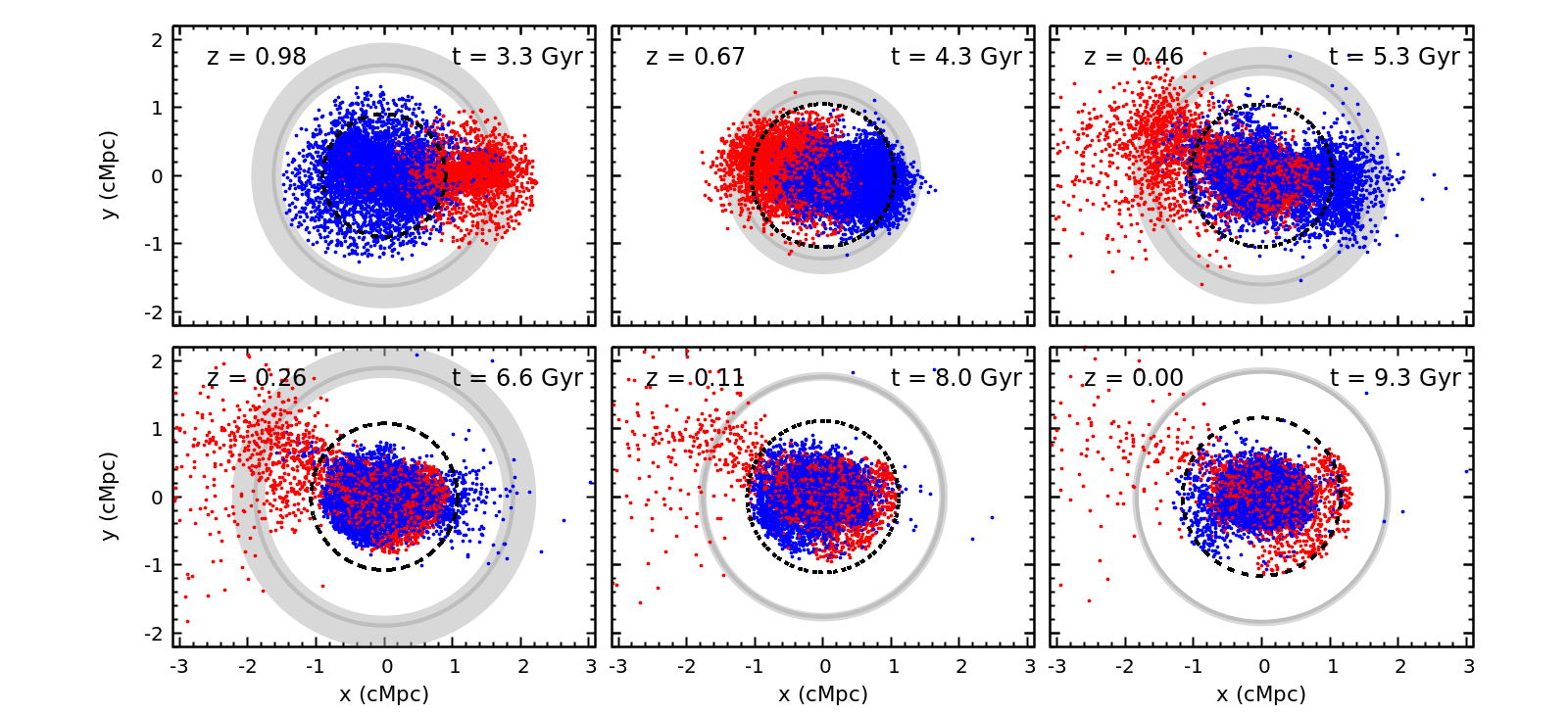}
\caption{Projected spatial distributions of DM particles (along the $h-$axis) in the simulation $\rm\xi2V10\Gamma1dm$. Blue and red colors indicate the particles belonging to the main and sub clusters in the initial condition, respectively. They are randomly selected (1 out of 100) from the simulation within $2r_{\rm vir}$ of the clusters. The black dashed circles indicate the cluster virial radii. The solid grey circles and the accompanying shaded annuli show the azimuthally-averaged splashback radii and their scatters, respectively. Note that we have shifted the particles in this figure (compared to that shown in Fig.~\ref{fig:slice_dm_pden}) so that the potential minimum of the merging system is at the origin of the coordinates. This figure shows how the DM particles mix in a merging process (see also Fig.~\ref{fig:part_dm_rv} for the evolution of their phase-space distribution). Some particles could reach very large cluster radii after the primary pericentric passage, even beyond the splashback radii (see Section~\ref{sec:results:dm}). }
\label{fig:part_dm_xy}
\end{minipage}
\hfill
\vspace{1mm}
\begin{minipage}[c]{\textwidth}
\centering
\includegraphics[width=0.9\linewidth]{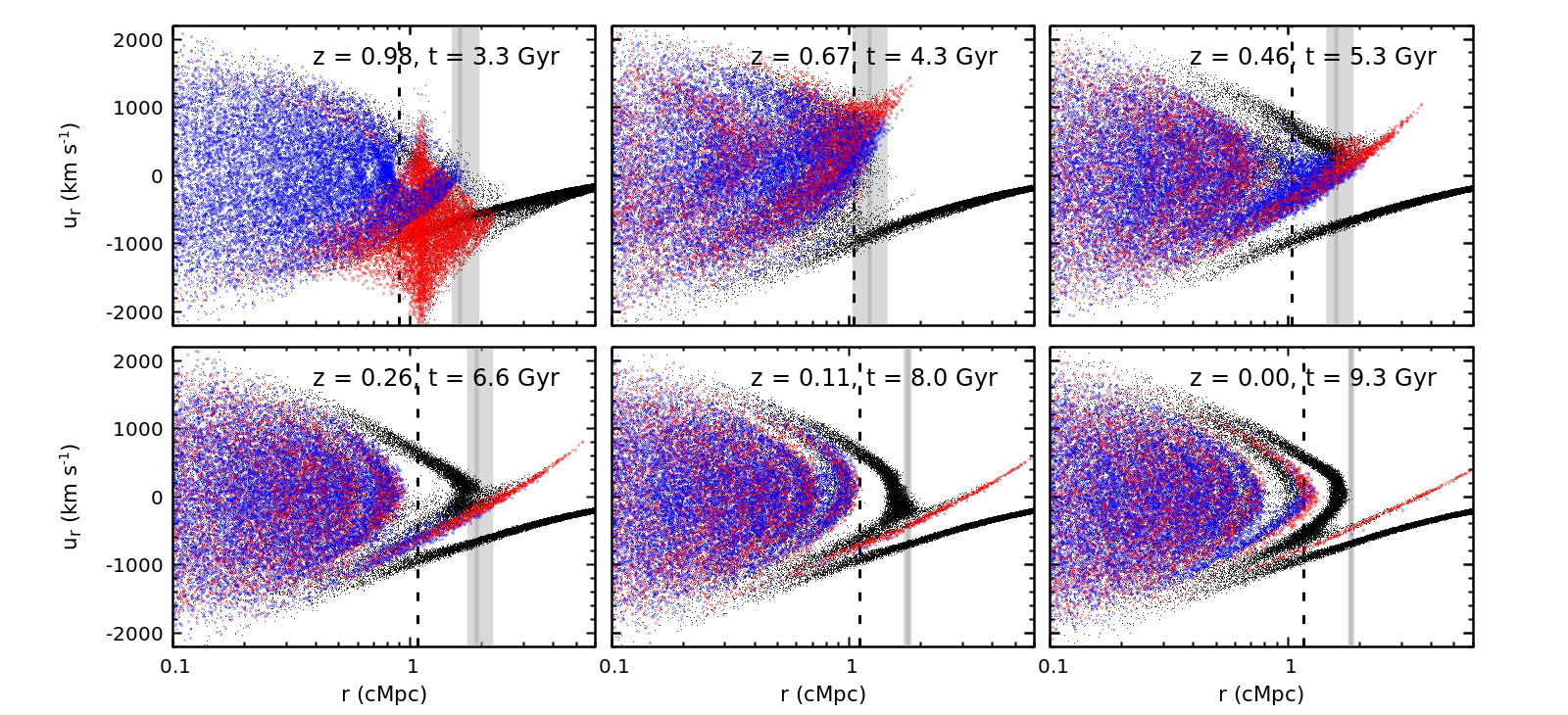}
\caption{Phase-space distributions (radius vs. radial velocity) of DM particles in the same simulation shown in Fig.~\ref{fig:part_dm_xy}. The black dots show the randomly-selected (1 out of every 100) DM particles from the whole system (i.e., including the merging clusters, accreting background particles, etc.). The red and blue particles belong to the main and sub clusters, respectively (similar to Fig.~\ref{fig:part_dm_xy}). They overlap with the black dots in the inner regions. The vertical dashed black lines indicate the cluster virial radii. The solid grey lines and the accompanying shaded areas show the azimuthally-averaged splashback radii and their scatters. This figure illustrates how the DM phase-space distribution evolves during a major merger. It further demonstrates that the splashback radii identified in the DM density profiles provide a good match to those inferred from the DM phase-space distribution in our model (see Section~\ref{sec:results:dm}).}
\label{fig:part_dm_rv}
\end{minipage}
\end{figure*}

Fig.~\ref{fig:slice_dm_pden} shows the time evolution of the DM density slices in our DM-only simulation $\rm\xi2V10\Gamma1dm$. The initial density distribution (at $z=2$) is shown in the first panel. As we mentioned in Section~\ref{sec:method:cluster}, the radial-orbit instabilities drive the DM cores into a bar shape rapidly. While the two dense ``bars'' approach each other, the gravitational torque tends to align them along the merger axis (see the second and third panels in order of time). However, at the moment of the core passage (i.e., primary pericentric passage, $t\simeq3.7\Gyr$), the alignment is sensitive to the initial orientations, orbits, and mass ratio of the two merging clusters (see also Fig.~\ref{fig:slice_m2_V1h_all}). After the pericentric passage, the DM halos penetrate each other. Their apocentric distance is much smaller than their initial separation due to the collective dynamical effects that lead to the dissipation of the kinetic energy and angular momentum of the merging clusters. They eventually merge into one elongated halo at $t\simeq5\Gyr$ ($z\simeq0.5$; see also Fig.~\ref{fig:part_dm_xy}), the relaxation of which lasts for a longer time \citep[see also, e.g.,][]{Poole2006,Nelson2012,Zhang2016}. The relaxation process mainly occurs within the virial radius, where potential fluctuations caused by the merger are the strongest.

\begin{figure*}
\centering
\includegraphics[width=0.9\linewidth]{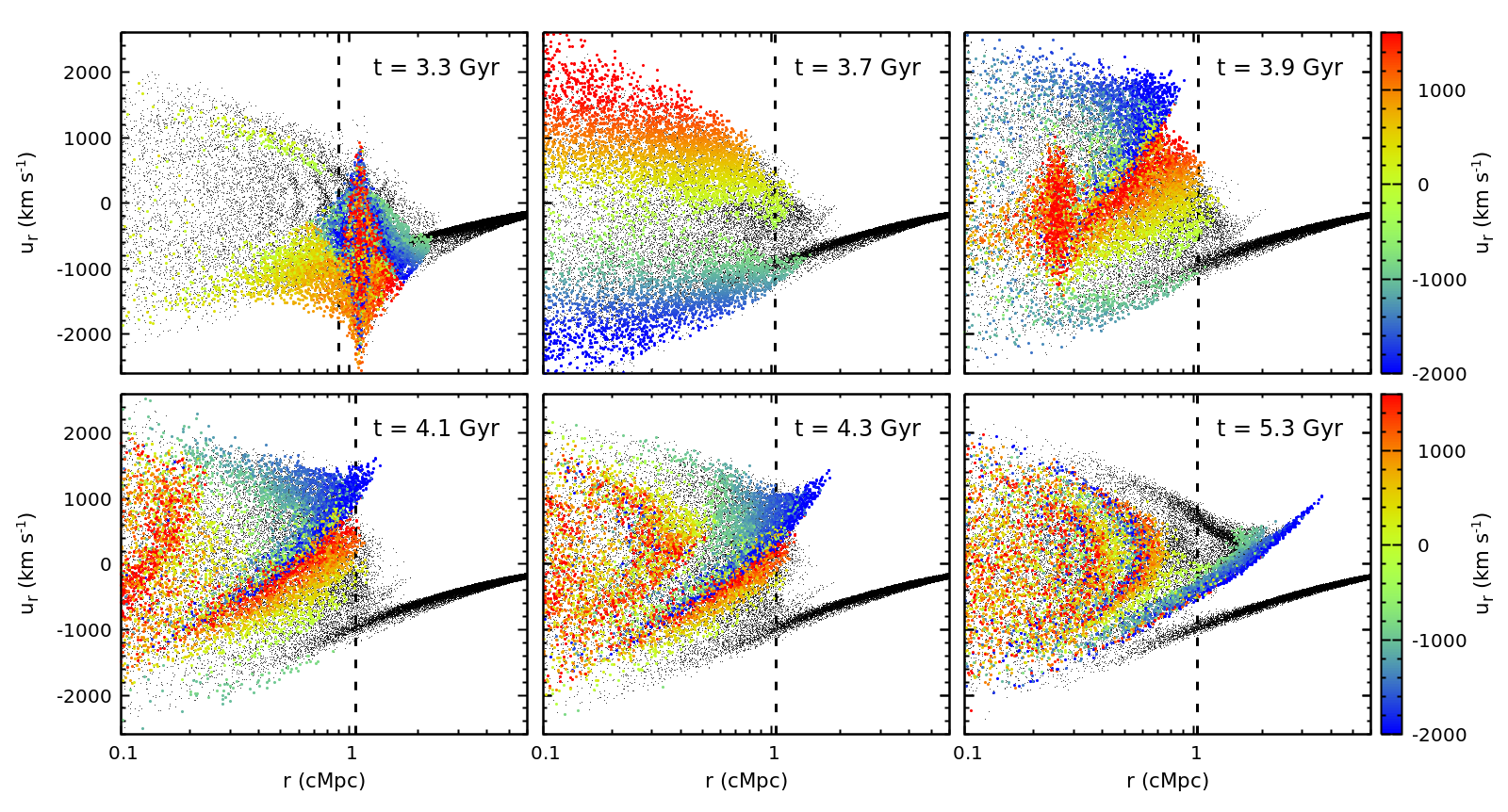}
\vspace{-1mm}
\caption{Similar to Fig.~\ref{fig:part_dm_rv} but highlighting with colors the DM particles that initially belong to the subcluster only. These particles are colored based on their radial velocities at $t=3.7\Gyr$ (the top-middle panel), close to the moment of the core passage. Black points show the distributions of all DM particles in the simulation (same as in Fig.~\ref{fig:part_dm_rv}). The vertical black dashed lines indicate the virial radii.
For some particles among the initially lagging ones from the infalling subhalo (the bluish particles), the gravitational accelerations from the main and sub halos add constructively during the first infall. After the core passage, these particles have high radial velocities and, therefore, they reach very large radii ($\gtrsim 3r_{\rm vir}$ of the merger remnant; see also Fig.~\ref{fig:part_dm_rv}). Meanwhile, the orbit of the subcluster decays fast due to the dissipative dynamical effects associated with the merger (see Section~\ref{sec:results:dm}).}
\label{fig:part_dm_rv_sub}
\end{figure*}

Fig.~\ref{fig:part_dm_xy} shows the projected spatial distributions of the DM particles in the same simulation as in Fig.~\ref{fig:slice_dm_pden}. These particles are randomly selected at the initial time within $2r_{\rm vir}$ of the main (blue) and sub (red) clusters. Their phase-space distributions (radius vs. radial velocity) are correspondingly shown in Fig.~\ref{fig:part_dm_rv} (see also Fig.~\ref{fig:part_dm_rv_sub}). In Figs.~\labelcref{fig:part_dm_xy,fig:part_dm_rv,fig:part_dm_rv_sub}, the origin of the coordinates is set at the position of the deepest potential of the system (different from that shown in Fig.~\ref{fig:slice_dm_pden}). Figs.~\labelcref{fig:part_dm_xy,fig:part_dm_rv} show that most DM particles from the two progenitor halos stay within $r_{\rm vir}$, shortly after the core passage (i.e., $t\gtrsim6\Gyr$). The matter distribution between $r_{\rm vir}$ and $r_{\rm sp}$, particularly in directions outside the merger axis, is mostly determined by the smooth mass accretion (characterized by the parameter $\Gamma_{\rm s}$).

Combining the spatial and velocity information of the particles, we see that some particles that escape to large radii reach the average splashback radius (the solid grey lines) and beyond. For example, in the third panel of Fig.~\ref{fig:part_dm_xy}, approximately $20$ and $4$ per cent of the particles exceed the cluster virial and splashback radii, respectively (see the corresponding panels in Figs.~\labelcref{fig:part_dm_rv,fig:part_dm_rv_sub} and also \citealt{Carucci2014}). In major mergers, bulk velocities of the merging halos are comparable to their internal velocity dispersion. Initially, these large-radii particles lag their host halos and are distributed near the merger axis. They move in the same direction as the bulk of the host halo and feel strong gravitational acceleration when approaching the other cluster. Some of them get ahead of their host (e.g., yellowish particles in the first panel in Fig.~\ref{fig:part_dm_rv_sub}) before the cluster pericentric passage and form a long stream in the phase-space distribution (see also \citealt{Kazantzidis2006,Valluri2007,Vogelsberger2011}). Particles that continue lagging their host until the pericentric passage (the bluish particles in Fig.~\ref{fig:part_dm_rv_sub}) could gain the largest radial velocities, reaching up to $\simeq 5r_{\rm vir}$ of the merger remnant at $t\gtrsim 5\Gyr$. Later, these particles  re-accrete onto the remnant in a new infall stream traced by red particles in the lower panels of Fig.~\ref{fig:part_dm_rv}. In Section~\ref{sec:results:dm:shape}, we will see that these stripped particles affect our measurements of the DM halos' shape.

\subsection{Splashback radii} \label{sec:results:dm:rsp}

The main goal of this study is to explore the evolution of the cluster's splashback radius during mergers. The head-on major merger in the run $\rm\xi2V10\Gamma1dm$ offers a representative example (see Figs.~\labelcref{fig:slice_dm_pden,fig:part_dm_xy,fig:part_dm_rv}). In the pre-merger stage, the cluster's outermost caustic surface is well-preserved. One can even see the intersections of the caustic surfaces from the two merging clusters (the second panel in Fig.~\ref{fig:slice_dm_pden}). When the two clusters are sufficiently close, they experience a rapid contraction due to the dramatic deepening of the gravitational potential (the third and fourth panels in Fig.~\ref{fig:slice_dm_pden}). Such rapid change is a specific feature of the merger process and does not occur in the self-similar evolution with the constant smooth accretion rate (in this case, $\Gamma_{\rm s}=1$).

Despite the significant and rapid dynamical evolution of DM distribution of the merging clusters, there is still a clear splashback boundary of the cluster in the DM density distribution, which becomes closer to the cluster's virial radius\footnote{The splashback at this stage of evolution also becomes close to the radius of the secondary outermost caustic of the DM halo in the self-similar model (see Fig.~\ref{fig:init_profs}), although the two outermost caustics of the merger remnant are well separated when the total MAR $\Gamma_{\rm vir}$ is low (see the last two panels in Fig.~\ref{fig:slice_dm_pden} and also \citealt{Diemer2014,Adhikari2014,Deason2020}).}. At this moment, a fraction of particles near this boundary has not yet reached their apocenters. These are the stripped particles from their parent merging halos, which have high positive radial velocities (the second panel in Fig.~\ref{fig:part_dm_rv}). As these particles move towards the apocenters of their orbits, the outer splashback boundary rapidly increases.

After $\sim1-2\Gyr$ from the core passage (approximately the cluster's dynamical time-scale $\tau_{\rm dyn}$), a well-defined outermost caustic surface of the halo starts to develop (see the fifth panel in Fig.~\ref{fig:slice_dm_pden}). It first becomes most prominent in the directions close to the merger axis since it is formed by the apocentric passages of the particles most accelerated during the merger (the third panel in Fig.~\ref{fig:part_dm_rv}). Shortly after that, a nearly spherical caustic surface emerges. Comparing the distributions of the black and colored points in Fig.~\ref{fig:part_dm_rv}, one can see that, after $t\simeq6\Gyr$, the outermost caustic is dominated by the newly smoothly accreted DM particles.

\begin{figure}
\centering
\includegraphics[width=0.9\linewidth]{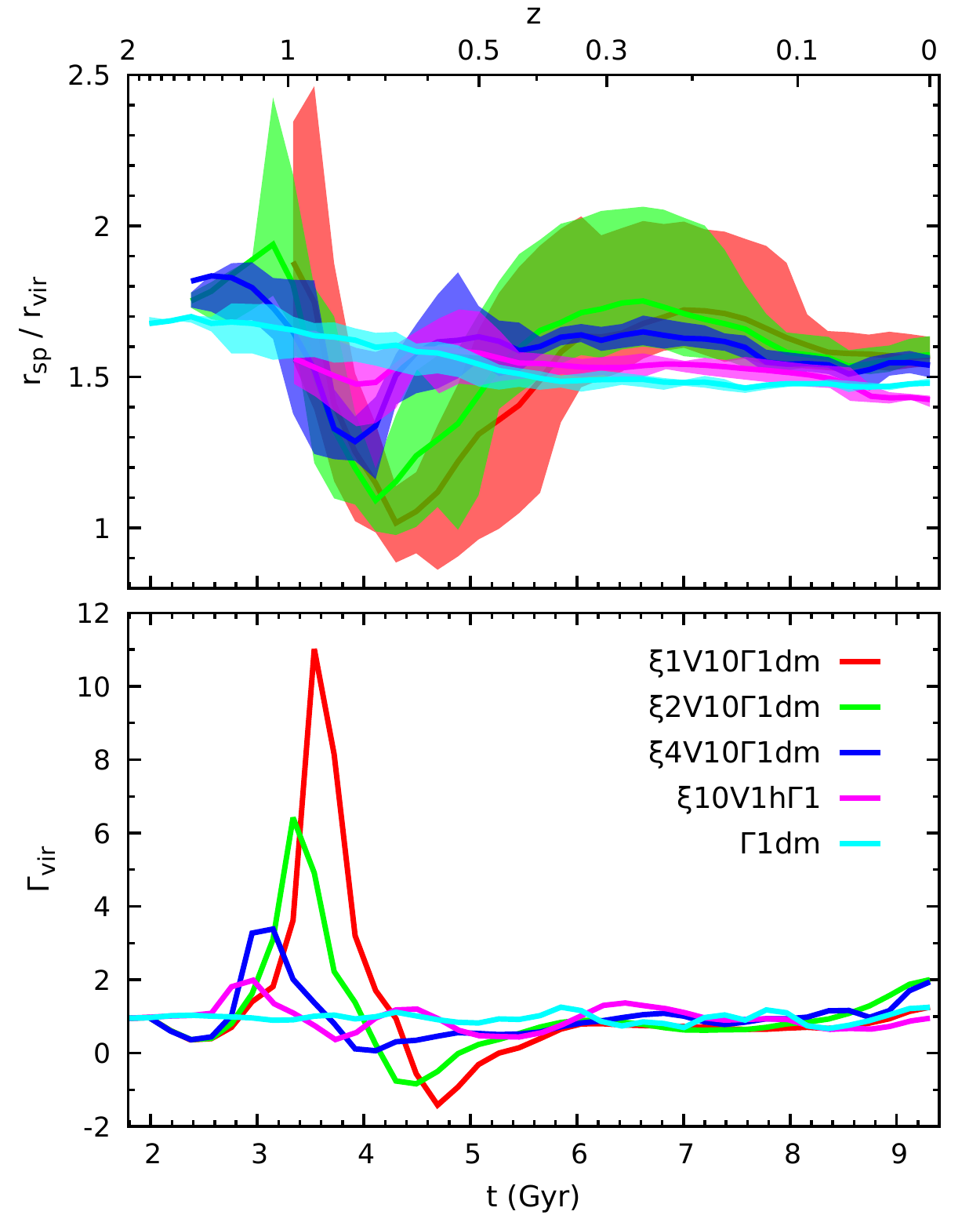}
\caption{\textit{Top panel:} Evolution of the splashback radii in units of $r_{\rm vir}$ in our DM-only simulations. The solid lines show the azimuthally-averaged splashback radii over $10^3$ uniformly distributed directions. The shaded areas indicate the scatters within the 10th--90th percentile range. \textit{Bottom panel:} cluster's MAR measured at the virial radius based on Eq.~(\ref{eq:mar_vir}). There is a strong correlation between $\Gamma_{\rm vir}$ and $r_{\rm sp}/r_{\rm vir}$ but with a $\simeq1\Gyr$ time offset between the moments of maximum $\Gamma_{\rm vir}$ and minimum $r_{\rm sp}/r_{\rm vir}$. This figure illustrates that the major mergers have a strong effect on both $r_{\rm sp}$ and $\Gamma_{\rm vir}$ that could last for a few Gyrs (see Section~\ref{sec:results:dm:rsp}). }
\label{fig:bnd_evo_dm_mar}
\end{figure}

Fig.~\ref{fig:bnd_evo_dm_mar} compares the evolution of the cluster splashback radii $r_{\rm sp}$ in units of $r_{\rm vir}$ in our major-merger simulations with the different merger mass ratio $\xi$. In the top panel of Fig.~\ref{fig:bnd_evo_dm_mar}, the solid lines show  $r_{\rm sp}$ averaged over $10^3$ directions. The 10th--90th percentile region of the splashback radii along these directions is shown by the shaded areas (see Section~\ref{sec:method:radii}). Comparing with Fig.~\ref{fig:part_dm_rv}, one can see that the measurements of the splashback radii based on the slope of the DM density profiles are consistent with those inferred from the DM phase-space distributions.

Major mergers affect the clusters' splashback radii in two main aspects. First, as seen in Fig.~\ref{fig:slice_dm_pden}, the sharp boundary of the DM halo shrinks prominently during the primary pericentric passage. This effect is more significant if $\xi$ is smaller and generally lasts for $1-2\tau_{\rm dyn}$. Secondly, the halo's outermost caustic surface noticeably deviates from a spherical symmetry when $\xi\lesssim2$, as reflected in the large scatters of the red and green curves \citep[see also][]{Mansfield2017}. It is elongated along the merger axis (see Fig.~\ref{fig:slice_dm_pden}), explaining why the averaged $r_{\rm sp}/r_{\rm vir}$ is closer to the lower bound of its scatter. This asymmetry of the splashback surface is the key reason that a fraction of matter and subhalos are outside of the spherically averaged splashback radius \citep[][]{Mansfield2020}. We have also checked the evolution of the splashback radius in simulations of mergers with the different initial orbital parameters. We find that when the initial impact parameter of the subcluster is not too large, as is in the case of most major mergers \citep{Sarazin2002}, the merger mass ratio is the main factor determining the evolution of the azimuthally-averaged $r_{\rm sp}$  (see the blue curves in Fig.~\ref{fig:bnd_evo_s1}).

\begin{figure}
\centering
\includegraphics[width=0.9\linewidth]{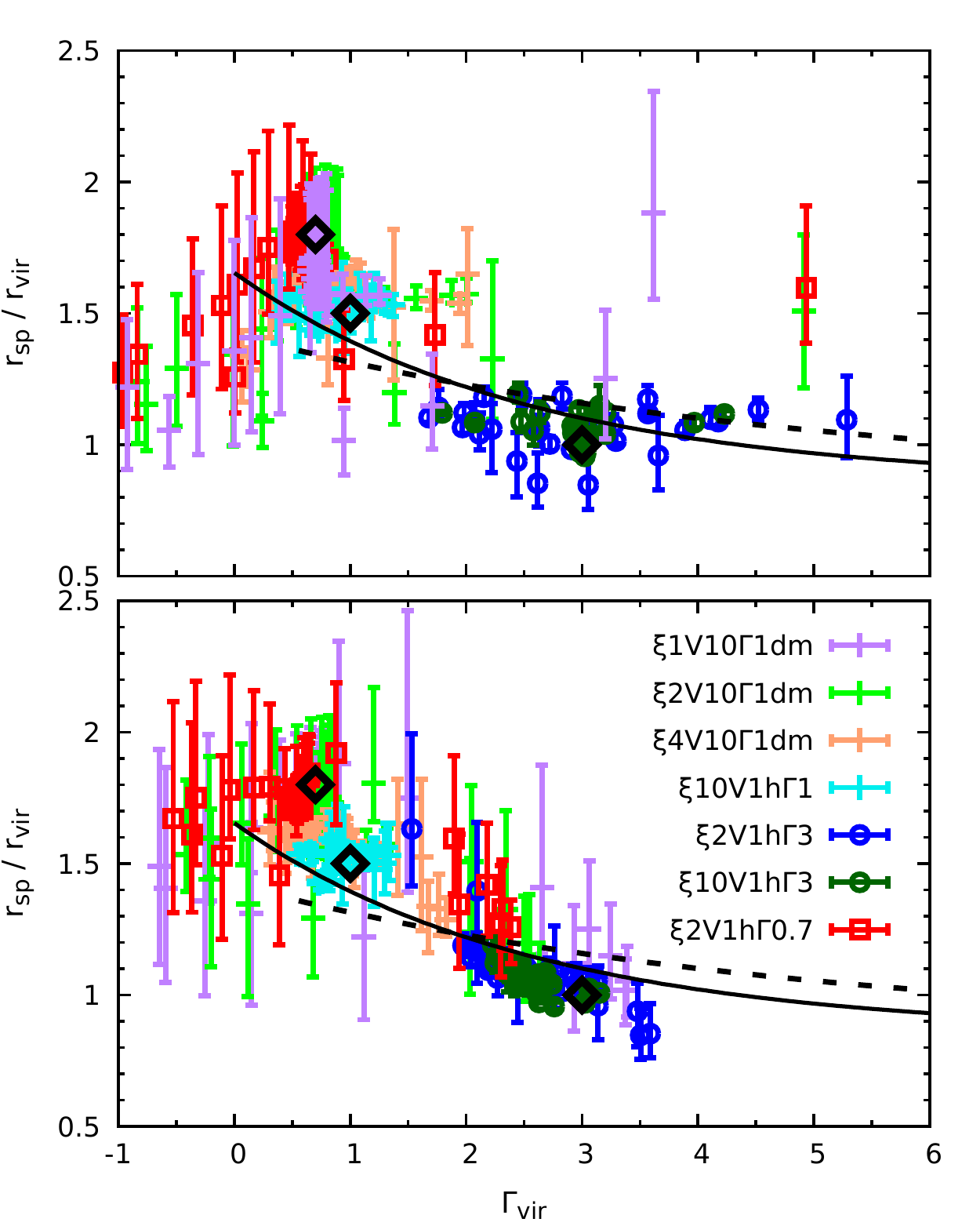}
\caption{$\Gamma_{\rm vir}-r_{\rm sp}/r_{\rm vir}$ diagram measured in our simulations. Each point shows the scaled splashback radius and its azimuthal scatter in one snapshot of the simulation. In the top panel, $\Gamma_{\rm vir}$ is estimated using Eq.~(\ref{eq:mar_vir}). In the bottom panel, $\Gamma_{\rm vir}$ is estimated using the two snapshots with a larger time interval, i.e., setting $a_2=a_1$ and $t(a_1)=t(a_0)+\Delta t$ in Eq.~(\ref{eq:mar_vir}), where $\Delta t=1.2\Gyr$. The three black diamonds show the self-similar solutions for $\Gamma_{\rm s}=0.7,\ 1,\ 3$. The solid and dashed black lines are the best-fit scaling relations from \citet{Diemer2017} and \citet{Mansfield2017}, respectively. This figure shows that major mergers with small $\Gamma_{\rm s}\simeq1$ contribute most to the scatter of the $\Gamma_{\rm vir}-r_{\rm sp}/r_{\rm vir}$ relation. Also, the shape of the correlation is sensitive to the way we estimate $\Gamma_{\rm vir}$ (see Section~\ref{sec:results:dm:rsp}). }
\label{fig:mar_rsp_corr}
\end{figure}

Finally, we discuss the dependence of $r_{\rm sp}/r_{\rm vir}$ on the MAR experienced by the main cluster during a merger. Recall that the $r_{\rm sp}/r_{\rm vir}$ ratio in the self-similar clusters is a function of only the mass accretion parameter $\Gamma_{\rm s}$ \citep{Adhikari2014,Shi2016}. For example, $r_{\rm sp}/r_{\rm vir}$ is $\simeq1.8,\ 1.5,\ 1.0$ when $\Gamma_{\rm s}=0.7,\ 1,\ 3$, respectively. However, during cluster mergers, the MAR generally becomes a function of time and radius. We follow the common choice in the literature and use the logarithmic rate of change of virial mass with scale factor as the measure of MAR \citep[e.g.,][]{Diemer2014,Mansfield2017}:
\be
\Gamma_{\rm vir}(a_1)=\frac{\log{\left[M_{\rm vir}(a_2)/M_{\rm vir}(a_0)\right]}}{\log{(a_2/a_0)}},
\label{eq:mar_vir}
\ee
where $a_0$, $a_1$ and $a_2$ are the scale factors of three successive snapshots\footnote{We emphasize that the evaluation of $\Gamma_{\rm vir}$ depends on the form of Eq.~(\ref{eq:mar_vir}), and is also sensitive to the time interval $\Delta t$ between the snapshots used to interpolate the accretion rate (see, e.g., Fig.~\ref{fig:mar_rsp_corr}). We use $\Delta t=t(a_2)-t(a_0)=0.4\Gyr\sim\tau_{\rm dyn}/4$ to resolve the main features that a merger imprints on the accretion rate curve. However, in full cosmological simulations, coarser $\Delta t\ (\gtrsim\tau_{\rm dyn})$ are usually adapted to suppress the scatter caused by the cluster's chaotic mass accretion history \citep[see, e.g.,][]{Diemer2017}. }.

The bottom panel in Fig.~\ref{fig:bnd_evo_dm_mar} shows the measurements of $\Gamma_{\rm vir}$ in our simulations. Merger leaves a prominent imprint in the evolution of $\Gamma_{\rm vir}$ in the form of a peak with the amplitude of up to $10$ times higher than the initial smooth $\Gamma_{\rm s}$. These peaks are further followed by troughs due to the contraction of the particle distribution during the core passage discussed above. The duration of the peak of $\Gamma_{\rm vir}$ and the subsequent trough last up to $1-2\tau_{\rm dyn}$. Comparison of the top and bottom panels of Fig.~\ref{fig:bnd_evo_dm_mar} shows a strong temporal correlation between $\Gamma_{\rm vir}$ and $r_{\rm sp}/r_{\rm vir}$ but with a $\simeq1\Gyr$ time offset between the moments of maximum $\Gamma_{\rm vir}$ and minimum $r_{\rm sp}/r_{\rm vir}$ (approximately half of the cluster crossing time-scale $\sim\tau_{\rm dyn}$). Interestingly, this correlation shows the same trend as that of the scaling relation $\Gamma_{\rm s}-r_{\rm sp}/r_{\rm vir}$ predicted in the self-similar model. When the cluster's mass growth rate ($\Gamma_{\rm s}$ or $\Gamma_{\rm vir}$) is higher, the ratio $r_{\rm sp}/r_{\rm vir}$ is smaller (compare Figs.~\labelcref{fig:bnd_evo_dm_mar,fig:bnd_evo_s3}). In other words, a small $r_{\rm sp}/r_{\rm vir}$ (e.g., $\simeq1$) could be caused by either a merger effect or a high smooth accretion rate $\Gamma_{\rm s}$ of the cluster or a combination of both.

Fig.~\ref{fig:mar_rsp_corr} shows the $\Gamma_{\rm vir}-r_{\rm sp}/r_{\rm vir}$ diagram, where each point shows $\Gamma_{\rm vir}$ and $r_{\rm sp}/r_{\rm vir}$ (and the scatter of the latter) measured in one simulation snapshot. In the top panel, the $\Gamma_{\rm vir}$ is estimated using Eq.~(\ref{eq:mar_vir}). In the bottom panel, however, we use a different version of $\Gamma_{\rm vir}$, measured from the two snapshots with $\Delta t=1.2\Gyr$ (i.e., setting $a_2=a_1$ and $t(a_1)=t(a_0)+\Delta t$ in Eq.~\ref{eq:mar_vir}). This definition is closer to that used in the analyses of cosmological simulations. The three black diamonds overlaid in the figure correspond to the self-similar solutions with $\Gamma_{\rm s}=0.7,\ 1,\ 3$.
This figure shows that the scaling relation $\Gamma_{\rm vir}-r_{\rm sp}/r_{\rm vir}$ measured in the simulations is sensitive to the definition of $\Gamma_{\rm vir}$. Major mergers mostly increase the scatter of the relation over a wide range of $\Gamma_{\rm vir}$. In the low MAR regime (e.g., $\Gamma_{\rm vir}\lesssim1$), the time lag between $\Gamma_{\rm vir}$ and $r_{\rm sp}/r_{\rm vir}$ (see Fig.~\ref{fig:bnd_evo_dm_mar}) plays an important role in shaping the scaling relation, which is no longer monotonic as expected in the self-similar model (see the top panel). However, if a sufficiently large $\Delta t$ (larger than the time lag) is adopted to estimate the $\Gamma_{\rm vir}$, the dependence is close to be monotonic (see the bottom panel). On the other hand, the scatter of the relation in the high MAR regime (e.g., $\Gamma_{\rm vir}\gtrsim3$), is mostly contributed by the major mergers with small $\Gamma_{\rm s}\ (\simeq1)$ rather than those with the large values (i.e., $\Gamma_{\rm s}\simeq\Gamma_{\rm vir}$). The bottom panel of the figure also shows that for sufficiently high $\Gamma_{\rm s}$ ($\gtrsim3$), the mergers, including the major mergers, only mildly affect the difference between $r_{\rm sp}$ and $r_{\rm vir}$ (see also the blue lines in Fig.~\ref{fig:bnd_evo_s3}), and the scatter in the $\Gamma_{\rm vir}-r_{\rm sp}/r_{\rm vir}$ diagram is relatively narrow.

The scaling relation $\Gamma_{\rm vir}-r_{\rm sp}/r_{\rm vir}$ has been extensively explored in cosmological simulations \citep[e.g.,][]{Diemer2014,More2015,Mansfield2017,Diemer2017}. In Fig.~\ref{fig:mar_rsp_corr}, we overplot the best-fit relations from \citet{Diemer2017} and \citet{Mansfield2017}\footnote{See eq.~(13) in \citet{Mansfield2017} and eqs.~(5)--(7) for $R^{87\%}_{\rm sp}/R_{\rm200m}$ in \citet{Diemer2017}, where we simply fix $\nu_{\rm 200m}=2$. The latter relation is estimated using \textsc{Colossus} python package \citep{Diemer2018}.} as the black solid and dashed lines, respectively. However, we have to emphasize that a quantitative comparison between our idealized and full cosmological simulations is non-trivial. First, the data points shown in Fig.~\ref{fig:mar_rsp_corr} are simply a collection of our simulation results. Their distribution does not include any information on the statistical properties of the mergers (e.g., cluster mass function and merger rate) inferred from the cosmological simulations.
Second, it is not easy to measure the intrinsic parameter $\Gamma_{\rm s}$ for individual clusters in the cosmological simulations. The measured $\Gamma_{\rm vir}$ in cosmological simulations always includes the smooth accretion, mergers, and pseudo-evolution over a long time-scale (see, e.g., \citealt{Diemer2013} and references therein). Nevertheless, we may still constrain the typical smooth MAR parameter $\Gamma_{\rm s}$ in galaxy clusters, as discussed below.

The dark blue and dark green points in Fig.~\ref{fig:mar_rsp_corr} correspond to simulations with a strong smooth accretion $\Gamma_{\rm s}$ (i.e., $\Gamma_{\rm vir}\approx \Gamma_{\rm s}$). The scatter of these points is smaller than those with the lower $\Gamma_{\rm s}$. This means that if $\Gamma_{\rm vir}\simeq \Gamma_{\rm s}\gtrsim3$, one might expect a tight correlation between $\Gamma_{\rm vir}$ and $r_{\rm sp}/r_{\rm vir}$. Cosmological simulations reveal a large scatter in this relation (see, e.g., Fig.~6 in \citealt{Mansfield2017}) that is more consistent with the scatter of the low-$\Gamma_{\rm s}$ cases in our simulations.  This indicates that mergers likely boost $ \Gamma_{\rm vir}$ in cosmological simulations while the smooth accretion $\Gamma_{\rm s}$ is small ($\sim 1$; see also \citealt{Diemand2008}, \citealt{Sugiura2020}, and more discussions in Section~\labelcref{sec:discussions:mar,sec:discussions:mar_eps}).

\subsection{Halo shape and orientation} \label{sec:results:dm:shape}

It is well-known that the shape of the relaxed (or closely relaxed) cluster halos in cosmological simulations can be well approximated by a triaxial ellipsoid \citep[e.g.,][]{Dubinski1991,Jing2002,Allgood2006}. The major axes of the ellipsoid tend to align with the directions of the large-scale filaments \citep[e.g.,][]{Splinter1997,Kasun2005ApJ,YZhang2009}. Cluster major mergers undoubtedly significantly affect halo's shape and orientation. Thus, it is of interest to check such process in our idealized merger simulations, where we could clearly see the impact from a single merger event on the remnant shape.

Following the method used in \citet{Zemp2011}, we find the best-fit ellipsoidal shape for the DM mass distribution at a given cluster radius (i.e., the length of the semi-major axis, $a$\footnote{For clarity, we redefine symbol $a$ to be the length of the major axis in Section~\ref{sec:results:dm:shape}. In the other sections of this paper, $a$ represents the cosmic scale factor.}, of the ellipsoid) by evaluating the eigenvalues and eigenvectors of the following normalized moment-of-inertia tensor over a narrow ellipsoidal shell $S_a$,
\be
\textbf{I}=\frac{\sum_{S_a}{m_{\rm DM}\textbf{r}\textbf{r}^{T}}}{\sum_{S_a}{m_{\rm DM}}},
\label{eq:inertia_tensor}
\ee
where $m_{\rm DM}$ and $\textbf{r}$ denote the mass and position vector of the DM particles within the shell. The shell radial width along its major-axis is fixed to $\Delta a=r_{\rm vir}/8$ in our estimation. The initial shape of the shell is spherical with the radius $a$, and then the shell's shape and orientation iteratively update based on the eigenvalues and eigenvectors of $\textbf{I}$ until the intermediate-to-major and minor-to-major axis ratios, $b/a$ and $c/a$, converge.

\begin{figure}
\centering
\includegraphics[width=0.95\linewidth]{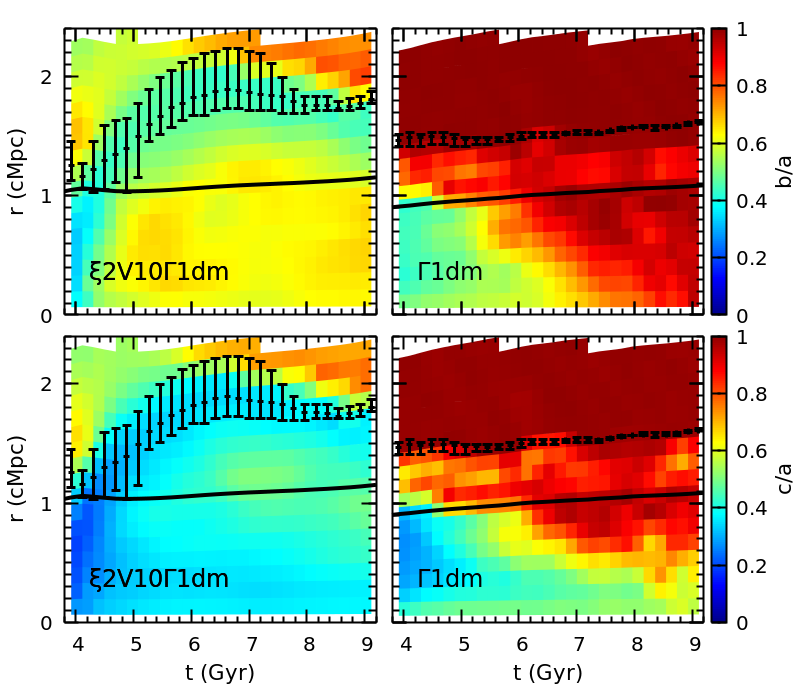}
\caption{Evolution of the ellipsoidal shape of the DM mass distribution in the simulations $\rm\xi2V10\Gamma1dm$ (left panels) and $\rm\Gamma1dm$ (right panels). The top and bottom panels show the intermediate-to-major ($b/a$) and minor-to-major ($c/a$) axis ratios, respectively. The overlaid solid lines and points with error bars indicate the virial and splashback radii of the clusters. This figure shows the significant role of a major merger in shaping the cluster's ellipticity (see Section~\ref{sec:results:dm:shape}).}
\label{fig:shape_dm}
\end{figure}

Fig.~\ref{fig:shape_dm} shows the evolution of the axis-ratio profiles $b/a$ and $c/a$ measured in the simulations $\rm\xi2V10\Gamma1dm$ (left panels) and $\rm\Gamma1dm$ (right panels). By comparing an isolated cluster with the one undergoing a major merger, we can see that the merger significantly affects the cluster shape, particularly in the outer regions. The core of the isolated cluster is aspherical, mainly due to the developed radial-orbit instabilities. As the mass accretes isotropically, the core gradually becomes rounder. Beyond $r_{\rm vir}$, the halo is nearly spherical throughout the isolated cluster simulation.
In a merging cluster (see the left panels in Fig.~\ref{fig:shape_dm}), the axis ratios are approximately constant at $r\lesssim r_{\rm sp}$ ($b/a\simeq0.6$ and $c/a\simeq0.4$) and do not significantly evolve with time  after $t\simeq5\Gyr$. The figure shows that the axis ratios are noticeably smaller near the splashback radius. Indeed, there is a marked  local minimum in both $b/a$ and $c/a$ around $r_{\rm sp}$ (shown by the points with error bars) that persists at all epochs. The distinct elongated shape of the DM distribution near the splashback boundary is mainly due to the DM particles experiencing the largest acceleration during the approach and pericentric passage of the merging halos, as these particles distribute predominantly along the merger axis (see Figs.~\labelcref{fig:slice_dm_pden,fig:part_dm_xy}). These results show that the sharp change around the splashback radius exists not only in the DM density profile and phase-space distribution but also in the shape of the mass distribution.

\begin{figure}
\centering
\includegraphics[width=0.95\linewidth]{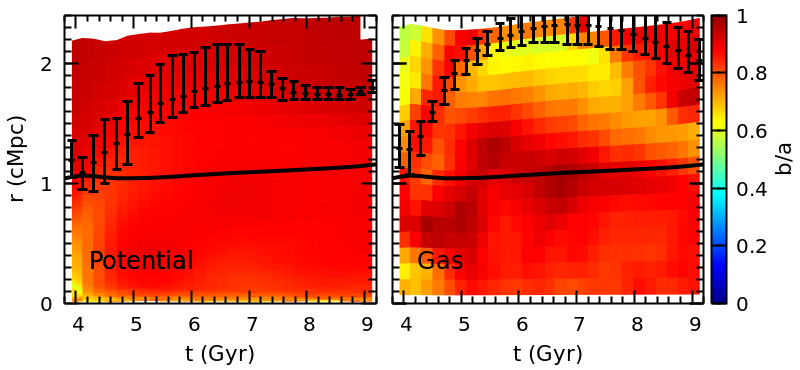}
\caption{Evolution of the intermediate-to-major ($b/a$) axis ratio of the gravitational potential (left panel) and the gas mass distribution (right panel) in the simulation $\rm\xi2V10\Gamma1$. The solid black lines indicate the cluster virial radius. The black points with error bars in the left and right panels show the splashback and MA-shock radii, respectively. This figure shows that the shape of the gravitational potential is much rounder than that of the DM halo in a merging cluster (cf., Fig.~\ref{fig:shape_dm}). The shape of the gaseous atmosphere to the first approximation reflects the shape of the potential (see Section~\ref{sec:results:dm:shape}). }
\label{fig:shape_pot}
\end{figure}

The shape of the stable outermost caustic surface is nearly spherical in the DM density slices (see the last two panels in Fig.~\ref{fig:slice_dm_pden}), and it is mainly determined by the shape of the gravitational potential rather than the mass distribution. The rounder shape of the potential compared to the DM density is not surprising, given that the latter is the Laplacian of the former. To measure the shape of the isopotential surfaces of a cluster, we bin its DM particles by their gravitational potential. The best-fit ellipsoidal shape for the particles in each bin is obtained through the inertia tensor $\textbf{I}$, Eq.~(\ref{eq:inertia_tensor}), with the replaced $m_{\rm DM}$ by the volume of the particles (i.e., $m_{\rm DM}/\rho_{\rm DM}$, where $\rho_{\rm DM}$ is the DM density at the particle position). Unlike the mass distributions, the gravitational potential has $b\simeq c$ and $b/a$ much closer to unity. The left panel in Fig.~\ref{fig:shape_pot} shows the evolution of the potential ellipsoid axis ratio $b/a$ in the run $\rm\xi2V10\Gamma1$. The almost spherical shape of the potential (i.e., $b\simeq c$ and $b/a\gtrsim0.8$) means that particles on different orbits during the merger have experienced similar potential change (and hence similar accelerations). This can explain why the DM outermost caustic surface in Fig.~\ref{fig:slice_dm_pden} is nearly spherical.

\begin{figure}
\centering
\includegraphics[width=0.95\linewidth]{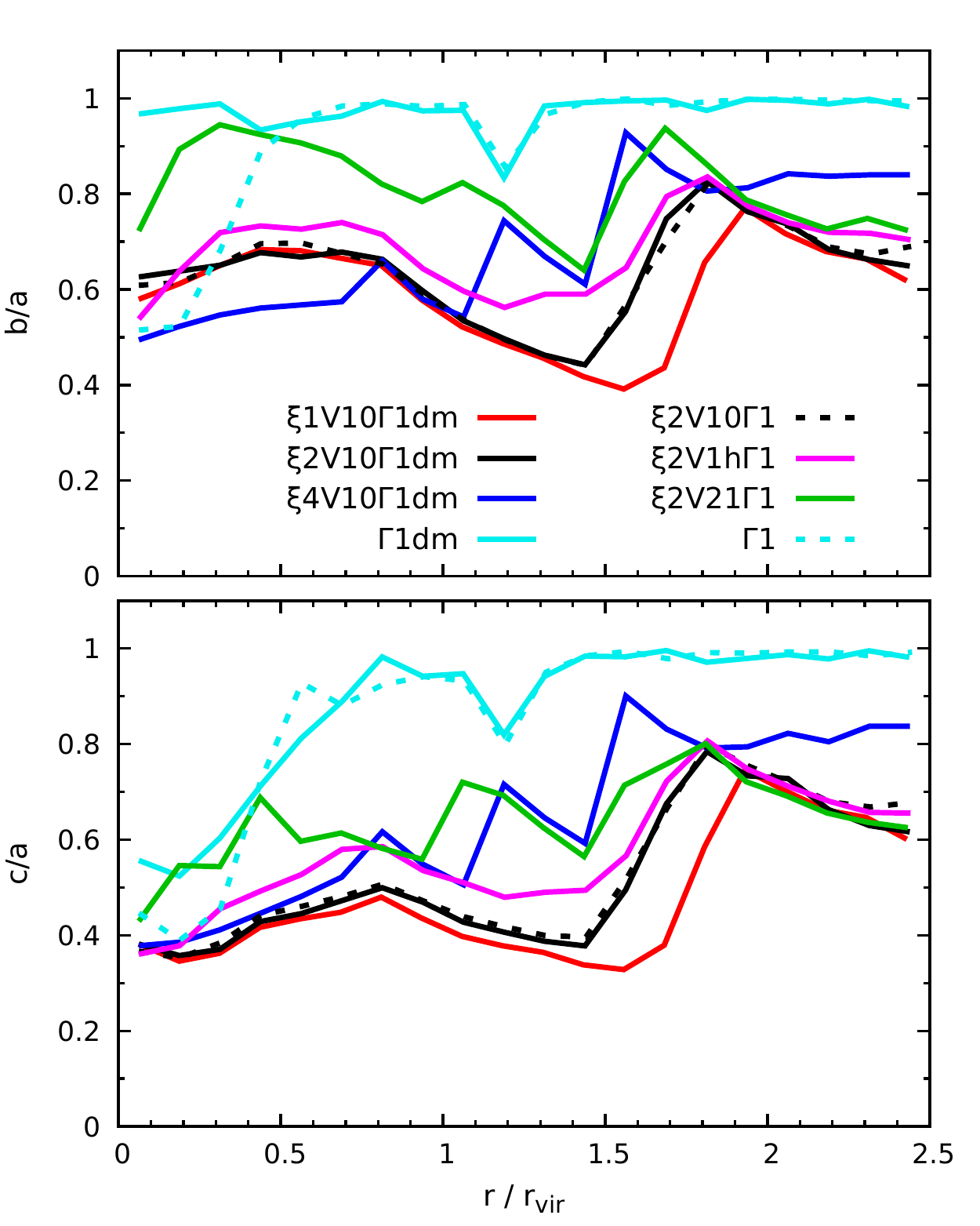}
\caption{Radial profiles of the axis ratios $b/a$ and $c/a$ of the DM halos at redshift $z=0$. They generally range $b/a\simeq0.6-0.8$ and $c/a\simeq0.4-0.6$ within the splashback radius, $r_{\rm sp}\simeq1.6r_{\rm vir}$. Long after the merger ($\sim5\Gyr$ after the core passage), the DM halos still have ellipsoidal shapes, which only mildly depend on the merger parameters (e.g., $\xi$ and $\textbf{v}_0$; see Section~\ref{sec:results:dm:shape}).}
\label{fig:shape_params}
\end{figure}

We also note that the simulations $\rm\xi2V10\Gamma1dm$ and $\rm\xi2V10\Gamma1$ show almost identical results in terms of the DM halo shape (see also Fig.~\ref{fig:shape_params}). Therefore, we can directly compare Figs.~\labelcref{fig:shape_dm,fig:shape_pot}. This also implies that the gaseous component in our cluster is dynamically unimportant to the DM. However, baryonic effects on the DM halo's shape have been discussed in the literature \citep[e.g.,][]{Kazantzidis2004a,Debattista2008}. They mainly work through the dissipative processes in the ICM (e.g., condensation of the cooling gas to the cluster center), which are not included in our simulations.

The distribution of the adiabatic gas should follow the shape of the gravitational potential as well unless the ICM significantly deviates from the hydrostatic equilibrium \citep[e.g.,][]{Lau2011}. This is clearly seen in Fig.~\ref{fig:shape_pot}. Here we measure the shape of the ICM the same way as that for the DM halo but replace $m_{\rm DM}$ by the gas cell mass in Eq.~(\ref{eq:inertia_tensor}). Only at the large cluster radius, $\gtrsim r_{\rm vir}$, the ellipticity of the ICM significantly deviates from that of the potential. This is mainly caused by the formation of MA-shocks and other non-equilibrium hydrodynamic structures in the cluster outskirts (see more discussions in Section~\ref{sec:results:gas}).

Fig.~\ref{fig:shape_params} compares the shape of the DM halos in the simulations with the different merger parameters (e.g., $\xi,\ \textbf{v}_0$) at redshift $z=0$\footnote{In Fig.~\ref{fig:shape_params}, one can see a noticeable difference on $b/a$ in the inner regions of the clusters ($<0.5r_{\rm vir}$) in the runs $\rm\Gamma1$ and $\rm\Gamma1dm$. This difference is caused by the radial-orbit instability rather than baryonic physics. To verify this, we have performed additional simulations with the same initial conditions as in the $\rm\Gamma1dm$ run but different random seeds. We find the shape of the halo's central region in these runs is rather random with $b/a$ ranging between $\simeq0.5-1$. This effect is not significant in our merging clusters since the DM particles in the cluster core have been violently re-distributed during the major merger process.}.
All halos experiencing major mergers have triaxial shapes within the splashback radius, i.e., $b/a\simeq0.6-0.8$ and $c/a\simeq0.4-0.6$. The shapes only mildly depend on the merger mass ratio and impact parameter. The larger the initial orbital angular momentum, the rounder the merger remnant (though the impact parameter in major mergers is usually small; see \citealt{Sarazin2002}). The latter, however, is true only long after the core passage of the merging clusters. During the merger and especially during the near core passage moment, the larger initial orbital angular momentum could lead to a larger asphericity of the halo (see the blue lines in Fig.~\ref{fig:bnd_evo_s1}).

In addition, we also find that the clusters are always elongated along the merger axis in our simulations (see, e.g., Figs.~\labelcref{fig:slice_dm_pden,fig:slice_m2_V1h_all}). This may provide a natural explanation for the alignment of the clusters' major axes with the surrounding large-scale structures \citep{Paz2011,Codis2012}. In general, our models are broadly consistent with the predictions from the cosmological simulations in terms of the halo's shape and orientation (see, e.g., \citealt{Lau2011}). Although a quantitative comparison here is non-trivial due to, e.g., the absence of filaments and large-scale tidal fields in our idealized model (see, e.g., \citealt{Gouin2021}), our simulations robustly reveal the significant role of major mergers in shaping galaxy clusters.

\section{Evolution of Gaseous Atmospheres} \label{sec:results:gas}

\begin{figure*}
\centering
\includegraphics[width=0.9\linewidth]{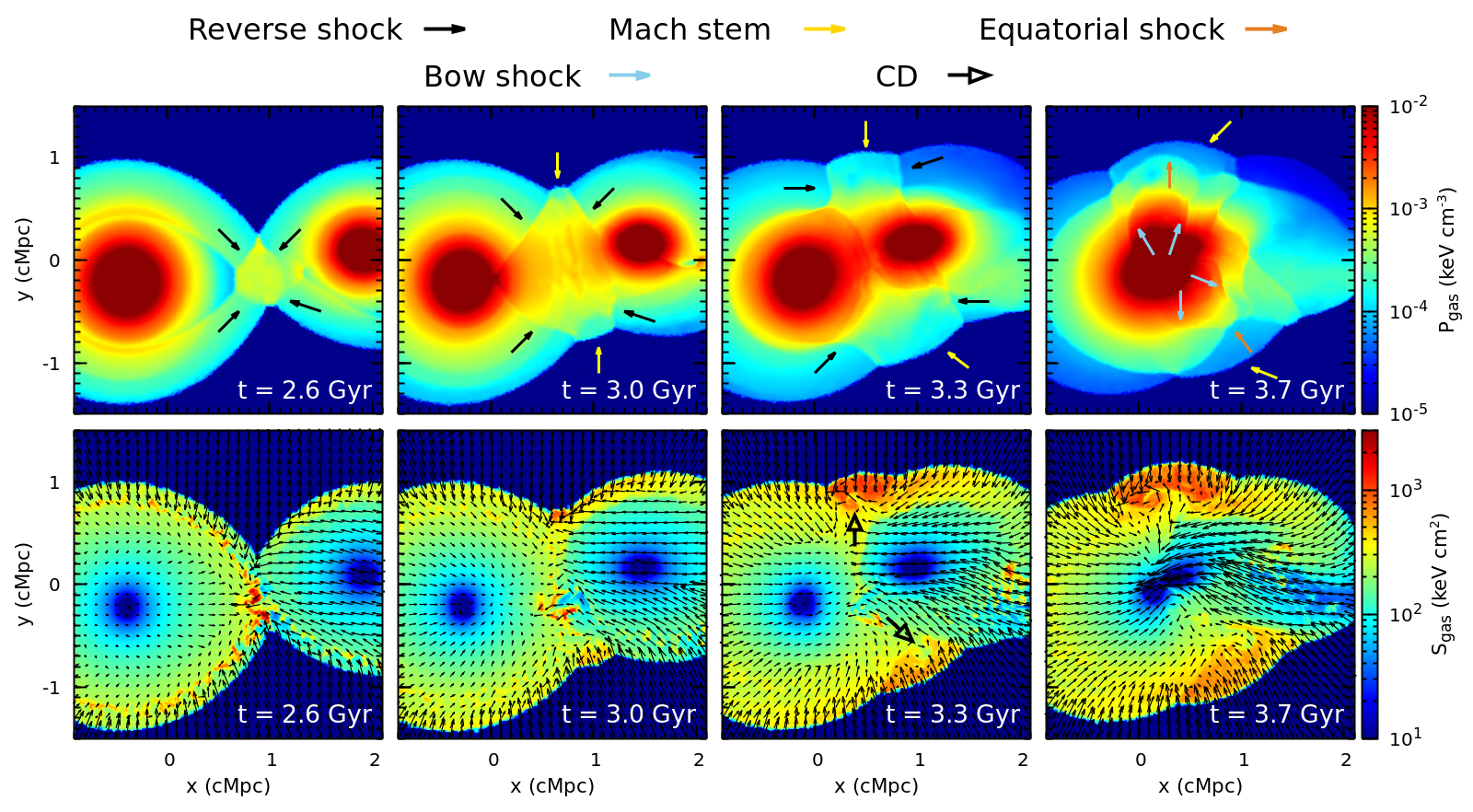}
\caption{Evolution of the gas pressure (top) and entropy (bottom) slices in the pre-merger stage of the cluster in the simulation $\rm\xi2V1h\Gamma1\_hres$. The overlaid arrows in the bottom panels show the gas velocity field in the $x-y$ plane. This figure illustrates rich shock structures formed in the cluster outskirts as the two merging clusters approach each other. The collision of the accretion shocks generates a pair of reverse shocks marked by the black arrows in the top panels. The interaction between these reverse shocks and the accretion shock further drives a Mach stem (yellow arrows). The gas shock-heated by the Mach stem is hotter and has higher entropy than the rest of the ICM. These two gas phases are separated by a CD, marked by the hollow black arrows in the bottom panel. When the clusters get closer, a pair of bow shocks emerges (blue arrows) as well as an equatorial shock (orange arrows), which propagates perpendicularly to the merger axis (see Section~\ref{sec:results:gas:pre-merger}). }
\label{fig:slice_zoom_pre_m2}
\end{figure*}

The hot gaseous atmospheres behave very differently from the collisionless DM in merging clusters. An example of an off-axis major merger in the run $\rm\xi2V1h\Gamma1\_hres$ is shown in Figs.~\labelcref{fig:slice_zoom_pre_m2,fig:slice_zoom_pre_m2b,fig:slice_zoom_post_m2}, which illustrate different stages of the merger process, respectively. Turbulence and discontinuities (including shocks and CDs) driven by the collisions of the atmospheres are clearly visible in the figures (see also Fig.~\ref{fig:slice_m2_V1h_all}).

In this section, we will mainly focus on the gaseous structures that emerge in the outskirts of the merging clusters. These structures are formed in two ways. One is by the direct interactions of the ICM from the two merging clusters, which mainly take place in the pre-merger stage before the primary pericentric passage (see Section~\ref{sec:results:gas:pre-merger}). The other is via the effect of the runaway merger shocks, which sweep through and heat the gas at the large cluster radii and reshape the morphology of the ICM (see Sections~\labelcref{sec:results:gas:ma-shock,sec:results:gas:cd}). This phase occurs near or after the primary apocentric passage of the subcluster.

\subsection{Pre-merger stage} \label{sec:results:gas:pre-merger}

Fig.~\ref{fig:slice_zoom_pre_m2} highlights the evolution of the cluster atmosphere during the pre-merge stage in the high-resolution run $\rm\xi2V1h\Gamma1\_hres$. The shocks and CDs formed in this stage are clearly seen in the pressure and entropy slices, respectively. As the two clusters are approaching each other, their atmospheres become tidally elongated. Subsequently, their accretion shocks contact and merge into one peanut-shape envelope, forming the baryonic boundary of the merged clusters\footnote{The accretion shocks of the two merging clusters likely merge at a much earlier time if a large-scale filament connects the clusters. In this case, the accretions shocks of the clusters may not collide with each other and remain connected by the accretion shock of the filament. Nevertheless, the following evolution (e.g., the formation of bow shocks and perpendicular outflow) would be similar to that described in our idealized simulations.}.
Meanwhile, two reverse shocks with small Mach numbers ($\mathcal{M}_{\rm s}\simeq1.5$) are formed and propagate in opposite directions (indicated by the black arrows in the top panels). A Mach stem then appears (yellow arrows; \citealt{Neumann1943}), resulting from the interaction between the accretion and reverse shocks. The gas shock-heated by the Mach stem is hotter and has higher entropy than the rest of the ICM. These two gas phases are separated by a CD (marked by the hollow black arrows in the entropy slice). In the full 3D picture, the CD and the ICM in general are susceptible to various hydrodynamical instabilities. For instance, the departure of shocks and the gas equi-density surfaces from spherical symmetry implies that a baroclinic mechanism of vorticity generation is at work \citep[e.g.,][]{Vazza2017}, along with the Kelvin-Helmholtz instability due to velocity shear, or the Richtmyer–Meshkov instability. The giant eddies formed near the interfaces of the CD are clearly seen in the entropy slices. As the two merging clusters get closer ($t\gtrsim3\Gyr$), the gas between them strongly compresses. Two bow shocks are formed ahead of the two clusters' cores in this stage (marked by the blue arrows; see also \citealt{Zhang2019a}). At the same time, the squeezed gas between the bow shocks drives an outflow (see the velocity vector field overlaid in the bottom-right panel), further exciting the so-called ``equatorial shocks'' marked by the orange arrows \citep[see, e.g.,][]{Ha2018}. The lifetime of the equatorial shocks, however, is relatively short ($<1\Gyr$). They eventually overtake the Mach stem and form MA-shocks (see the second panel in Fig.~\ref{fig:slice_zoom_pre_m2b} and more discussions in Section~\ref{sec:results:gas:ma-shock}). In X-ray observations, such hot outflows along the directions perpendicular to the merger axis have been discovered in several pre-merging clusters \citep[e.g.,][]{Akamatsu2017,Gu2019}.

\begin{figure}
\centering
\includegraphics[width=0.9\linewidth]{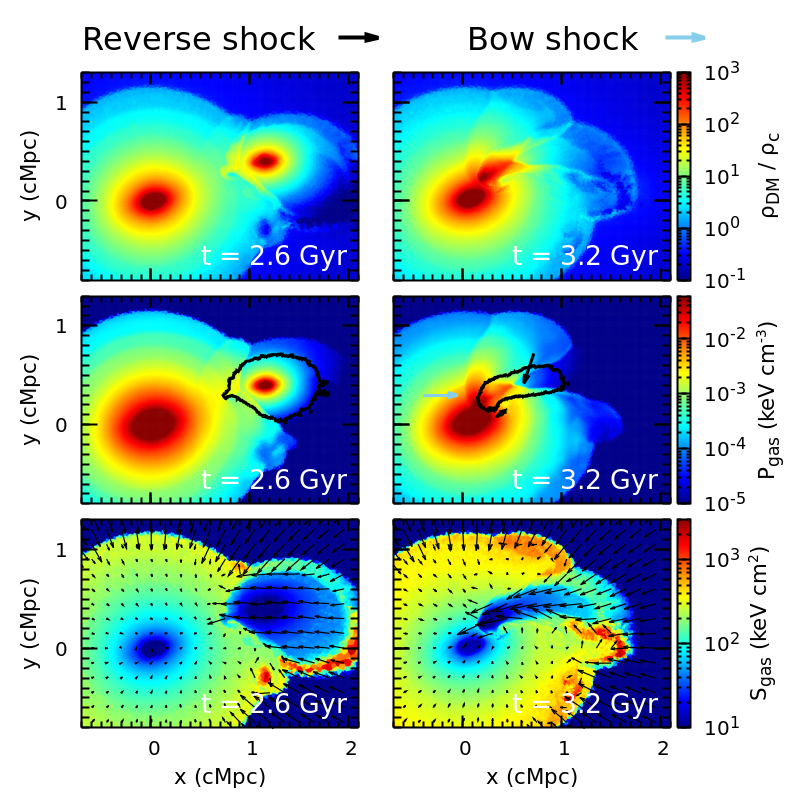}
\caption{Similar to Fig.~\ref{fig:slice_zoom_pre_m2} but for the pre-merger stage of a minor merger in the simulation $\rm\xi10V1h\Gamma1\_hres$. The contours overlaid in the pressure slices indicate the regions where $f_{\rm dye}>0.9$. The blue and black arrows mark the bow shock formed in front of the subcluster and the reverse shock propagating in the subcluster, respectively. The atmosphere of the infalling subcluster is quickly deformed into a cylindrical shape before reaching the pericenter (see Section~\ref{sec:results:gas:pre-merger}). }
\label{fig:slice_zoom_pre_m10}
\end{figure}

\begin{figure*}
\centering
\includegraphics[width=0.9\linewidth]{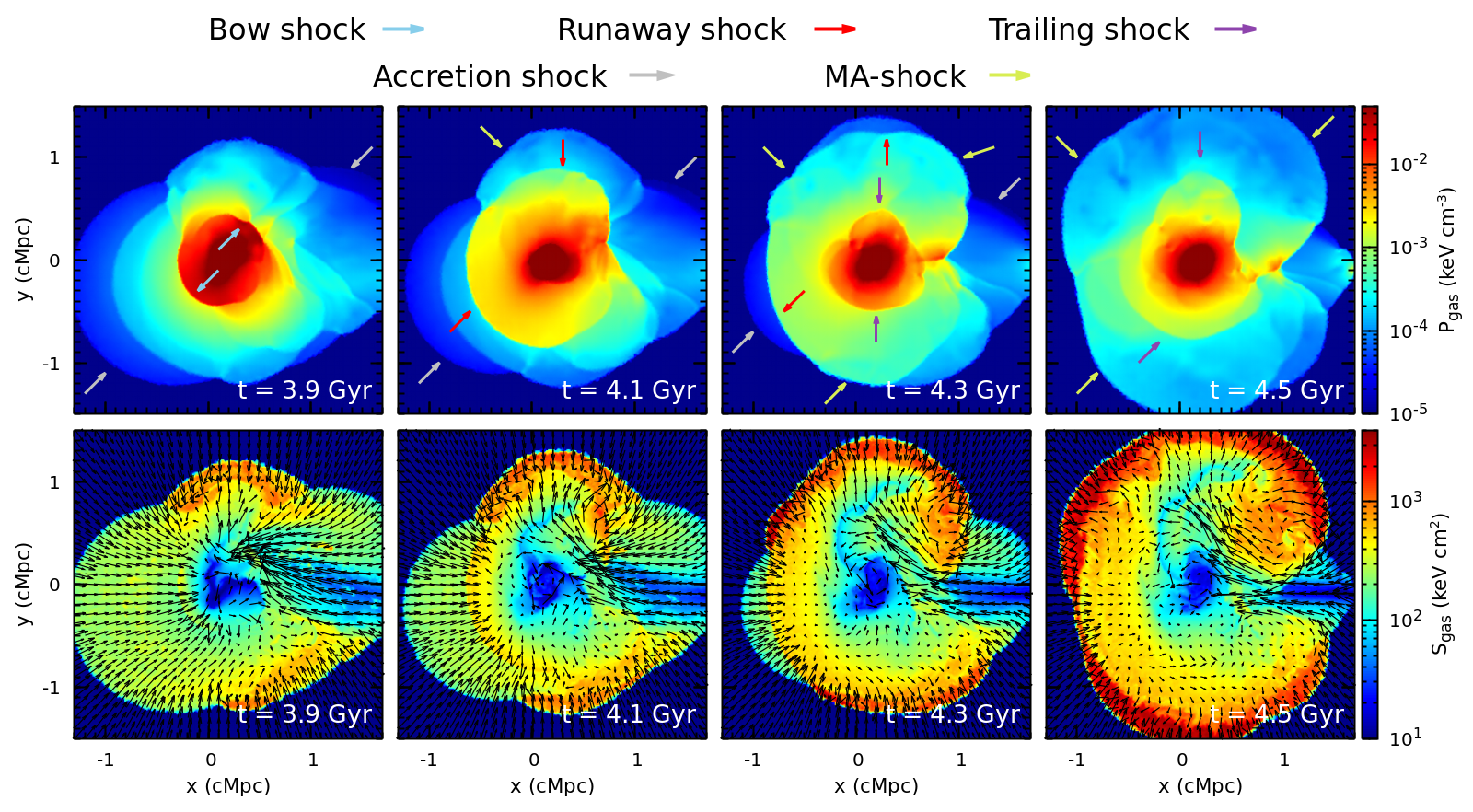}
\caption{Similar to Fig.~\ref{fig:slice_zoom_pre_m2} but highlighting the moments shortly after the core passage in the simulation $\rm\xi2V1h\Gamma1\_hres$. The bow shocks (blue arrows) driven by the merging clusters evolve rapidly into the runaway shocks (red arrows), which then collide with the accretion shock (grey arrows) at $t\simeq4.3\Gyr$. The MA-shocks (lime arrows) are consequently formed in such shock collision, becoming the new gaseous boundary of the merger remnant (see Section~\ref{sec:results:gas:ma-shock}). They firstly form in some directions (see the snapshot at $t=4.3\Gyr$) and eventually cover most of the remnant. Also, a pair of trailing-shock edges (purple arrows) appear in the central region at $t\simeq4.3\Gyr$. Together with the runaway shocks, they constitute the \textit{N}-wave structures discussed by \citet{Zhang2021}. }
\label{fig:slice_zoom_pre_m2b}
\end{figure*}

Minor mergers, in contrast, show a very different picture from that described above since the sizes of the two merging clusters are very different (see an example in Fig.~\ref{fig:slice_zoom_pre_m10}). The atmosphere of the infalling subcluster is compressed and stripped off from its original shallower gravitational potential as it penetrates the main cluster \citep[see also, e.g.,][]{Roediger2015,Vijayaraghavan2015}. Before reaching the pericenter, the subcluster's atmosphere deforms into a cylindrical shape (see the black contours of $f_{\rm dye}$; and also Fig.~\ref{fig:slice_m10_V21_all} for the merger with a larger initial impact parameter). Only a bow shock formed in front of the subcluster and a reverse shock propagating in the subcluster atmosphere are prominently seen in the pressure slices.

\subsection{Formation and evolution of MA-shocks} \label{sec:results:gas:ma-shock}

After the first pericentric passage, the bow shocks driven during the pre-merger stage gradually evolve into runaway merger shocks \citep{Zhang2019b}. They eventually overtake the accretion shock and reshape the morphology of the ICM \citep{Zhang2020a}. Such a process is illustrated in Fig.~\ref{fig:slice_zoom_pre_m2b}, where the runaway and accretion shocks collide around $t\simeq4.3\Gyr$. In this collision, a MA-shock forms a new baryonic boundary of the cluster and subsequently moves (radially) outwards. Note that, in contrast to the 1D models presented in \citet{Zhang2020a}, our 3D merger simulations show that the MA-shock is first formed near the directions perpendicular to the merger axis and, therefore, is no longer spherically-symmetric. At the same time, a CD appears near the initial position of the accretion shock, which is the product of the shock collision as well (see also Fig.~\ref{fig:slice_m2_V1h_all} and Section~\ref{sec:results:gas:cd}). The gas between the CD and MA-shock front is heated only by the MA-shock and has higher entropy than the rest of the ICM that is heated successively by the pristine accretion shock and runaway shock (see more discussions in \citealt{Zhang2020a}).

Due to the ongoing mass accretion, the MA-shock front would eventually stall at a large cluster radius and then fall backwards. 1D simulations by \citet{Zhang2020a} have pointed out two key parameters that determine the evolution of MA-shocks at the peripheries of the clusters. One parameter is the Mach number of the runaway shock, while the other one is the smooth MAR parameter $\Gamma_{\rm s}$. The MA-shock could survive for a longer time if the runaway shock is stronger and/or $\Gamma_{\rm s}$ is smaller. We here check their findings in more realistic 3D merger simulations and investigate the relative positions between the MA-shock front and the outermost DM caustic surface.

\begin{figure*}
\centering
\includegraphics[width=0.9\linewidth]{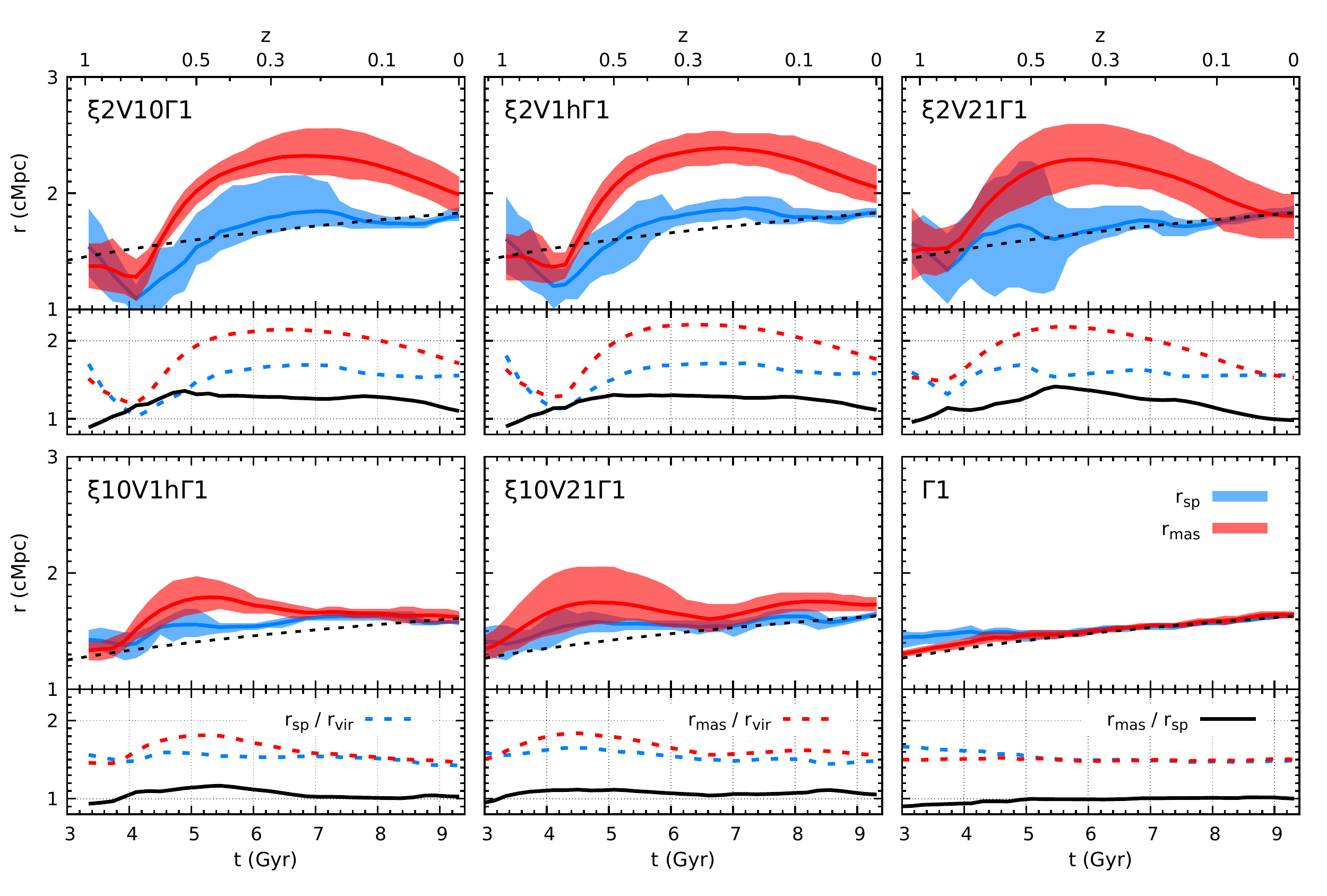}
\caption{Evolution of the cluster MA-shock (red) and splashback (blue) radii in the simulations with the same $\Gamma_{\rm s}\ (=1)$ but different merger parameters. In the upper subplot of each panel, the solid lines and their shaded areas show the azimuthally-averaged $r_{\rm sp}$ (or $r_{\rm mas}$) and the scatters within the 10th--90th percentile range. The dashed black line shows the self-similar scaling relation $r\propto t^{2\Gamma_{\rm s}/9}$ with the amplitude normalized to $r_{\rm sp}$ at $z=0$. The evolution of the ratios $r_{\rm sp}/r_{\rm vir}$, $r_{\rm mas}/r_{\rm vir}$ and $r_{\rm mas}/r_{\rm sp}$ is shown in the lower subplots. One can see that the cluster mergers significantly affect the shock and splashback radii. The ratio $r_{\rm mas}/r_{\rm sp}$ is typically above $1$ and depends on the merger mass ratio $\xi$ (see Section~\ref{sec:results:gas:ma-shock}). }
\label{fig:bnd_evo_s1}
\end{figure*}

Fig.~\ref{fig:bnd_evo_s1} compares the evolution of the splashback and MA-shock radii ($r_{\rm sp}$ and $r_{\rm mas}$) in the simulations with the same $\Gamma_{\rm s}\ (=1)$ but different merger parameters. As a baseline, the evolution of an isolated cluster is shown in the bottom-right panel. For each cluster, we measure its splashback and MA-shock radii in $10^3$ directions. The MA-shock front is identified as the outermost jump in the gas-entropy profile (see Section~\ref{sec:method:radii}). The azimuthally-averaged $r_{\rm sp}$ and $r_{\rm mas}$ are shown as the solid blue and red lines in the upper subplot in each panel. The shaded areas indicate their scatters ($10-90$ percentiles).
This figure shows that the major mergers affect both $r_{\rm mas}$ and $r_{\rm sp}$ much more significantly than the minor ones. However, in either case, the MA-shock front always reaches a larger cluster radius than the DM caustic surface. We show the ratios between the averaged MA-shock and splashback radii in the lower subplot of each panel. They generally range from $r_{\rm mas}/r_{\rm sp}\simeq1.3-1.4$ to $1.0-1.2$ in major ($\xi=2$) and minor ($\xi=10$) mergers, respectively, and show only mild dependence on the merger's initial relative velocity and impact parameter. The largest $r_{\rm mas}/r_{\rm sp}$ ratios could generally last for a few Gyrs in the major mergers. To further illustrate the deviation of our merging clusters from the self-similar model, we overlaid the self-similar scaling relation $r\propto t^{\delta}$ as the dashed black lines in Fig.~\ref{fig:bnd_evo_s1}, where $\delta=2\Gamma_{\rm s}/9$ and $r$ is in the comoving coordinates \citep{Shi2016}. Their amplitudes are normalized to $r_{\rm sp}$ at $z=0$. We can clearly see that both the locations of the baryonic and DM boundaries of galaxy clusters do not evolve self-similarly during major mergers. The evolution of DM splashback, however, approaches the self-similar prediction shortly after the merger completes, while the MA-shock does it at a much later epoch.

\begin{figure*}
\centering
\includegraphics[width=0.9\linewidth]{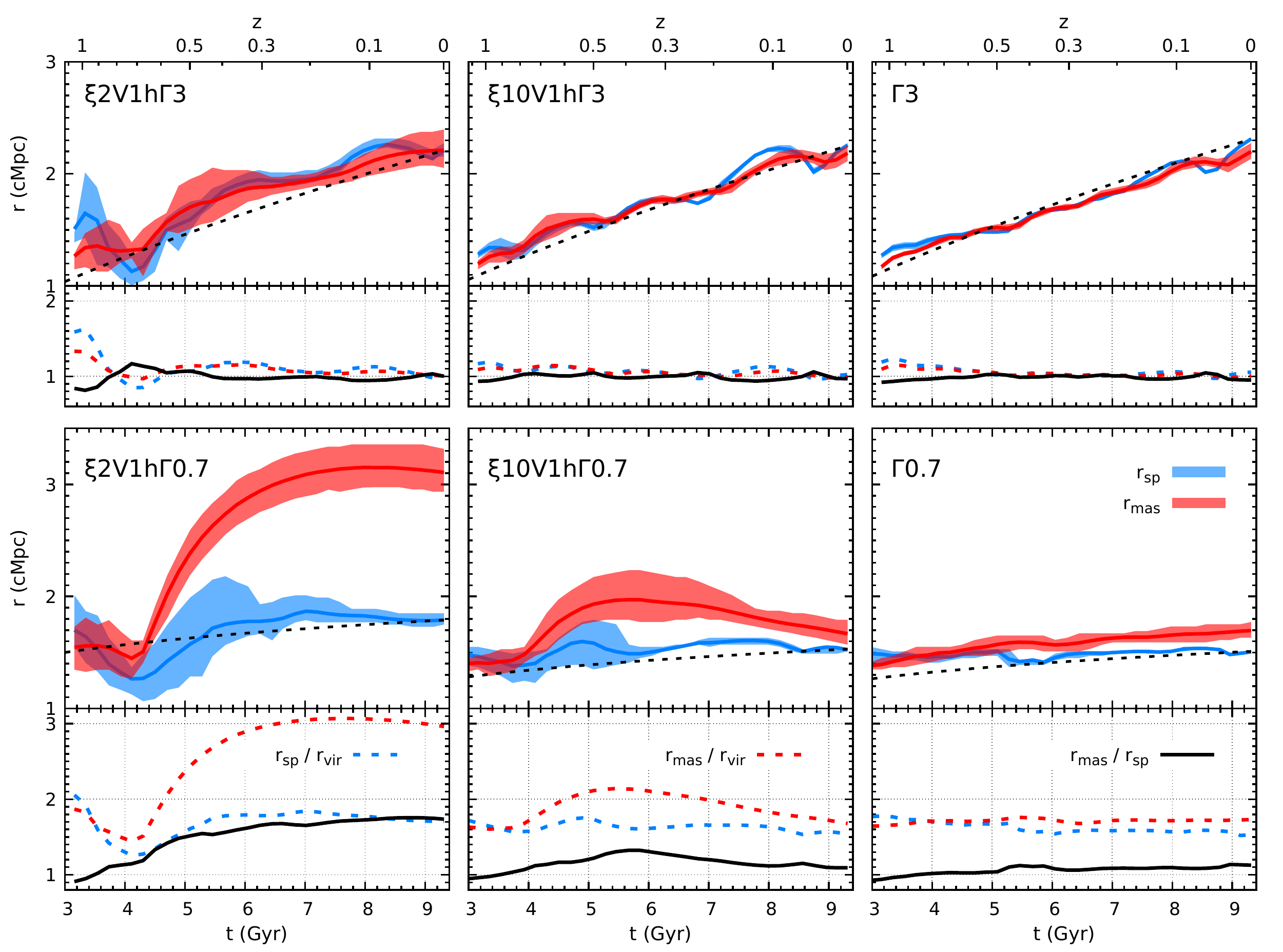}
\caption{Similar to Fig.~\ref{fig:bnd_evo_s1} but for the simulations with the different smooth MAR $\Gamma_{\rm s}=3$ (top panels) and $0.7$ (bottom panels). This figure demonstrates that $\Gamma_{\rm s}$ is a key parameter to determine $r_{\rm mas}/r_{\rm sp}$ in galaxy clusters (see Section~\ref{sec:results:gas:ma-shock}).}
\label{fig:bnd_evo_s3}
\end{figure*}

Fig.~\ref{fig:bnd_evo_s3} further shows the dependence of the radii $r_{\rm sp}$ and $r_{\rm mas}$ on the smooth MAR $\Gamma_{\rm s}$. When $\Gamma_{\rm s}$ is smaller, the MA-shocks survive for a longer time and reach larger cluster radii. This is in line with the previous findings by \citet{Zhang2020a}. In this study, we further quantify the dependence of the ratio $r_{\rm mas}/r_{\rm sp}$ on $\Gamma_{\rm s}$, namely, $r_{\rm mas}/r_{\rm sp}\simeq1$ when $\Gamma=3$ and $r_{\rm mas}/r_{\rm sp}\simeq1.7\ (\xi=2)$ and $1.3\ (\xi=10)$ when $\Gamma=0.7$. A larger smooth MAR pushes the MA-shock front closer to the splashback radius.

To directly compare with the 1D models presented in \citet{Zhang2020a}, we measure the Mach number $\mathcal{M}_{\rm rs}$ of the runaway merger shocks (driven by the subcluster) shortly before colliding with the accretion shock in the simulations presented in Figs.~\labelcref{fig:bnd_evo_s1,fig:bnd_evo_s3}. We estimate their Mach numbers approximately along the merger axis and find that $\mathcal{M}_{\rm rs}\simeq3-4.5$ and $\simeq2-2.5$ in the major and minor mergers, respectively. They mildly depend on $\Gamma_{\rm s}$ (i.e., the smaller $\Gamma_{\rm s}$, the larger $\mathcal{M}_{\rm rs}$) and are almost independent of the subclusters' trajectories. These results partially explain why MA-shocks are more significant in major mergers (see \citealt{Zhang2020a} for more discussions on the role of $\mathcal{M}_{\rm rs}$ in the evolution of MA-shocks). However, in our 3D simulations, the runaway merger shock powering a MA-shock depends not only on $\mathcal{M}_{\rm rs}$ but also on the shock geometry (shape, size). The shock strength is highly non-uniform on the shock front. We find that \citet{Zhang2020a} quantitatively overestimated the dependence of MA-shocks on the parameter $\mathcal{M}_{\rm rs}$ (see their figs.~4 and 5), which is mainly caused by the spherical symmetry assumed in their 1D models. Given the complexity of the 3D mergers, we suggest that the merger mass ratio $\xi$ is a more natural parameter to quantify the evolution of MA-shocks than $\mathcal{M}_{\rm rs}$.

A prominent departure of the accretion shock location from the splashback radius in galaxy clusters has been noticed in cosmological simulations, namely, $r_{\rm mas}/r_{\rm sp}\simeq1.5-2.5$ with a large scatter (see fig.~3 in \citealt{Aung2020} and also \citealt{Molnar2009,Lau2015,Zinger2018a}). Fig.~\ref{fig:bnd_evo_s3} shows that the large $r_{\rm mas}/r_{\rm sp}$ arises only in mergers set up with $\Gamma_{\rm s}\lesssim1$, while $r_{\rm mas}$ and $r_{\rm sp}$ are always comparable when $\Gamma_{\rm s}$ is high ($\gtrsim3$, i.e., the smooth accretion dominates the cluster mass growth rather than the merger). Given that the $r_{\rm mas}/r_{\rm sp}$ measured in cosmological simulations is consistently large ($\sim 1.5-2.5$) for clusters in different evolutionary stages and at different ages, one may na\"{\i}vely conclude that smooth MAR in cosmological clusters is always small. However, this is not true. A key ingredient of clusters forming in a self-consistent cosmological environment is the highly anisotropic mass accretion with generally high accretion rates along filamentary directions, and much lower MAR and smoother accretion along other directions. Strong MA-shocks around clusters forming in cosmological simulations are detected primarily along directions away from prominent filaments where the accretion rate of gas is low. Thus, the large $r_{\rm mas}/r_{\rm sp}$ ratios for the external shocks around clusters in such simulations is mainly due to the low local gas accretion rate along non-filamentary directions, in which the merger shocks propagate easily and quickly (see Section~\ref{sec:discussions:mar} for additional discussion on this issue).

In our MA-shock scenario, the mergers, particularly major mergers, play an important role. Given the major merger rate of galaxy clusters is about one per halo per unit redshift (e.g., \citealt{Fakhouri2008,Fakhouri2010}) and the typical lifetime of the MA-shocks is a few Gyrs if $\Gamma_{\rm s}\simeq1$, the high Mach number shocks observed at the cluster peripheries that separate the ICM and intergalactic medium (usually referred to as ``accretion shocks'' in the literature) are mostly MA-shocks.

One thing to note is that the $r_{\rm mas}/r_{\rm sp}$ in our model (see, e.g., Figs.~\labelcref{fig:bnd_evo_s1,fig:bnd_evo_s3}) only reflects the effect of a single merger event. A strong cumulative effect of multiple runaway shocks on MA-shock is expected if several major/minor mergers take place consecutively within a relatively short time period (e.g., a few Gyrs; see an example in figs.~6 and 7 in \citealt{Zhang2020a} from the cosmological simulation), which would further increase $r_{\rm mas}/r_{\rm sp}$. This effect is even seen in our run $\rm\xi10V21\Gamma1$, where the infalling subcluster has a large initial impact parameter. In the bottom-middle panel of Fig.~\ref{fig:bnd_evo_s1}, we can see that the MA-shock was driven twice, corresponding to the first and second orbits of the subcluster, respectively (compare to Fig.~\ref{fig:traj_halo}), although the secondary runaway shock is rather weak. In this sense, the location of the MA-shock front depends on the merger history of the cluster. Our simulations of single merger events provide only a lower limit of the ratio $r_{\rm mas}/r_{\rm sp}$. The effects from multiple mergers will accumulate, becoming sufficient to explain the $r_{\rm mas}/r_{\rm sp}$ ratio seen in the cosmological simulations if the local smooth mass accretion is low. \cite{Aung2020} showed the effect of mergers on radially expanding atmospheres along the non-filamentary directions (see their fig.~7). However, it is not trivial to quantify this effect in cosmological simulations. The large-scale filaments connecting the merging clusters may significantly bias the measurements of $r_{\rm mas}/r_{\rm sp}$. Also, the cumulative effect of multiple mergers could blur the measured effect of a single major-merger event on $r_{\rm mas}/r_{\rm sp}$ (see, e.g., their fig.~8).

\subsection{Shock-driven CDs} \label{sec:results:gas:cd}

Cluster mergers drive not only shocks but also CDs, which can be generally classified into three types according to their origin. The type-\Romannum{1} CD appears at the interface of the atmosphere of a small cluster moving into the ICM of a larger system (see, e.g., the entropy slices in Fig.~\ref{fig:slice_zoom_pre_m10} and also \citealt{Vikhlinin2001,Markevitch2002} and \citealt{Nagai2003} for examples in observations and cosmological simulations). The type-\Romannum{2} CDs arise when the gas core of a cluster is perturbed by either a passing subcluster \citep[e.g.,][]{Ascasibar2006} or shock \citep[e.g.,][]{Churazov2003}. The gas layers with different entropies are offset from their equilibrium positions. They are also known as the ``sloshing cold fronts'' in galaxy clusters (see \citealt{Markevitch2007,Zuhone2016} for reviews). The type-\Romannum{3} CDs, namely \textit{shock-driven} CDs, are formed in the collisions of shocks \citep{Birnboim2010,Zhang2020a,Zhang2020b}, which are specifically described by the Riemann problem in the 1D situation \citep{Landau1959}. Our simulations show that all three types of CDs co-exist in a merging cluster.

\begin{figure}
\centering
\includegraphics[width=0.9\linewidth]{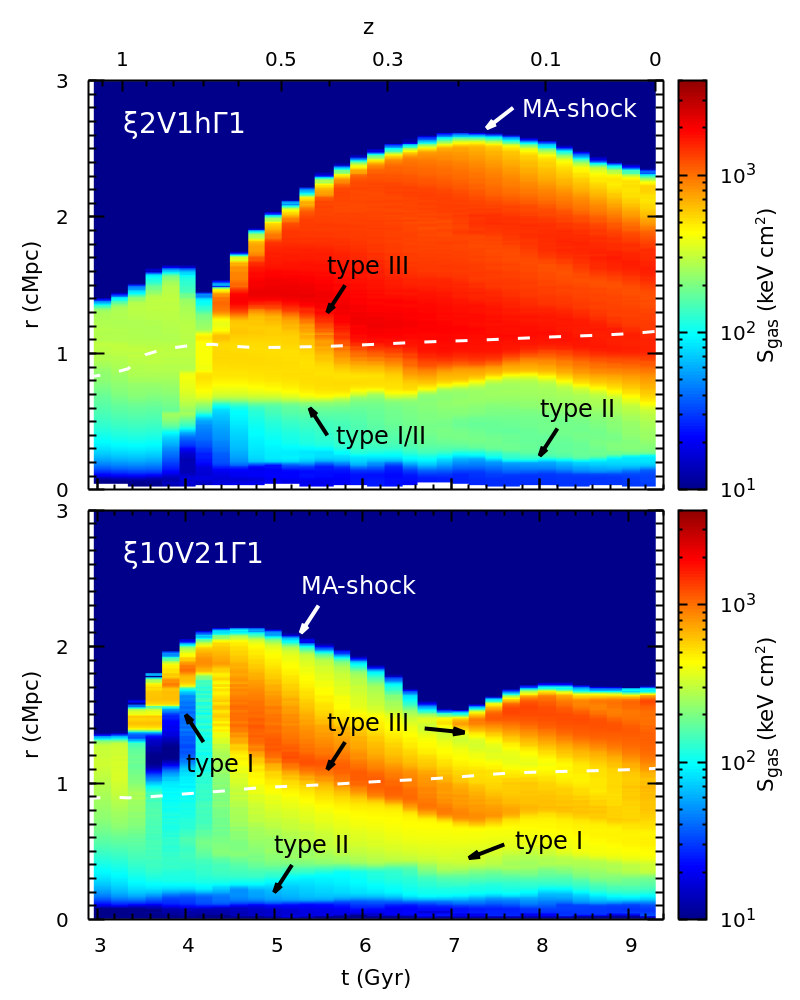}
\caption{Evolution of the gas-entropy radial profiles in the simulations $\rm\xi2V1h\Gamma1$ (top panel) and $\rm\xi10V21\Gamma1$ (bottom panel). The profiles are measured in the cones along the direction $\textbf{n}=(-1,\,-1,\,0)$ and with the opening angles $20^{\circ}$ (see the cross-sections of the cones in the $x-y$ plane in Figs.~\labelcref{fig:slice_m2_V1h_all,fig:slice_m10_V21_all}). The dashed white lines show the cluster virial radii. The black arrows indicate different types of CDs driven in the mergers. The type-\Romannum{3} CDs are formed in the collisions between the runaway and accretion shocks, accompanying the MA-shocks. They could survive for a few Gyrs in our simulations (see Section~\ref{sec:results:gas:cd}). }
\label{fig:prof_cd_entr}
\end{figure}

Fig.~\ref{fig:prof_cd_entr} shows the evolution of the gas entropy profile in our simulations $\rm\xi2V1h\Gamma1$ and $\rm\xi10V21\Gamma1$. The profile is measured within a cone centered at the position of the deepest potential. The cone axis is along the direction $\textbf{n}=(-1,\,-1,\,0)$, and the opening angle is $20^{\circ}$ (see the cross-sections of the cones in the $x-y$ plane in Figs.~\labelcref{fig:slice_m2_V1h_all,fig:slice_m10_V21_all}). In this figure, we mark the CDs formed during the major/minor mergers. In the inner region ($r\lesssim200\ckpc$), sloshing patterns (type-\Romannum{2} CDs) appear after the cluster's primary pericentric passage. They are more prominent in $\rm\xi10V21\Gamma1$ (see also Fig.~\ref{fig:slice_m10_V21_all}), where the subcluster has a larger impact parameter. Unlike those in previous idealized merger simulations \citep[e.g.,][]{Ascasibar2006}, the sloshing features in our model always stay in the cluster inner region even a few Gyrs after the core passage, due to the continuous mass accretion of the system. In the same run, the type-\Romannum{1} CDs driven by the subcluster are also formed. They enter the conic region (where we measured the profiles) between $t\sim3.5-4.5\Gyr$. In $\rm\xi2V1h\Gamma1$, the situation is a bit more complicated. The low-entropy gas that initially resides in the center of the main cluster is pushed outwards by the infalling subcluster during the core passage (see Fig.~\ref{fig:slice_m2_V1h_all}). The formed CD ($r\sim700\ckpc$) is thus a blend of types-\Romannum{1} and \Romannum{2}. We find that such a CD is prominent only in our major mergers and indicates the maximum radius that the gas from the subcluster can reach (along the cone's direction $\textbf{n}$). The atmosphere of the main cluster beyond this radius is solely affected by the runaway or MA- shocks.

The shock-driven CD (type-\Romannum{3}) is formed when the runaway shock overtakes the accretion shock, as is clearly seen in Fig.~\ref{fig:prof_cd_entr}. It separates the gas heated by the MA-shock and the pristine accretion shock on the radially outer and inner regions. This explains the formation of a high-entropy gas shell between the CD and MA-shock front. Though such features have been well captured in the 1D model presented in \citet[][see their fig.~2 and also \citealt{Zhang2020a}]{Zhang2020b}, there are several differences between the 3D and 1D results. First, in the 3D runs, the shock-driven CDs radially move inwards shortly after their formation, particularly, the discontinuities formed in the minor mergers. They eventually stall within the cluster virial radius, slightly closer to the cluster center than predicted in the 1D model. Second,  since a series of runaway shocks and their trailing edges are generated during the pericentric passages of the subcluster (see Fig.~\ref{fig:slice_zoom_pre_m2b} and also \citealt{Zhang2021}), we could see multiple type-\Romannum{3} CDs at large cluster radii in our simulations. Third, the CD lifetime in the 3D simulations is shorter than that in the 1D models since the instabilities are absent in the latter. In our simulations, the CDs are gradually disrupted by the instabilities developed near the interfaces after approximately $3-5\Gyr$ (see, e.g., Figs.~\labelcref{fig:slice_m2_V1h_all,fig:slice_m10_V21_all}). However, we stress that many other factors may influence these values, including the Reynolds number of the atmosphere, pre-existing turbulence/substructures in cluster outskirts \citep[e.g.,][]{Nagai2011,Vazza2013}, and effects of magnetic fields \citep[e.g.,][]{Berlok2020}, and need to be explored in the future.

\begin{figure*}
\centering
\includegraphics[width=0.9\linewidth]{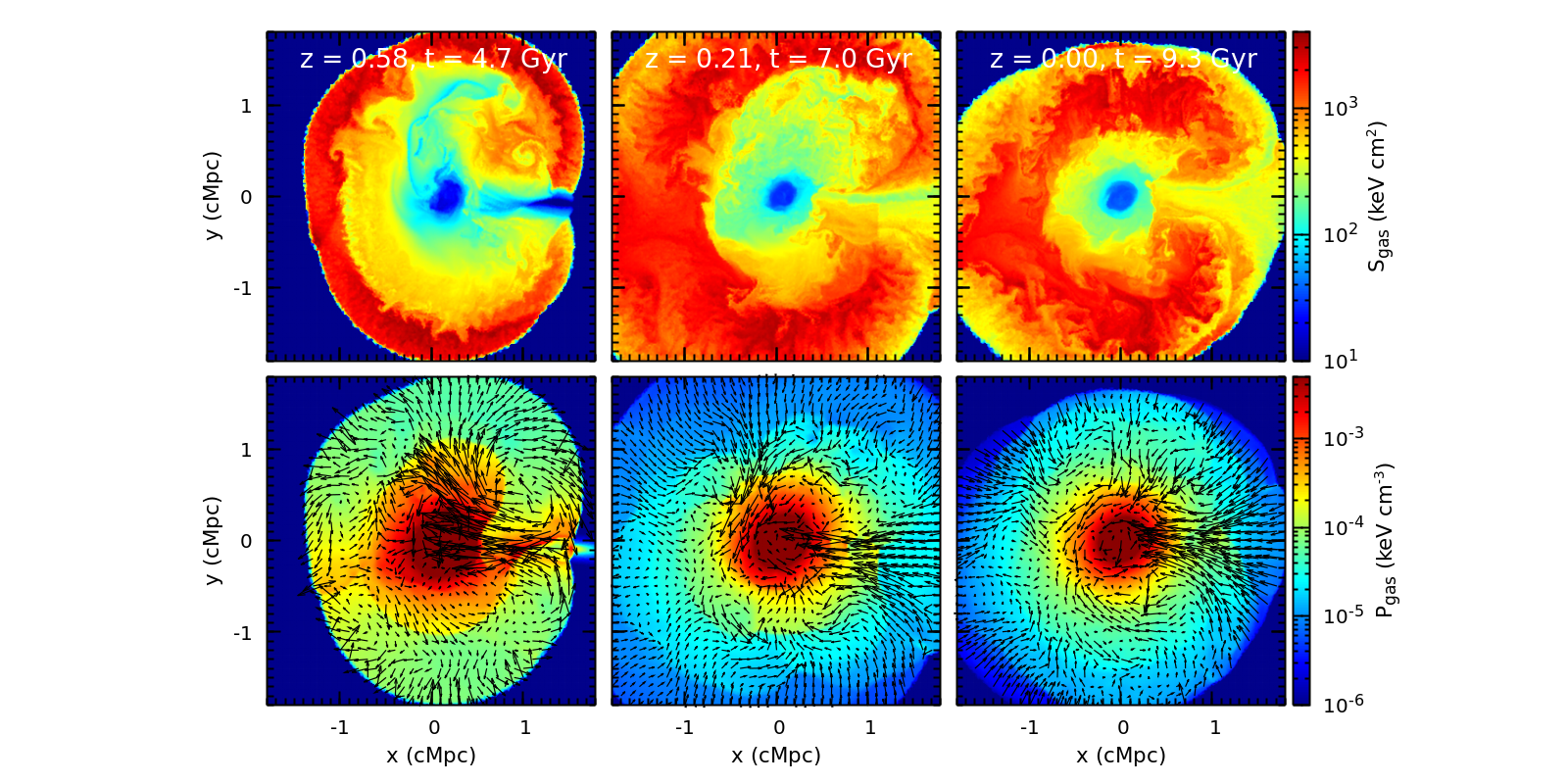}
\caption{Similar to Figs.~\labelcref{fig:slice_zoom_pre_m2,fig:slice_zoom_pre_m2b}, but for the post-merger stage of the cluster major merger in the simulation $\rm\xi2V1h\Gamma1\_hres$. The gas velocity in the ICM in the $x-y$ plane is shown as vectors in the pressure slices. One can see turbulence in the cluster outskirts driven by the major merger and associated with the formation of the MA-shock, type-\Romannum{3} CDs, and high-entropy gas shell (see Section~\ref{sec:results:gas:cd}). }
\label{fig:slice_zoom_post_m2}
\end{figure*}

We have compared our simulations $\rm\xi2V1h\Gamma1$ (see Fig.~\ref{fig:slice_m2_V1h_all}) and $\rm\xi2V1h\Gamma1\_hres$ (see Fig.~\ref{fig:slice_zoom_post_m2}) to further check if the spatial resolution of the simulations affects the evolution of the CDs. We find that the gaseous structures formed in these two runs are broadly consistent. It is not surprising, though, that there are more small-scale eddies and other turbulent substructures near the CDs in $\rm\xi2V1h\Gamma1\_hres$. These differences almost do not affect the lifetimes of the shock-driven CDs.

Recently, \citet{Oneil2020} and \citet{Aung2020} reported an edge in the stacked gas-density profiles of galaxy clusters near the virial radius in the cosmological simulations. We speculate that the type-\Romannum{3} CDs could be partially responsible for this feature since they are commonly seen in galaxy clusters near the virial radii, and the gas-density profile between the shock-driven CDs and MA-shock fronts is relatively shallow (see Fig.~\ref{fig:slice_m2_V1h_all} and also fig.~3 in \citealt{Zhang2020b}).

In Fig.~\ref{fig:slice_zoom_post_m2}, we highlight the turbulence in the cluster outskirts driven by a major merger (see the gas-velocity field overlaid in the bottom panels). Its roughly isotropic distribution implies the generation of turbulence has a tight relation with the formation of the MA-shock, type-\Romannum{3} CDs, and high-entropy gas shell in-between. Instabilities are excited near the interfaces of the CDs. They are more prominent close to the directions perpendicular to the merger axis, where the velocity shear is stronger (see the top-left panel). As predicted in the 1D model, the entropy profile in the high-entropy shell decreases with the radius (see Fig.~\ref{fig:prof_cd_entr} and also \citealt{Zhang2020a}) and is thus convectively unstable.
The negative entropy gradient is, however, rather small, implying long time scales of the instability growth $\simeq5-20\Gyr$.
It needs further investigations to figure out how efficient this process would drive the turbulence, which is beyond the scope of this study. We defer a detailed analysis of the instabilities formed in our merger simulations and their driven mechanisms to our follow-up paper.

We finally briefly discuss the shock-driven CDs in observations. Since these CDs are generated in either major or minor mergers and could survive for a few Gyrs, they would be commonly seen in the cluster outskirts. They also have distinguishable features from other types of CDs. For example, they are usually Mpc-scale and tangential to the cluster radial direction\footnote{Note that a type-\Romannum{1}/\Romannum{2} (or type-\Romannum{1}) CD formed in our simulation has an approximately spherical geometry as well. In contrast to the shock-driven CD, it is located at smaller radii and only prominent in major mergers and minor mergers with large impact parameter (see Fig.~\ref{fig:prof_cd_entr}).}. Although detecting these CDs is challenging with existing X-ray observatories as they are located beyond the virial radius, one observational candidate has been identified in the Perseus cluster near the virial radius \citep{Walker2020,Zhang2020b}. Such CDs could also show up in the SZ maps, provided that the electron and ion temperatures are not equal.

It is worth mentioning A2142 -- another interesting candidate \citep{Markevitch2000} with a giant CD found at $\sim0.5r_{\rm vir}$ southeast from the cluster center \citep{Rossetti2013}. An infalling group is at a similar radius but in the south of the cluster \citep{Eckert2014}. The morphology of A2142 is similar to the merging cluster modeled in our run $\rm\xi10V21\Gamma1$. The infalling group is around its primary or secondary apocenter. In such a scenario, the CD could be explained by either a type-\Romannum{1} or type-\Romannum{3} CD illustrated in the bottom panel of Fig.~\ref{fig:prof_cd_entr} (see also Fig.~\ref{fig:slice_m10_V21_all}). A more careful comparison with the observations, however, is required to distinguish these two possibilities. The distribution of metallicity might be a useful tracer since the type-\Romannum{1} CD separates the ICM from two different clusters (unlike type-\Romannum{3} CDs). The same tracer might work for type-II CDs too, if there is a gradient of metallicity in the ICM of the main halo.

\section{Discussion} \label{sec:discussions}

\subsection{Anisotropic smooth mass accretion in galaxy clusters} \label{sec:discussions:mar}

The spherical self-similar model, albeit simplified, captures several key features of galaxy clusters, including DM caustic surfaces and accretion shocks. In this work, we try to bridge this model and cosmological simulations of galaxy clusters. A key assumption in the self-similar model that the smooth matter accretion is isotropic obviously deviates from the reality, where the accretion is highly anisotropic even when the merger effects are excluded. The smooth mass accretion is supposed to be high along the filaments and low in directions without filaments \citep[e.g.,][]{Molnar2009,Zinger2016}. Therefore, in this section, we consider a two-accretion-mode model  to qualitatively discuss the implications of our simulation results for more realistic situations.

In this model, the cluster is assumed to grow via two independent (low and high) accretion modes. Each mode is described by a self-similar solution but with different values of $\Gamma_{\rm s}$. They represent the smooth mass accretions along the non-filamentary and filamentary directions, respectively. Note that in the self-similar model the higher $\Gamma_{\rm s}$ reflects the larger amplitude of the gas and DM density profiles in the infall region outside the halo  \citep{Shi2016}. The evolution of the total cluster mass $M(z)$ can then be written as
\be
M(z) = M(0)\Big[\chi a(z)^{\Gamma_{\rm s,l}} + (1-\chi)a(z)^{\Gamma_{\rm s,h}}\Big],
\label{eq:mar_l_h}
\ee
where $\chi$ is the mass fraction of the cluster at $z=0$ contributed by the low-accretion mode; $\Gamma_{\rm s,l}$ and $\Gamma_{\rm s,h}$ are the smooth MAR parameters of the low and high accretion modes ($\Gamma_{\rm s,l}<\Gamma_{\rm s,h}$), respectively. The averaged smooth MAR parameter of the cluster is then
\be
\langle\Gamma_{\rm s}\rangle=\Gamma_{\rm s,l}+\frac{(1-\chi)(\Gamma_{\rm s,h}-\Gamma_{\rm s,l})a^{\Gamma_{\rm s,h}-\Gamma_{\rm s,l}}}{\chi+(1-\chi)a^{\Gamma_{\rm s,h}-\Gamma_{\rm s,l}}},
\label{eq:mar_mean}
\ee
based on the definition given in Eq.~(\ref{eq:mar}). Collisionless DM particles are subject to the deepening of the gravitational potential. Therefore, the azimuthally-averaged $r_{\rm sp}$ is mostly determined by the total $\langle\Gamma_{\rm s}\rangle$ rather than $\Gamma_{\rm s,l}$ or $\Gamma_{\rm s,h}$. For the gaseous atmospheres, the evolution of $r_{\rm sp}/r_{\rm mas}$ depends on the local $\Gamma_{\rm s}$ (see, e.g., Figs.~\labelcref{fig:bnd_evo_s1,fig:bnd_evo_s3}). In non-filamentary directions with likely low smooth MAR, i.e., $\Gamma_{\rm s,l}\sim1$, one expects strong accretion/MA-shock and an offset between $r_{\rm sp}$ and $r_{\rm mas}$. This is consistent with the predictions from cosmological simulations. In direction along the filaments, if the mass accretion is high (e.g., $\Gamma_{\rm s,h}\gtrsim3$ and also $\Gamma_{\rm s,h}>\langle\Gamma_{\rm s}\rangle$), the ratio $r_{\rm mas}/r_{\rm sp}$ would  always be $\lesssim1$, as shown in the top panels of Fig.~\ref{fig:bnd_evo_s3}.

\citet{Molnar2009} noted two groups of shocks existing in cluster outskirts in the cosmological simulations. They have high and low Mach numbers, respectively. The latter are near the cluster virial radius (albeit with large scatter, $\sim0.5r_{\rm vir}$) and sometimes referred to as ``virial'' shocks. They could be either the runaway merger shocks in the non-filamentary directions \citep{Zhang2019b,Zhang2021} or the accretion (or sometimes MA-) shocks in the filamentary directions. Our simulations have demonstrated that, when $\Gamma_{\rm s}=3$, $r_{\rm mas}\simeq r_{\rm vir}$ (see Fig.~\ref{fig:bnd_evo_s3} and also \citealt{Shi2016}), and major mergers hardly change it. If we assume that the ``virial''  shock is formed in the filamentary directions, we could estimate $\Gamma_{\rm s,h}\sim 3$ in the filamentary directions.

\subsection{Predicting smooth MAR from extended Press--Schechter formalism} \label{sec:discussions:mar_eps}

We have shown that the smooth MAR $\Gamma_{\rm s}$ plays an important role in determining the splashback and MA-shock radii of galaxy clusters in our simulations. In reality, the situation is more complicated. We need at least to distinguish $\Gamma_{\rm s,l}$, $\Gamma_{\rm s,h}$, and $\langle\Gamma_{\rm s}\rangle$ (see Eqs.~\ref{eq:mar_l_h} and \ref{eq:mar_mean}) to consider the anisotropy of the smooth accretion in clusters. It is also important to directly measure such quantities in cosmological simulations, which is non-trivial\footnote{\citet{Sugiura2020} proposed a way to measure the $\Gamma_{\rm s}$ parameter for the DM halos in cosmological simulations by fitting the self-similar model to the halo's phase-space distribution. However, since the particles in the inner regions of the halos are accreted earlier and reflect the MAR at a higher redshift, the best-fit $\Gamma_{\rm s}$ tends to be biased to a higher value than the present one (see their section~5). The redistribution of DM particles in (major) mergers may also affect the fitting results.}.

Alternatively, we could determine $\Gamma_{\rm s}$ (or $\langle\Gamma_{\rm s}\rangle$) in the $\Lambda$CDM universe in a semi-analytical way, i.e., by modeling the MAR of galaxy clusters using a Monte-Carlo approach based on the extended Press--Schechter (EPS) formalism \citep{Bond1991,Lacey1993}. Following the method developed in \citet{Parkinson2008}, we plant halo merger trees for galaxy clusters from redshift $z=0$ to $4$. Note that, instead of the Einstein-de Sitter universe assumed in our merger simulations, we use the same cosmological parameters (a flat $\Lambda$CDM cosmology) as those by \citet{Parkinson2008} in this subsection. The mass assembly history of a cluster is assumed to track the tree's main branch, which starts from the host halo at the root and always walks downwards along the branch of the most massive progenitor until reaching the leaf node.

\begin{figure}
\centering
\includegraphics[width=0.9\linewidth]{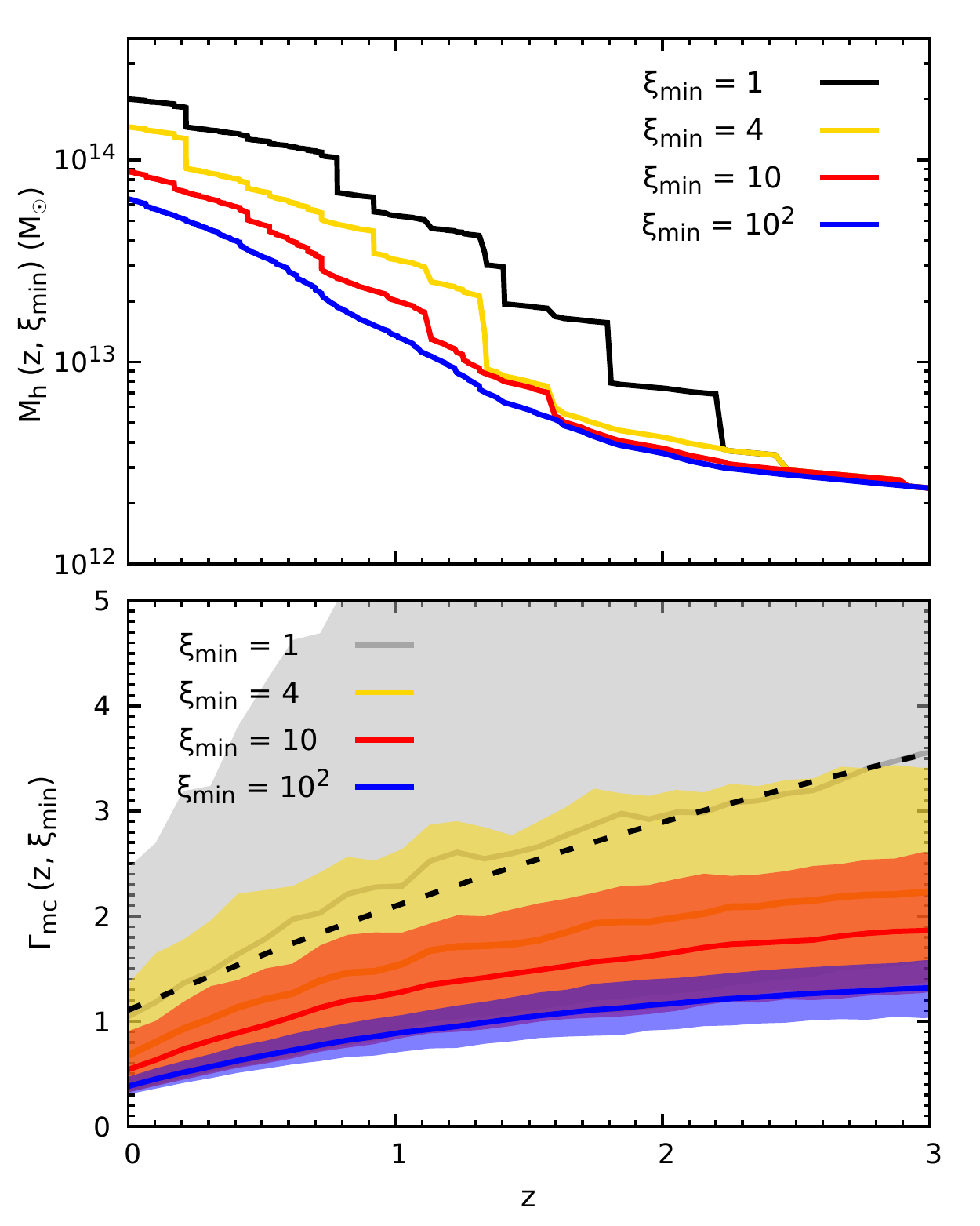}
\caption{\textit{Top panel:} Mass assembly history of a halo with the final mass $2\times10^{14}\msun$ at $z=0$ modeled by the Monte Carlo method (black line; see \citealt{Parkinson2008}). The yellow, red, and blue lines show the evolution of the halo mass while excluding all contributions from the mergers with the mass ratio $\xi\leq\xi_{\rm min}=4,\ 10$, and $10^2$, respectively. \textit{Bottom panel:} Evolution of the MAR parameter $\Gamma_{\rm mc}$ when $\xi_{\rm min}=1,\ 4,\ 10,\ 10^2$ (see Eq.~\ref{eq:mar_mc}). The solid lines and shaded areas show the mean rates and their scatters (10th--90th percentiles) over $10^3$ halo assembly histories, respectively. The dashed black line represents the best-fit mean MAR from the Millennium cosmological simulations \citep{Fakhouri2010}. This figure shows that the clusters' smooth MAR is $\simeq1$ when $z<3$, prominently smaller than their full MAR (see Section~\ref{sec:discussions:mar_eps}).}
\label{fig:mar_mtrees}
\end{figure}

An example of such assembly histories is exhibited in the top panel of Fig.~\ref{fig:mar_mtrees} (black line). One can see that the cluster experiences multiple major and minor mergers since $z=3$. To further quantify the mass contributions from the mergers by different merger mass ratio $\xi$, we estimate the evolution of the halo mass $M_{\rm h}(z,\,\xi_{\rm min})$ excluding all the contributions from the mergers with $\xi\leq\xi_{\rm min}$ later than redshift $z_{\rm s}$ (i.e., $M_{\rm h}(z_{\rm s},\,\xi_{\rm min})\equiv M_{\rm h}(z_{\rm s},\,1)$). In the top panel of Fig.~\ref{fig:mar_mtrees}, the yellow, red, and blue lines illustrate the evolution of $M_{\rm h}(z,\,\xi_{\rm min})$ when $\xi_{\rm min}=4,\ 10$ and $10^2$ ($z_{\rm s}=3$), respectively. One can see that about one third of the cluster's final mass comes from the smooth accretion and small minor mergers ($\xi\gtrsim10^2$; see \citealt{Genel2010}).

We then estimate the halo's MAR as
\be
\Gamma_{\rm mc}(z,\,\xi_{\rm min})=\frac{\log{[M_{\rm h}(z,\,\xi_{\rm min})/M_{\rm h}(z+\Delta z,\,\xi_{\rm min})]}}{\log{[a(z)/a(z+\Delta z)]}}\frac{M_{\rm h}(z,\,\xi_{\rm min})}{M_{\rm h}(z,\,1)},
\label{eq:mar_mc}
\ee
where $\Delta z = |z(t) - z(t-t_{\rm dyn})|$ is the redshift interval that is separated by the halo's dynamical time-scale $t_{\rm dyn}\equiv(4\pi G\Delta_{\rm vir}\rho_{\rm c}(z)/3)^{-1/2}$. This definition is close to those widely used in the literature when $\xi_{\rm min}=1$ \citep[e.g.,][]{Diemer2017}. We note that it is meaningless to choose a much shorter time interval here (cf., Eq.~\ref{eq:mar_vir}) since the mergers are completed instantaneously in the merger trees. In this work, we fix the final mass of the cluster $M_{\rm h}(0,\,1)=2\times10^{14}\msun$ and randomly generate $10^3$ assembly histories.

The bottom panel of Fig.~\ref{fig:mar_mtrees} shows the evolution of the modeled $\Gamma_{\rm mc}(z,\,\xi_{\rm min})$ when $\xi_{\rm min}=1,\ 4,\ 10,\ 10^2$. The solid lines and shaded areas show the mean MARs and the 10--90th percentile ranges, respectively. As we expect, the MAR is a strong function of $\xi_{\rm min}$. The full MAR (i.e., $\xi_{\rm min}=1$) ranges $\sim1-4$ from $z=0$ to $3$, which has been well known from the previous studies \citep[see, e.g.,][]{Zhao2009,Jiang2014}. Its large scatter is mostly due to the (major) mergers. For comparison, we show the best-fit mean MAR from the Millennium cosmological simulations as the dashed black line in the figure \citep{Fakhouri2010}. The good agreement here is not surprising because our merger trees are calibrated by the Millennium simulations as well \citep{Parkinson2008}. When the $\xi_{\rm min}$ gets larger, the corresponding MAR and its scatter become smaller. The $\Gamma_{\rm mc}$ spans $\simeq0.5-2$ ($z=0-3$) when $\xi_{\rm min}\leq10$, which is mostly contributed by the smooth accretion and mildly increases with redshift. This result is broadly in line with our constraint on $\Gamma_{\rm s}$ discussed in Section~\ref{sec:results:dm:rsp}.

We emphasize here again that it is crucial to separate the smooth MAR $\Gamma_{\rm s}$ from $\Gamma_{\rm vir}$ (see Eqs.~\ref{eq:mar} and \ref{eq:mar_vir}) when analyzing the mass assembly of a cluster. The former is a key factor in characterizing the mass distributions beyond $r_{\rm vir}$ of galaxy clusters. We also suggest that the method to estimate $\Gamma_{\rm mc}(z,\,\xi_{\rm min})$ used in this subsection might be a robust way to constrain the $\Gamma_{\rm s}$ parameter for individual galaxy clusters in cosmological simulations. However, we also stress that the merger scenario in the EPS formalism is still highly simplified because (1) no structure of the merging halos is resolved, (2) mergers complete instantaneously, and (3) possible correlations among different merger events and correlations between mergers and periods of increased smooth mass accretion are missing in this ansatz. These simplifications may partially affect the scatter of $\Gamma_{\rm mc}(z,\,\xi_{\rm min})$ shown in Fig.~\ref{fig:mar_mtrees}.

\subsection{The effect of non-radial orbits of galaxies on the splashback location in observed clusters} \label{sec:discussions:obs_rsp}

Observationally, the splashback radii of galaxy clusters have been detected in the stacked member-galaxy surface density \citep[e.g.,][]{More2016,Chang2018,Murata2020} and weak lensing measurements \citep{Chang2018,Contigiani2019}. Some observations measured a lower $r_{\rm sp}/r_{\rm vir}$ in the optically-selected clusters than the theoretical expectations from the cosmological simulations (see \citealt{Busch2017} for a discussion on the selection and projection effects on this measurement). Such systematic bias, however, disappears if the SZ cluster samples are used instead \citep[e.g.,][]{Shin2019,Adhikari2020}. Theoretically, there is no guarantee that the backsplash galaxies share the same outer boundary with the smoothly accreted DM because of the dynamical friction \citep{Adhikari2016,Oneil2020} and the non-radial orbits of the infalling objects. The latter is rarely mentioned in the literature, and we consider it below.

\begin{figure}
\centering
\includegraphics[width=0.9\linewidth]{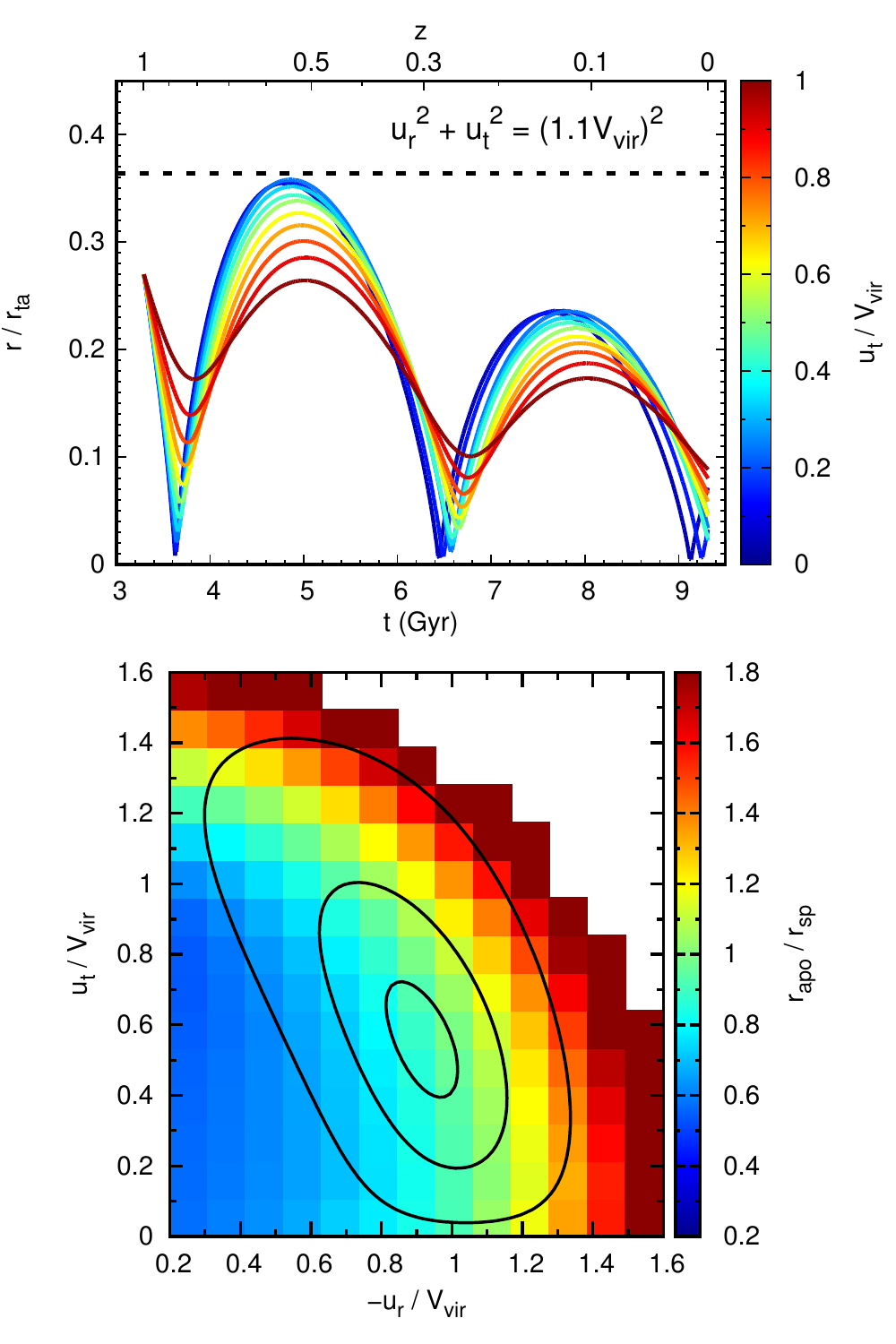}
\caption{\textit{Top panel:} Trajectories of massless test particles in a self-similar cluster ($\Gamma_{\rm s}=1$). These particles are initially set at the virial radius at redshift $z=1$ and have an initial speed of $1.1V_{\rm vir}$. The line color encodes the particles' initial tangential velocities. The horizontal dashed line shows the splashback radius of the self-similar cluster $r_{\rm sp}$. This figure shows that the apocentric distance of galaxies depends on their non-radial orbits. \textit{Bottom panel:} Primary apocentric distance of the test particles in units of $r_{\rm sp}$ as a function of the particles' initial radial and tangential velocities. The overlaid black contours show the best-fit 2D distribution of orbital velocities of the infalling DM subhalos at the host halo's virial radius from cosmological simulations \citep{Benson2005}. These three contours indicate the regions covering 90, 50, and 10 per cent of the subhalo sample (from the outer to inner), respectively. The averaged $r_{\rm apo}/r_{\rm sp}$ within the regions enclosed by the contours is smaller than the unity, which implies that the non-radial orbits of the clusters' member galaxies may affect the measurements of the splashback radius (see Section~\ref{sec:discussions:obs_rsp}).}
\label{fig:galaxy_orbits}
\end{figure}

To illustrate the effect of the galaxies' non-radial orbits on their apocenters, we estimate the trajectories of the massless test particles in the self-similar cluster with different initial velocities. The top panel in Fig.~\ref{fig:galaxy_orbits} shows an example of test particles that are initially set at the virial radius at redshift $z=1$ and have an initial speed of $1.1V_{\rm vir}$, where $V_{\rm vir}\equiv\sqrt{GM_{\rm vir}/r_{\rm vir}}$ is the virial velocity. All test particles fall into the halo for the first time. These particles show different trajectories due to their various initial radial ($u_{\rm r}$) and tangential ($u_{\rm t}$) velocities. Their primary apocentric distance $r_{\rm apo}$ is smaller if the initial velocity has a larger tangential component. The galaxies' non-radial orbits, therefore, impact the splashback radius of galaxy clusters determined from their member galaxy distributions, even if the galaxies are so small that the dynamical friction is negligible.

The bottom panel of Fig.~\ref{fig:galaxy_orbits} further shows the primary apocentric distance of the test particles in units of $r_{\rm sp}$ as a function of the initial radial and tangential velocities. For comparison, we overplot the best-fit two-dimensional (2D) distribution of the infalling velocities of the subhalos $f(u_{\rm r},\,u_{\rm t})$ at the host halo's virial radius from the cosmological simulations (black contours; see \citealt{Benson2005} and their eqs.~2--4 and parameters for $z=1$). The contours (from the outer to inner) indicate the regions covering 90, 50, and 10 per cent of the total sample, respectively. This figure shows that the splashback radius inferred by the distribution of small subhalos is close to the real $r_{\rm sp}$. We simply estimate the mean ratio $r_{\rm apo}/r_{\rm sp}$ within the three regions marked by the contours, i.e., $\int{f(u_{\rm r},\,u_{\rm t})(r_{\rm apo}/r_{\rm sp})\dd u_{\rm r}\dd u_{\rm t}}=0.96,\ 0.92$, and $0.90$ (from the outer to inner). It implies that, besides the dynamical friction, the non-radial orbits of galaxies could lead to an underestimation of the splashback radius as well. However, this conclusion depends on the distribution of the infalling galaxies' velocities \citep[e.g.,][]{Vitvitska2002,Benson2005,Jiang2015} and the angular momentum of the DM particles smoothly accreted into the cluster. The latter is assumed to be zero in this discussion.

We suggest that a detailed comparison of the splashback radii measured in the weak lensing and galaxy distribution may provide an opportunity to explore the statistics of the galaxies' motions (e.g., distributions of the relative velocities and orbital ellipticity) before they enter the cluster. Such measurements for galaxies with different properties (e.g., star formation, color, morphological types, etc.) can indicate if galaxy orbits in clusters play a role in shaping these properties \citep[e.g.,][]{Adhikari2020}.

\begin{figure}
\centering
\includegraphics[width=0.9\linewidth]{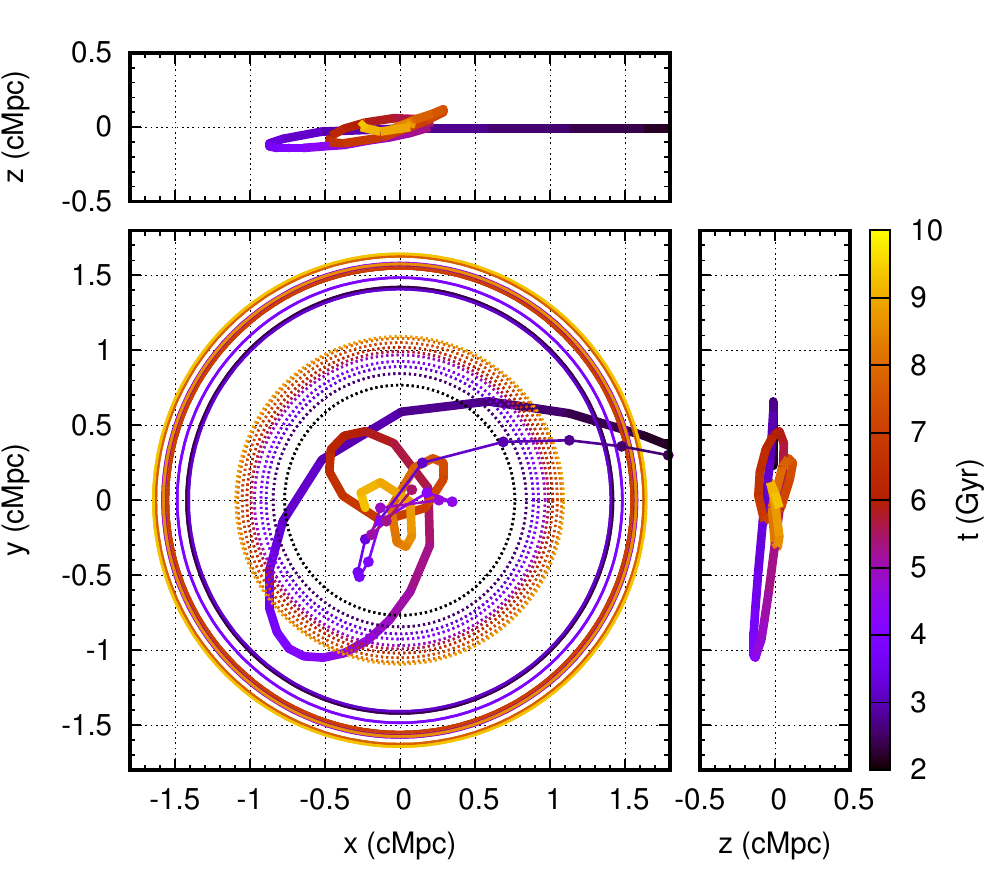}
\caption{Three orthographically-projected views of the orbital trajectory of the infalling subcluster in the simulation $\rm\xi10V21\Gamma1$ (thick solid lines). The line color encodes the cosmic time. The center of the main cluster is at the origin of the coordinates. The thin solid and dotted circles show the evolution of the azimuthally-averaged splashback radii and cluster virial radii, respectively. For comparison, the projected trajectory of the subcluster in the $x-y$ plane in the run $\rm\xi10V1h\Gamma1$ is overlaid in the bottom-left panel as the points connected by thin lines. This figure shows that the orbital trajectories of the infalling subclusters are sensitive to the initial merger parameters (e.g., $\mathbf{v}_0$; see Section~\ref{sec:discussions:obs_rsp}).  }
\label{fig:traj_halo}
\end{figure}

For massive galaxies, the dynamical friction and tidal effects are non-negligible, making the situation more complicated \citep[see][]{Adhikari2016}. Fig.~\ref{fig:traj_halo} shows the orbital trajectory of the infalling subcluster in our simulation $\rm\xi10V21\Gamma1$ (thick solid line). The line color encodes the cosmic time. The solid and dashed circles show the evolution of $r_{\rm sp}$ and $r_{\rm vir}$, respectively. For comparison, we also show the projected trajectory of the subcluster in the run $\rm\xi10V1h\Gamma1$ in the bottom-left panel (points connected by the thin lines). This figure shows that the apocentric distance of the subclusters is sensitive to the infalling orbits but very different from that predicted in Fig.~\ref{fig:galaxy_orbits}, since the dynamical friction is a function of the object's mass, velocity, and the DM density of the environment. In such a situation, the infalling object along a purely radial orbit may have a smaller apocentric distance than that in a non-radial orbit.

\subsection{Role of MA-shocks in ram-pressure stripping infalling galaxies}  \label{sec:discussions:quenching}

MA-shocks prominently extend the gaseous atmospheres of galaxy clusters up to $\simeq2-3r_{\rm vir}$. The infalling field galaxies would therefore feel the ram pressure even when they are far beyond the cluster virial radius \citep[e.g.,][]{Book2010,Bahe2013,Zinger2018a,Arthur2019}. Once the galaxies cross the MA-shock front, their hot gaseous atmospheres feel a drastic increase of the ram pressure because (1) the gas density of the ambient medium raises by a factor of $4$ due to a high Mach number of a MA-shock; and (2) the newly accreted gas stops falling together with the galaxies within the $r_{\rm mas}$. In our simulations, we find that the ICM gas density between $r_{\rm vir}$ and $r_{\rm mas}$ is approximately uniform $\rho_{\rm icm}\simeq10\rho_{\rm c}\Omega_{\rm b}/\Omega_{\rm m}$ (see Fig.~\ref{fig:slice_m2_V1h_all}) and its radial velocity is smaller than $V_{\rm vir}/4$ (either negative or positive).

We quantify the ram pressure role in striping galaxies in cluster outskirts with the method described in \citet[][see also \citealt{Gunn1972}]{McCarthy2008}. The galaxy's hot atmosphere is peeled off beyond the radius $r$ if the ram-pressure force is larger than the gravitational force, i.e.,
\be
\rho_{\rm icm}u_{\rm rel}^2>\alpha\frac{GM_{\rm gal}(r)\rho_{\rm gal,gas}(r)}{r},
\label{eq:ram_pres}
\ee
where $M_{\rm gal}(r)$ and $\rho_{\rm gal,gas}(r)$ are the galaxy's enclosed total mass and gas density at the radius $r$, $u_{\rm rel}$ is the relative velocity between the galaxy and the ICM, $\alpha$ is a geometric parameter. \citet{McCarthy2008} suggested that $\alpha=2$ provides consistent results between their analytical and numerical models, which we adapt in this study as well. To qualitatively understand the efficiency of the ram pressure, we specifically evaluate Eq.~(\ref{eq:ram_pres}) at the galaxy's virial radius ($r_{\rm gal,vir}$) and $r_{500}$ (referred to as $r_{\rm gal,500}$ hereafter), and simply assume $\rho_{\rm icm}=10\rho_{\rm c}\Omega_{\rm b}/\Omega_{\rm m}$, $u_{\rm rel}=V_{\rm vir}$, and $M_{\rm gal}(r_{\rm gal,500})=M_{\rm gal}(r_{\rm gal,vir})/2$ in the calculation. The Eq.~(\ref{eq:ram_pres}) could thus be re-written as
\be
\xi_{\rm gal}\equiv\frac{M_{\rm vir}}{M_{\rm gal}(r_{\rm gal,vir})}\gtrsim\Big(\frac{\alpha\Delta}{10}\Big)^{3/2},
\label{eq:ram_pres_xi}
\ee
where $\Delta=\Delta_{\rm vir}$ and $500$ at $r_{\rm gal,vir}$ and $r_{\rm gal,500}$, respectively. Note that the gas fraction of galaxies is assumed to be equal to that of the ICM in Eq.~(\ref{eq:ram_pres_xi}). We finally find that the ICM could remove galaxies' atmosphere at $r_{\rm gal,vir}$ if $\xi_{\rm gal}\gtrsim10^2$ -- the condition for the ICM to have an effect on the infalling galaxies beyond the virial radius. Moreover, a large fraction of the galaxies' atmospheres would be stripped, namely, reaching at least $r_{\rm gal,500}$, if $\xi_{\rm gal}\gtrsim10^3$. While the cluster-galaxy merger mass ratio largely determines the efficiency of the stripping process, this may not work for the galaxies entering the cluster along filamentary directions. Though the gaseous environment is denser in the filaments, the relative velocity between the galaxies and the surrounding atmosphere in the filaments might be much smaller than the virial velocity.

\section{Conclusions} \label{sec:conclusions}

In this study, we perform idealized cosmological simulations to explore how mergers shape the DM and gas distributions in the outskirts of galaxy clusters. Specifically, we simulate mergers between two galaxy clusters. Each cluster is initialized using the spherical self-similar collapse model with the accretion shock and DM splashback feature self-consistently included. Comparing the results of the merger simulations with the self-similar baseline model allows us to investigate the effects of a merger on the properties of the merger remnant, particularly the locations of the splashback boundary and the external shock. Results of this study can be used to bridge the cluster's self-similar model and the cluster formation simulations in a full cosmological context.

We investigate in detail the evolution of the outer DM caustic (splashback) during and after the merger, development and propagation of the merger-driven shocks (including the external shocks, namely MA-shocks) and CDs in the gas atmosphere, and evolution of the shape of the DM, gas, and gravitational potential. Our simulation results support the interpretation that mergers drive the prominent offsets between the external shock and splashback radii in galaxy clusters measured in cosmological simulations. We further discuss the effects of smooth accretion and mergers on the evolution of clusters, the effects of orbital parameters of member galaxies on the location of the observational splashback feature, and the role of the MA-shock in stripping infalling galaxies' atmospheres. Our main findings are summarized as follows.
\begin{itemize}
  \item Major mergers strongly affect the splashback radii of galaxy clusters with a small smooth MAR parameter ($\Gamma_{\rm s}\lesssim1$) used to set up the self-similar initial condition. During the core passage, the matter experiences a rapid contraction. The splashback radius of the main cluster shrinks significantly. For example, in mergers with the mass ratio $\xi\lesssim 2$, the splashback radius decreases to $\simeq r_{\rm vir}$, while in larger mass-ratio mergers, the decrease is still significant and reaches the minimum value of $r_{\rm sp}/r_{\rm vir} \simeq 1.3-1.4$ (see Fig.~\ref{fig:bnd_evo_dm_mar}). The splashback radius restarts to increase shortly after the core passage and approaches the self-similar prediction after about $1-2\tau_{\rm dyn}$.
  \item We show that a fraction of particles belonging to the subcluster is in resonance with the merger. They experience increases in their radial velocities while approaching the main cluster, achieve large kinetic energies, and could reach up to $\simeq5r_{\rm vir}$ of the merger remnant (see Figs.~\labelcref{fig:part_dm_rv,fig:part_dm_rv_sub}). At later epochs, the outermost caustic surface of the merger remnant is mostly constituted by the newly smoothly accreted DM particles.
  \item The total MAR $\Gamma_{\rm vir}$ experienced by the main cluster during the merger (see Eq.~\ref{eq:mar_vir}) can be much larger than the smooth MAR parameter $\Gamma_{\rm s}$. We find that the $r_{\rm sp}/r_{\rm vir}$ ratio exhibits a strong temporal correlation with the evolution of $\Gamma_{\rm vir}$ but with a $\simeq 1\Gyr$ time offset (see Fig.~\ref{fig:bnd_evo_dm_mar}). The form of the correlation, however, is sensitive to the way we define $\Gamma_{\rm vir}$ (see Fig.~\ref{fig:mar_rsp_corr}).
  \item The major mergers of the clusters with low initial $\Gamma_{\rm s}\ (\simeq1)$ make dominant contribution to the scatter of the relation $\Gamma_{\rm vir}-r_{\rm sp}/r_{\rm vir}$ on the high $\Gamma_{\rm vir}\ (\gtrsim3$) end. Therefore, the large scatter of this scaling relation measured in the cosmological simulation indicates that clusters generally have a low rate of smooth accretion (i.e., $\Gamma_s\sim 1$; see Section~\ref{sec:results:dm:rsp}). We show that this conclusion is consistent with the direct measurements of the smooth accretion rate from the mass assembly histories generated using the EPS ansatz (see Section~\ref{sec:discussions:mar_eps}). Our results imply that it is important to distinguish the $\Gamma_{\rm s}$ and $\Gamma_{\rm vir}$ rates of galaxy clusters when exploring the evolution of their splashback radii.
  \item We measure triaxial-ellipsoid axis ratios of cluster's DM halo, gas halo, and gravitational potential at different radii during and after mergers. As expected, we find the merger remnant is significantly aspherical, with the major axis of the ellipsoid closely aligned with the merger axis. The axis ratios of the DM halos are generally $b/a\simeq0.6-0.8$ and $c/a\simeq0.4-0.6$ within the splashback radius and only mildly depend on the merger parameters (e.g., merger mass ratio, impact parameter). However, they quickly drop to lower values around $r\simeq r_{\rm sp}$ (see Fig.~\ref{fig:shape_dm}), implying that mergers leave an imprint in the strongly aspherical shape of the DM distribution near the splashback surface.
  \item The hot atmospheres of galaxy clusters could significantly expand by the merger-driven MA-shocks, which form when runaway shocks generated by mergers overtake the pristine accretion shocks. Our simulations show that the smooth MAR $\Gamma_{\rm s}$ and merger mass ratio $\xi$ are the key parameters that determine the evolution of the MA-shocks during mergers (see Figs.~\labelcref{fig:bnd_evo_s1,fig:bnd_evo_s3}). When $\Gamma_{\rm s}\lesssim1$, the MA-shock fronts reach radii much larger than the DM splashback boundaries, e.g., $r_{\rm mas}/r_{\rm sp}\simeq1.7\ (\Gamma_{\rm s}=0.7)$ and $1.3\ (\Gamma_{\rm s}=1)$ when $\xi=2$. In contrast, when the smooth MAR is high (e.g., $\Gamma_{\rm s}\gtrsim3$), the MA shock does not propagate beyond $\simeq r_{\rm sp}$, even during a major merger.
  \item Given the dependence of the maximum radius of the MA-shock on $\Gamma_{\rm s}$ in our merger simulations and the fact that the external shocks of galaxy clusters found in cosmological simulations are generally at $r_{\rm mas}/r_{\rm sp}\simeq1.5-2.5$, we conclude that (1) these external shocks are mostly driven by the runaway shocks formed in the mergers. Namely, they are MA-shocks rather than the ordinary accretion shocks (see Section~\ref{sec:results:gas:ma-shock}). (2) The MAR in the directions away from filaments, where the shock extends the most, is small (see Section~\ref{sec:discussions:mar}).
  \item Multiple shock structures are developed in the cluster pre-merger stage. In major mergers, the interaction of the accretion shocks of the two merging clusters drives a pair of reverse shocks and further a strong Mach stem. When the two clusters get closer, bow shocks are formed in front of them. The compressed gas between the clusters generates a hot outflow propagating perpendicularly to the merger axis (see Fig.~\ref{fig:slice_zoom_pre_m2}).
  \item We classify the CDs formed in cluster mergers into three types based on their formation scenarios. They could co-exist in a single merger event and are well captured in our idealized simulations (see Section~\ref{sec:results:gas:cd}). In particular, the shock-driven CDs, formed in the collisions between the runaway and accretion shocks, are located in the cluster outskirts (from $\simeq0.5r_{\rm vir}$ to $r_{\rm mas}$) and survive for $\simeq3-5\Gyr$ in our simulations. These CDs and the MA-shocks may further drive turbulent gas motions in the outer cluster regions.
  \item We show that the location of the splashback feature traced by cluster galaxies depends on their orbital parameters during the infall. The bias is small if the galaxies have similar orbital-parameter distributions with the smoothly accreted DM. However, it can become substantial and bias the measured splashback radius low for subsets of galaxies with more tangential orbits (see Section~\ref{sec:discussions:obs_rsp}). Conversely, such bias detected for subsets of the cluster galaxies with specific properties (e.g., color, star formation, morphology) can be useful to probe the connections between these properties and galaxy orbits.
  \item In galaxy clusters, the gas between $r_{\rm vir}$ and $r_{\rm mas}$ can substantially boost the ram-pressure stripping efficiency of the hot gaseous atmospheres of the infalling galaxies. We argue that the cluster-to-galaxy mass ratio $\xi_{\rm gal}$ is a key factor that determines the efficiency of this process. The effect is present when $\xi_{\rm gal}\gtrsim10^2$ and becomes efficient when $\xi_{\rm gal}\gtrsim10^3$ (see Section~\ref{sec:discussions:quenching}).
\end{itemize}

\section*{Acknowledgments}

CZ and IZ would like to thank Chihway Chang for useful discussions on the splashback radius observations with galaxy surveys. We are also grateful to Neal Dalal for useful discussions related to dynamics of mergers and splashback. Part of the simulations presented in this paper were carried out using the Midway computing cluster provided by the University of Chicago Research Computing Center. IZ is partially supported by a Clare Boothe Luce Professorship from the Henry Luce Foundation. AK was supported by the NSF grants AST-1714658 and AST-1911111 and NASA ATP grant 80NSSC20K0512.

\section*{Data Availability}

The data underlying this article will be shared on reasonable request to the corresponding author.

\appendix

\section{Simulation method} \label{sec:appendix:simulation}

In this section, we provide additional technical details of our simulations that are not fully covered in Section~\ref{sec:method}.

\subsection{Modeling of DM halos} \label{sec:appendix:simulation:dm}

We generate the DM particles for a self-similar cluster in the following way. We first select a cosmic time $t_{\rm ita}$ much earlier than the starting time of our simulations, i.e., $t_{\rm ita}=10^{-4}t(z_{\rm ini})$. The turnaround radius of the halo at $t_{\rm ita}$ is, therefore, $r_{\rm ita}=r_{\rm ta}(z_{\rm ini})\left[t_{\rm ita}/t(z_{\rm ini})\right]^{2(1+\Gamma_{\rm s}/3)/3}$. We then split the volume between radii $r_{\rm ita}$ and $r_{\rm ta}(z_{\rm ini})$ into a number of spherical shells and calculate shell masses at the moments when they reach their turnaround radii. Each shell here corresponds to one DM particle of the halo within $r_{\rm ta}(z_{\rm ini})$ at $t(z_{\rm ini})$. We properly select the inner and outer boundaries of the shells so that they approximately have the same mass. Since the self-similar solution provides the trajectories of these shells \citep{Bertschinger1985,Shi2016}, we can estimate their radial positions and velocities at $t(z_{\rm ini})$. To convert these shells into particles, we randomly assign them 3D positions based on their radial positions.

Generating the DM particles outside of the halo's turnaround radius $r_{\rm ta}(z_{\rm ini})$ is more straightforward based on the self-similar solution. In the same way, we split the volume between $r_{\rm ta}(z_{\rm ini})$ and $\kappa_{\rm max}r_{\rm ta}(z_{\rm ini})$ into spherical shells. The $\kappa_{\rm max}$ determines the outer boundary of the matter distribution surrounding the cluster (see Fig.~\ref{fig:ic_sketch}). To save the simulation run time, we logarithmically increase the shell's mass (from the inner to outer) by a factor of $1+2\times10^{-7}$, so that the mass of the outermost shell is approximately $3-5$ times higher than the innermost one. The latter is set to be the averaged particle mass within $r_{\rm ta}(z_{\rm ini})$. The total number of DM particles in the simulation is reduced by $\sim50$ per cent in such a setting. The bottom panel in Fig.~\ref{fig:init_profs} shows an example of the radial positions and velocities of the DM particles in the cluster with $\Gamma_{\rm s}=1$.

\subsection{Setting-up merger initial conditions} \label{sec:appendix:simulation:ic}

\begin{figure}
\centering
\includegraphics[width=0.9\linewidth]{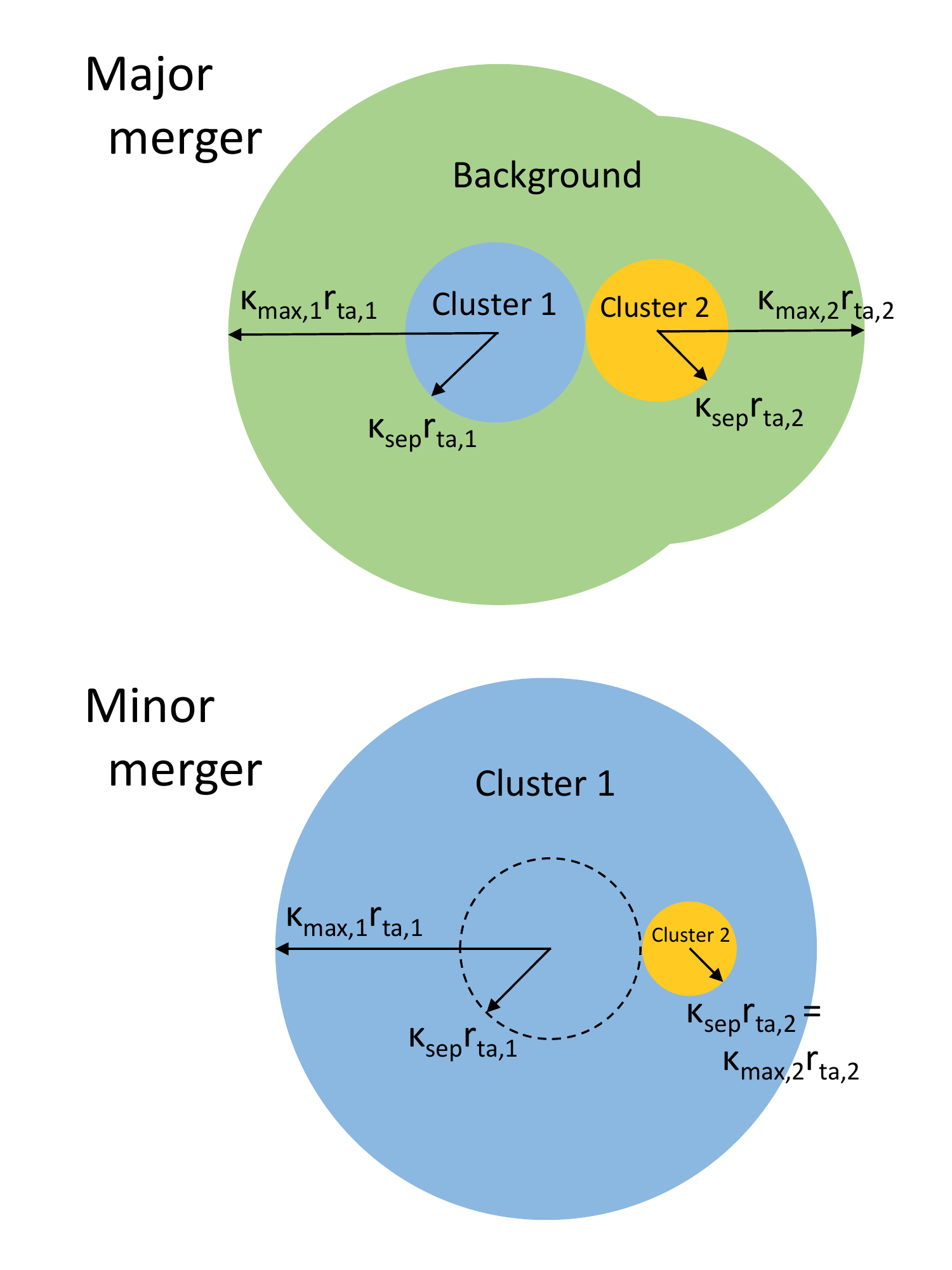}
\caption{A sketch illustrates how we set two self-similar clusters in the initial conditions for major (top) and minor (bottom) merger cases, respectively. In the major mergers, we retain the inner regions of the two clusters (within $\kappa_{\rm sep}r_{{\rm ta},i}$, $i=1,\ 2$; see the blue and yellow regions) and define a `background' area (green) where the matter distribution is a combination of those of two individual clusters. In minor mergers, we simply embed the inner region of the subcluster in the main cluster (see Section~\ref{sec:method:ic} and Appendix~\ref{sec:appendix:simulation:ic}). }
\vspace{-3mm}
\label{fig:ic_sketch}
\end{figure}

Fig.~\ref{fig:ic_sketch} illustrates our strategy to set two self-similar clusters in the initial conditions. Their initial separation is $\kappa_{\rm sep}(r_{\rm ta,1}+r_{\rm ta,2})$. In major mergers, the density and velocity distributions of the DM and gas are a linear combination of those of the individual clusters in a ``background'' region (green area), i.e.,
\be
\rho_{\rm X}(\textbf{r})=\sum_{i=1,\,2}\rho_{{\rm X},i}(|\textbf{r}-\textbf{r}_{{\rm c},i}|) - \rho_{\rm c},
\label{eq:ic_rho}
\ee
\be
\textbf{u}_{\rm X}(\textbf{r})=\frac{\sum_{i=1,\,2}\rho_{{\rm X},i}(|\textbf{r}-\textbf{r}_{{\rm c},i}|)\textbf{u}_{{\rm X},i}(|\textbf{r}-\textbf{r}_{{\rm c},i}|)}{\sum_{i=1,\,2}\rho_{{\rm X},i}(|\textbf{r}-\textbf{r}_{{\rm c},i}|)},
\label{eq:ic_vel}
\ee
where the subscript $\rm X$ represents `DM' and `gas' for the DM and gaseous components, respectively; $\textbf{r}_{{\rm c},i}$ is the vector of the cluster $i$'s center; $\rho_{{\rm X},i}$ and $\textbf{u}_{{\rm X},i}$ are the density and velocity profiles of the individual cluster. In the region $|\textbf{r}-\textbf{r}_{{\rm c},i}|<\kappa_{\rm sep}r_{{\rm ta},i}$, the density and velocity distributions are fully determined by the profiles of the cluster $i$ (i.e., $\rho_{{\rm X},i}$ and $\textbf{u}_{{\rm X},i}$; see the blue and yellow regions). We note that, in such setups, the gas distribution is initially discontinuous at the boundaries $|\textbf{r}-\textbf{r}_{{\rm c},i}|=\kappa_{\rm sep}r_{{\rm ta},i}$, which leads to weak gaseous structures in the early phase of the simulations (see, e.g., the left panels in Fig.~\ref{fig:slice_zoom_pre_m2}). However, it would not affect any conclusions made in this study. In the minor merger cases, we simply embed the subcluster (only the region within $\kappa_{\rm sep}r_{\rm ta,2}$) in the main cluster (see the illustration in Fig.~\ref{fig:ic_sketch}).

\section{Illustration of major and minor mergers} \label{sec:appendix:illustration}

\begin{figure*}
\centering
\includegraphics[width=0.9\linewidth]{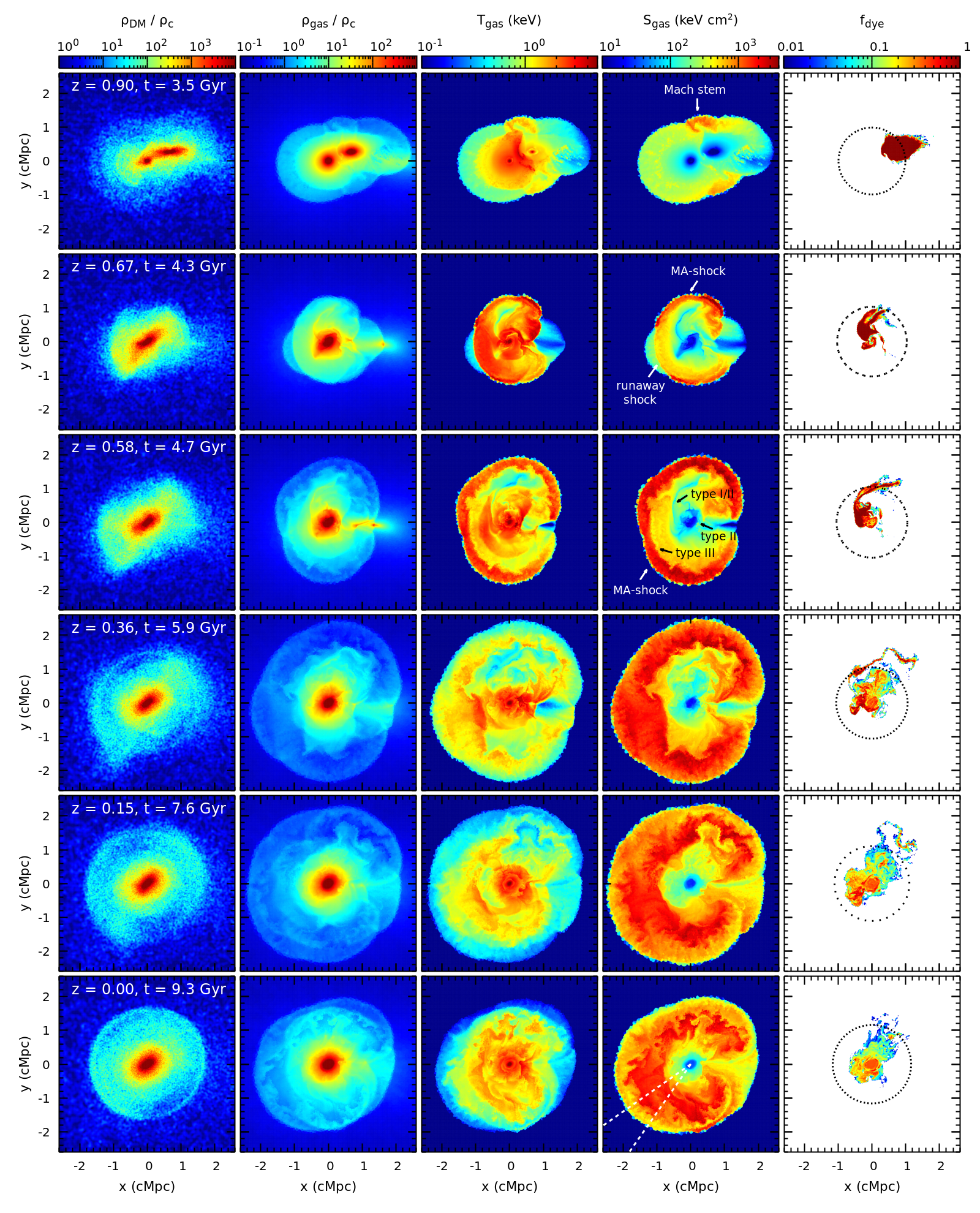}
\caption{An off-axis major merger modeled in the simulation $\rm\xi2V1h\Gamma1$. Its primary pericentric passage occurs at $t\simeq4\Gyr$. The panels show the evolution of the slices of the DM density, gas density, temperature, entropy, and scalar field $f_{\rm dye}$ in the $x-y$ plane (from left to right). The black dotted circles overlaid in the rightmost column show the cluster virial radius. The white dashed lines in the bottom entropy panel indicate the conic region where the entropy profile shown in Fig.~\ref{fig:prof_cd_entr} is measured. Different types of CDs and shock fronts formed during the merger are indicated by the black and white arrows in the entropy slices (see also Figs.~\labelcref{fig:slice_zoom_pre_m2,fig:slice_zoom_pre_m2b,fig:prof_cd_entr}). This figure illustrates an entire major-merger process and its effects on the DM and gas distributions of galaxy clusters (see Sections~\labelcref{sec:results:dm,sec:results:gas} and Appendix~\ref{sec:appendix:illustration}).}
\label{fig:slice_m2_V1h_all}
\end{figure*}

\begin{figure*}
\centering
\includegraphics[width=0.9\linewidth]{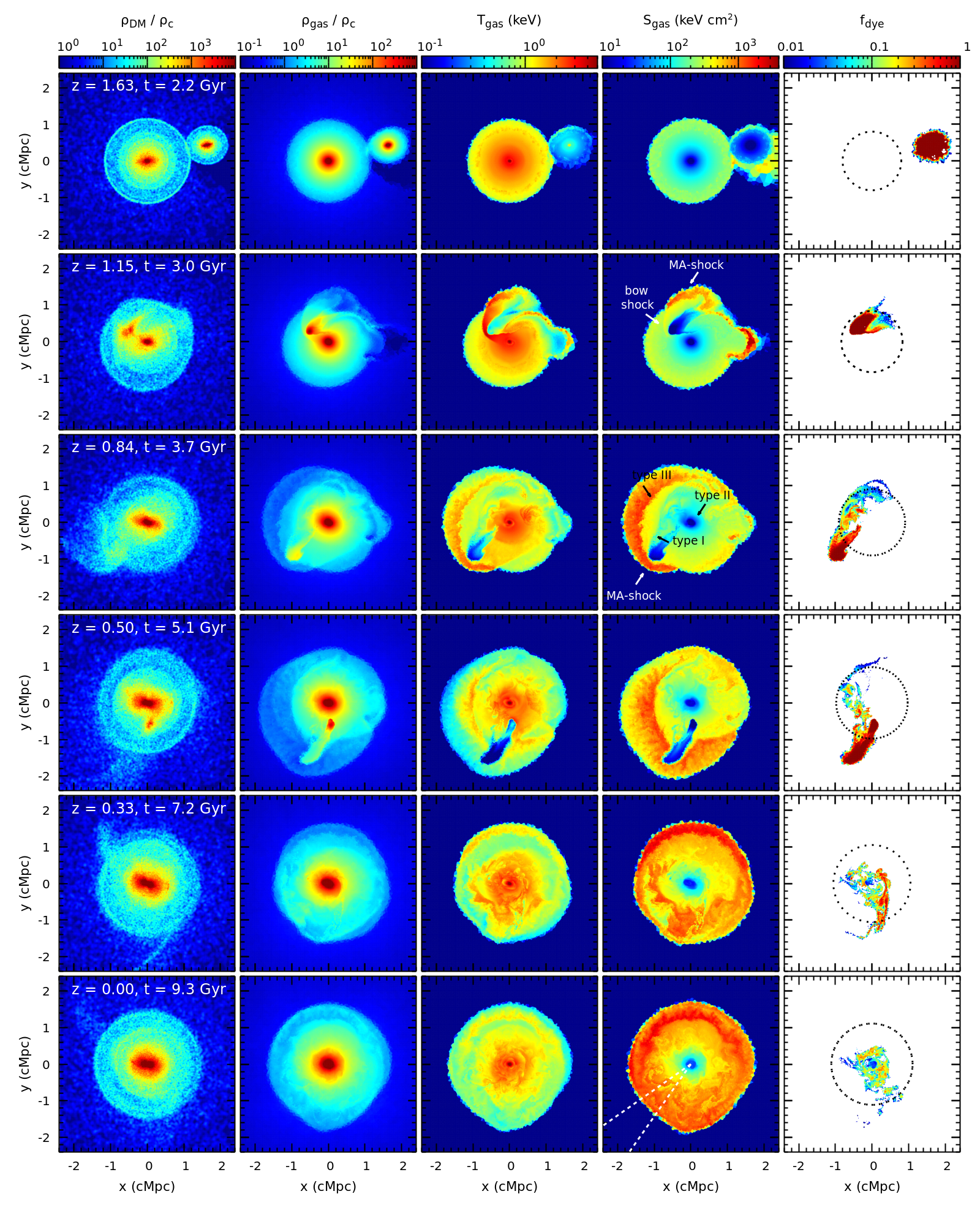}
\caption{Similar to Fig.~\ref{fig:slice_m2_V1h_all} but for a minor merger in the simulation $\rm\xi10V21\Gamma1$ (see Section~\ref{sec:results:gas} and Appendix~\ref{sec:appendix:illustration}). }
\label{fig:slice_m10_V21_all}
\end{figure*}

Figs.~\labelcref{fig:slice_m2_V1h_all,fig:slice_m10_V21_all} show examples of major and minor mergers of our idealized cosmological simulations, respectively. They exhibit the evolution of the slices of the DM density, gas density, temperature, entropy, and scalar field $f_{\rm dye}$ in the $x-y$ plane (from the left column to the right) during the merger. Due to the gravitational torque, the merging clusters do not always move in the $x-y$ plane but stay close to it (see also Fig.~\ref{fig:traj_halo}). These figures show that mergers significantly disturb both the DM and gas distributions in galaxy clusters (see detailed discussions in Sections~\labelcref{sec:results:dm,sec:results:gas}). Various shocks and CDs are driven during the merger process (marked with the white and black arrows, respectively).

\bsp	
\label{lastpage}
\end{document}